\documentclass[aps,pre,twocolumn]{revtex4}
\usepackage{amsmath,bm,epsfig}
\usepackage[dvips]{color}
\usepackage[normalem]{ulem}
\definecolor{g-blue}{rgb}{0.83,0.95,1}
\definecolor{g-yellow}{rgb}{1,1,0.7}
\definecolor{g-green}{rgb}{0.9,1,0.9}
\definecolor{green}{rgb}{0,0.6,0}
\definecolor{cyan}{rgb}{0,0.7,0.7}
\definecolor{black}{rgb}{0,0,0}
\definecolor{grey}{rgb}{0.4 ,0.4 ,0.4 }

\newcommand{\B}[1]{{\bm{#1}}}
\newcommand{\C}[1]{{\mathcal{#1}}}    

\renewcommand{\sb}[1]{_{\text {#1}}}  
\renewcommand{\sp}[1]{^{\text {#1}}}  
\newcommand{\Sp}[1]{^{^{\text {#1}}}} 
\def\Sb#1{_{\scriptscriptstyle\rm{#1}}}

\def\Fbox#1{\vskip1ex\hbox to 8.5cm{\hfil\fboxsep0.3cm\fbox{%
  \parbox{8.0cm}{#1}}\hfil}\vskip1ex\noindent}  
\def\<{\left\langle}    \def\>{\right\rangle}
\def\({\left(}          \def\){\right)}
\def \[ {\left [} \def \] {\right ]}
\def\~{\widetilde}
\begin{document}
\title{{Structure   of  quantum vortex tangle} in   $^4$He  counterflow turbulence}  \author{Luiza Kondaurova$^{1}$,
Victor L'vov$^{2}$, Anna Pomyalov$^{2}$ and Itamar Procaccia$^{2}$}
\today

\affiliation{$^1$Institute of Thermophysics, Novosibirsk, Russia\\
$^2$Department
of Chemical Physics, The Weizmann Institute of Science, Rehovot
76100, Israel
}

\begin{abstract}
 The main goal of this paper is to present a comprehensive characterization of
well developed vortex tangles in a turbulent counterflow in quantum fluids (with a laminar normal fluid component). We analyze extensive numerical simulations using the vortex filament method, solving the full Biot-Savart equations for the vortex dynamics in a wide range of temperatures and counter-flow   velocities.  In addition to a detailed analysis of traditional characteristics such as vortex line density, anisotropic and curvature parameters of the vortex tangle, we stress other dynamical and statistical characteristics which are either much less studied or even unstudied. The latter include reconnection rates, mean mutual friction forces, drift velocities and the probability distribution functions of various tangle parameters: the loop length, the line curvature, the mean curvature of loops with a given length, etc.
During these studies we compare the three main reconnection procedures which are widely used in the literature, and
identify which properties are strongly affected by the choice of
  the reconnection criteria and which of them are practically
  insensitive to the reconnection procedure.  The conclusion is that  the vortex
  filament method in the framework of the Biot-Savart equation sufficiently robust and well suited for the
  description of the steady state vortex tangle in a quantum counterflow. The
Local-Induction Approximation to this equation may be successfully used to analytically establish
relationships between mean characteristics of the stochastic vortex tangle.

\end{abstract}
\maketitle
\tableofcontents
\section*{{Introduction}}

 The term ``quantum turbulence" or ``superfluid turbulence" refers to a
tangle of interacting {quantized} vortex lines, which are formed, for
example, in superfluid $^4$He, $^3$He or in Bose-Einstein condensates of
ultra-cold atoms.  The vorticity in superfluids is
restricted to a set of vortex lines around which the
circulation is quantized to multiples of the circulation quantum
$\kappa=h/m$. Here $h$ is Plank's constant and $m$ is mass of either
atoms with integer spin, like $^4$He or Cooper pairs of $^3$He
atoms.  The creation of sustained quantum turbulence can be achieved by either mechanical excitations \cite{exp1,exp2,exp3,exp4,exp5,exp6,exp7,exp8}, or by heat currents  (so-called counterflow turbulence). Experimental studies of thermal counterflow, initiated almost
sixty years ago by Vinen \cite{Vinen57,Vinen58}, became the most extensively
studied forms of quantum turbulence ~\cite{Wang87,ChildersTough76,Tough82,counter1,counter2,counter3}

In the context of the popular two-fluid model of superfluids the phenomenon of thermal counterflow may be considered as consisting of two {interpenetrating} fluid flows - a normal viscous component flowing in the direction of the
temperature gradient and carrying the heat flux, and an inviscid superfluid component flowing in the opposite direction to keep a zero total mass transfer. These two components may have different velocity and density fields.
 More sophisticated arrangements~\cite{Ladik-2012} allow one to realize (mechanically driven) pure super-flows in a relatively wide (7-10 mm) channel, with the normal fluid component practically at rest.
In both arrangements, a dense vortex tangle is excited under the influence of the velocity difference between the two components quantum turbulence.

Some statistical properties of quantum vortex tangles in  counter- and super-flows
were studied experimentally and numerically
\cite{Vinen57,Vinen58,Wang87,ChildersTough76,Tough82,counter1,counter2,counter3,Ladik-2012,Schwarz85,Schwarz88,AdachiTsubota10,deWaeleAarts94,Aarts94,Ashton81,GM41},
Among seminal contributions to
these studies we should mentioned again a pioneering
work by Vinen \cite{Vinen57,Vinen58}, in which he  also suggested a
phenomenological description of counterflow turbulence,
Eq. (17) and by Schwarz \cite{Schwarz88}, who established some important bridge-relationships between mean characteristics
of the vortex tangle and pioneered numerical simulations
of the counterflow turbulence mainly in the Local
Induction Approximation (see Sec. IIA2). Later Adachi,
Fujiyama, and Tsubota \cite{AdachiTsubota10} demonstrated that for an adequate
numerical study of the counterflow turbulence one
has to relax this approximation and to use the Biot-Savart
equation (10a), according to which each point of the vortex
line is swept by the velocity field produced by the
entire tangle. The development of the field was recently reviewed in \cite{SS-2012,TsubotaReview2013,NemReview2012}.

The intensity of quantum turbulence is usually characterized by the vortex line density per unit
volume, sometime referred as VLD and denoted by $\cal L$.  Another related characteristic is the intervortex distance $\ell\equiv 1/\sqrt{\cal L}$. Other properties of the tangle, such as the mean curvature, scale with $\ell$~\cite{Schwarz88}.

During temporal evolution vortex lines can collide and reconnect changing the tangle's topology. Thus vortex loops  can merge or break up into smaller loops. These reconnections occur on scales comparable with the vortex
core radius and were studied in the approximation of the
Gross-Pitaevskii
equation \cite{Samuels92,KoplicLevine93,BarenghiRec1,BarenghiRec2,Nazarenko2003,Ogawa2002}.
This approximation is adequate
when the vortex core radius  exceeds significantly the interatomic distance, like in $^3$He, but not
in the $^4$He. Recently the reconnection events were visualized experimentally \cite{PaolFisherLath10}.
Reconnections play a crucial role in the vortex dynamics. In particular they directly affect
the steady state value of $\ell$. In typical experimental conditions the vortex tangle is dense in the
sense that the intervortex distance $\ell$ is much smaller than the
characteristic size of the experimental cell $H$, which is about 1
cm. At the same time the tangle is sparse enough such that $\ell$ is
much larger than the vortex core radius $a_0$, which is of the order
of $10^{-8}$~cm in $^4$He.  To follow the evolution of the vortex
tangle at the intermediate scales $a_0\ll \ell \ll H $, Schwarz
\cite{Schwarz85,Schwarz88} proposed to use a vortex filament method (sometime referred to as VFM)
in which minor core variations of the quantized vortices are  ignored and the vortices are
approximated as directional lines with a predefined core
structure. If so, the time evolution of these vortex lines is governed
by the Biot-Savart equation ~\eqref{BSE}. Numerically vortex lines may be approximated as a set of small
straight vortex filaments described by a  directional set of
connected points placed at distances much smaller than  $\ell$.

 The  Biot-Savart equation for quantized vortices does not describe the vortex
reconnections.  They are included in vortex filament methods as an additional artificial
procedure that changes the connectivity of pairs of
points according to  some reconnection criterion.  Reconnection criteria
are based on a physical intuition and the results of numerical
simulations. Since the method introduction by Schwarz
\cite{Schwarz85,Schwarz88}, a number of different criteria
\cite{deWaeleAarts94,SamKiv1999,KivotidiesRec,Samuels92,KondaurovaNemir05,TsubotaNemir00,AdachiTsubota10}
were introduced and modified over time. Currently, three criteria are
frequently used to trigger the reconnections during the evolution of the
vortex tangle.  These are based either on geometrical proximity
\cite{TsubotaNemir00,AdachiTsubota10,Samuels92} or on
the dynamics of vortex filaments~\cite{KondaurovaNemir05,KondaurovaAndrNemir08,KondaurovaAndrNemir10},
leading to a different number of reconnections and various
changes in  the vortex tangle  topology.

 The presence of variety of artificial reconnection procedures in vortex filament methods
and the spread in values of basic characteristics of the tangle, such
as the vortex reconnection rates and steady-state vortex line density $\C
L$, resulted in the superfluid community sharing an opinion that was made
explicit recently by Skrbek and
Sreenivassan~\cite{SS-2012} : ``While it is clear that the full
Biot-Savart approach is certainly better [than the Local-Induction
Approximation (LIA), see below], there are still other aspects such as
approach to vortex reconnections and influence of possible normal
fluid turbulence that make the predictive power of these simulations
limited at the best."

In this paper we consider this strong statement as a research challenge,
turning it to our main question: ``To what extent can one state  that the statistical properties of the
developed vortex tangles obtained by vortex filament methods (in a wide range
of parameters) are robust under changes of the
reconnection procedures and other implementation details?". To answer this question we report in this paper the results of
comprehensive numerical simulations of counterflow turbulence for a wide
range of parameters: at low, medium and high temperatures $T$ (1.3,
1.6 and 1.9 K) and the counterflow velocities $V\sb{ns}$ ranging from
$0.3$ to $ 1.2$~cm/s. For all combinations of temperatures and
velocities we compared results of vortex filament method  with three different
reconnection
criteria~\cite{AdachiTsubota10,BarenghiRec2,KondaurovaAndrNemir10}.  We
found which properties are strongly affected by the choice of the reconnection
criteria (e.g. the reconnection rate differs more than in
order of magnitude for different criteria), which properties
only relatively weakly depend on this choice (such as mean properties of the tangle) and which are insensitive to
it (such as probability distribution functions of local properties).  Our results partially agree with preliminary observations by
Baggaley~\cite{Baggaley2012} who recently
compared the values of the vortex line density calculated with
a number of reconnection criteria for $T=1.6K$ and $0.35<V\sb{ns}<0.65\,$cm/s and concluded that the values of $\cal{L}$ are
insensitive to the choice of the criterion for these parameters.

The paper is organized as follows: in Section~\ref{s:Stat} we describe
mean and local statistical characteristics of vortex tangle beginning with a summary of the main notations and
 abbreviations used in the paper.

 Section~\ref{s:NM} is devoted to a brief overview of the vortex filament method. In particular,  in Subsec.~\ref{ss:WE} we  present the basic equation of the vortex line motion. In Subsec.~\ref{ss:CVR} we discuss the reconnection criteria and clarify in Subsec.~\ref{ss:ID} the implementation details.

 Our results are presented and discussed in Secs.~\ref{s:RD1}, \ref{s:stat} and \ref{ss:loc}.  In Sec.~~\ref{s:RD1} we consider the dynamics of the vortex tangle, including its evolution toward steady state and reconnection dynamics with different reconnection criteria. Here we also show the typical tangle configurations for different reconnection criteria.

 In Sec.~\ref{s:stat}
 we describe the mean characteristics of vortex tangle, starting in
 Subsec.~\ref{sss:VLD} with a detailed discussion of the vortex line density and its dependence on the temperature and counterflow velocity in comparison with results of other simulations and laboratory experiments. We also discuss the mean tangle anisotropy, the mean and RMS vortex line curvatures, the mean friction force between normal and superfluid components, the drift velocity of the vortex tangle and the mean and most probable loop lengths.
In some sense this level of
description is similar to the thermodynamical approach
to gases and fluids that deals with the mean characteristics
such as temperature, pressure, density, etc., averaged
over finite (physical) volume.

A more advanced description
of continuous media was reached in statistical
physics and kinetics in the framework  of probability
distribution functions, PDFs, (e.g. Maxwell-Boltzmann
PDF of atomic velocities) and correlation
functions (e.g. of atomic positions). Similarly, the measurable mean
characteristics of the vortex tangle  provide important but very limited
information on the tangle properties.

Clearly, PDFs and correlation functions are much
more informative and the theoretical description of quantum
turbulence definitely calls for such a knowledge, see,
e.g. review by Nemirovski~\cite{NemReview2012}. Unfortunately there is
not much chance that these details can be subject to experimental
study. Therefore numerical characterization
of detailed local vortex tangle statistics addressed in Sec. V is
important and timing. In particular we show that
the core of the PDF of vortex loop length, Eq~.\eqref{PDFsA}, and PDF
of line curvature, Eq.~\eqref{PDFsB}, have exponential form with linear
prefactor  [$ \propto x \exp{(-x)}$], while PDF of the mean-loop
curvature has a Gaussian form Eq.~\eqref{PDF} [$\propto  \exp{(-  x^2)}$].

We discuss also the correlation between
loop length and their mean curvature. Finally we
find the autocorrelation of the vortex-line orientation.

The concluding Sec.~\ref{s:Con} summarizes our view and results on the physical picture of $^4$He counterflow turbulence. It begins in Sec.~\ref{ss:idea} with a short discussion of
standard idealizations and their realizability that determine the set of relevant physical parameters of the problem.
In Sec.~\ref{ss:Vns}, employing dimensional reasoning and (where required) some simple physical arguments, we use the latter to describe the
dependence of the basic physical characteristics of the problem on the counterflow velocity. Next we present a detailed summary of our numerical results and list the actual numerical values of the corresponding dimensionless parameters which, according to na\"ive dimensional reasoning are expected to be of the order of unity.

 In the short Sec.~\ref{ss:Br} we recall relations that stem from the local induction approximation ~\cite{Schwarz88} that bridge the vortex line density, the mutual friction force  and the tangle drift velocity with the anisotropy and curvature parameters of the tangle. We summarize our results on realizability of these relations in numerical simulation with the full Biot-Savart equations.

In the next Sec.\,\ref{ss:PDFs} we summarize our results on various PDFs that characterize different aspects of the tangle statistics. The last Sec.~\ref{ss:Res} deals with the dependence of our numerical results on the reconnection criteria and  culminates with the optimistic statement: \\

At the current level of understanding of the vortex
statistics in counterflow turbulence, the vortex filament method numerical method in the framework of full Biot-Savart equation \eqref{eq:BSE} provides adequate
qualitative and reasonably accurate quantitative information on the quantum
vortex dynamics in superfluid turbulence. This information is required for the further
development of an adequate physical model of this intriguing phenomenon.

Although local induction approximation to Eq. \eqref{eq:BSE}
fails to reproduce accurately vortex
tangle properties in numerical studies \cite{AdachiTsubota10}, we demonstrate
that analytical relationships between different mean characteristics
of the vortex tangles, found in \cite{Schwarz88} within the LIA framework, are well obeyed in our Biot-Savart similatons.
Therefore we think  that the Local Induction Approximation may be effectively used in analytical theory of counterflow turbulence.

\section{\label{s:Stat} Statistical description  of the vortex tangle}

\begin{figure}
\includegraphics[width=8 cm]{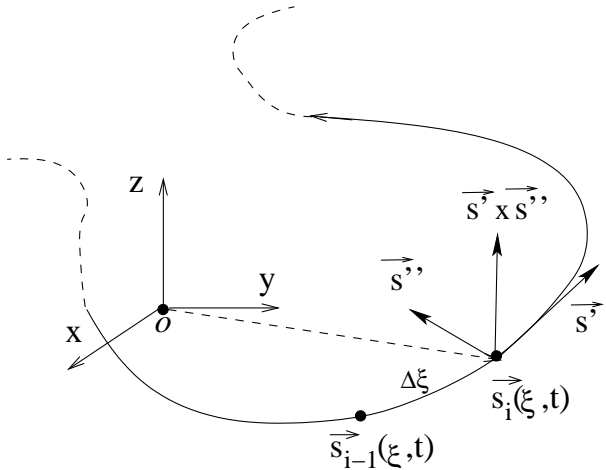}
\caption{\label{f:1}The coordinate system. The origin of the
Cartesian system is placed at the center of the computational
box. Each vortex line point ${\bm s}_i$ is defined by Cartesian coordinates
$x_i,y_i,z_i$ and a label $\xi$ along the line. Vectors ${\bm s'}$,
${\bm s''}$ and ${\bm s'} \times {\bm s''}$ are the tangential, the
local curvature vector and the direction of the local induced
velocity, associated with the point ${\bm s}(\xi)$ of the vortex
filament, respectively. Primes denote differentiation with respect to
the instantaneous arclength $\xi$. }
\end{figure}

\subsection{\label{ss:abbr} Abbreviations and main notations}
\begin{description}
\item LHS \&  RHS -- left-  and right-hand side of equation;

 \item BSE -- Biot-Savart equation~\eqref{BSE};

\item VFM -- vortex filament method, Sec.~\ref{s:NM}

\item LIA -- local induction approximation, Sec.~\ref{sss:LIA}, Eqs.~\eqref{eq:LIA};
\item GC --  Geometrical reconnection criterion, Sec.~\ref{sss:G},   Eq.\,\eqref{G-BSE};
    \item GEC -- Geometric-energetic reconnection criterion, Sec.~\ref{sss:G};
\item DC --  Dynamical  reconnection criterion, Sec.~\ref{sss:D};

\item $l_j$ --  length of particular $j$-loop;
\item $L\sb{tot}=\sum\limits _{j=1}^N l_j$ and $\overline L =L\sb{tot}/N$ -- total and mean  length of   the vortex tangle, consisting of $N$ loops;
 \item VLD -- vortex line density of the tangle, occupying volume $\C V$,   $\C L=L\sb{tot}/\C V$, Eq.\,\eqref{defL};
\item $\ell\equiv 1/ \sqrt{\C L}$ -- mean intervortex distance, Eq.\,\eqref{defL};

\item  ${\bm V}_{\rm s}$, ${\bm V}_{\rm n}$ and ${\bm V}_{\rm ns}={\bm
V}_{\rm n}-{\bm V}_{\rm s}$-- the macroscopic  velocities of super-fluid, normal fluid and counterflow, respectively;

\item $\B V\sb{vt}$ -- mean drift velocity of the vortex tangle with respect to the superfluid rest frame;

\item $\B s(\xi)$ -- Cartesian coordinate of the vortex line, parameterized with the arc-length $\xi$,  Fig.\,\ref{f:1};
    \item      $\B s'(\xi)\equiv d \B s/ d\xi$ -- local direction of the vortex line, Fig.\,\ref{f:1};
     \item   $\B s''(\xi)\equiv d^2 \B s/ d\xi^2$ -- local curvature vector,  Fig.\,\ref{f:1};

\item $\delta(t)$-- smallest distance between two vortex lines, approaching reconnection;
\item PDF  -- probability  distribution function $\C P(x)$ of some local characteristics $x$, normalized such that $\int \C P(x)\, dx=1$;
\item $\C P(l)$ \& $\C P(|s''|)$ --  PDFs of the loop length and  of the line curvature
\item  $\overline X \equiv \langle x \rangle =  \int x \C P(x)\, dx$--Mean value of $x$;
\item $\widetilde X\equiv \sqrt{\langle x^2\rangle}$--RMS, Root-mean-square value of $x$ ;

\item $L_*$ -- Most probable value of the loop length, \\ $d\, \C P(l)/ dl \big \vert_{l=L_*}=0$.
\item $ \overline S\equiv \langle |s''| \rangle$, \quad $ \widetilde S\equiv \sqrt{\langle |s''|^2 \rangle}$ -- mean and RMS  vortex curvature, Eqs.~\eqref{curv};
\item  $R\equiv 1/\widetilde S$  --   characteristic  radius of curvature of the vortex tangle;
\item $a_0=1.3 \times 10^{-8}$ cm -- effective core radius of $^4$He;
\item $\kappa=9.97 \times 10^{-4}$ cm$^2$/s-- circulation quantum;
\item $\Lambda=\ln \Big[c R/a_0\Big]\,, c=O(1),\widetilde \Lambda=\Lambda/4
\pi $,- defines the stiffness of the vortex line with respect to bending, Eq.~\eqref{eq:LIA};
\item $\beta=\kappa \widetilde  \Lambda$- quantify the local contribution to the line point velocity, Eq.~\eqref{eq:LIA}.
\end{description}
\subsection{\label{ss:glob}  Statistical characteristics of  the vortex tangle}
  \subsubsection{\label{sss:VLD} The vortex line density $\C L$ and the parameter $\gamma$}
Denote the total vortex length in the tangle occupying a volume $\C V$ as $L\sb{tot}$. Then the vortex line density $\C L$ and mean intervortex distance $\ell$ can be found as:
 \begin{eqnarray}\label{defL}
L\sb{tot}=\int \limits _{\C C}d\xi\,, \quad \C L\equiv L\sb{tot}/\C V\,, \quad \ell\equiv 1/\sqrt{\C L}\ .
 \end{eqnarray}
Here the integral is taken over the whole vortex configuration $\C C$.

Asserting that $\kappa$ ($[\kappa]=$cm$^2$/s) is the only relevant parameter in the problem and $[\C L]=$1/cm$^{-2}$ one can employ the counterflow velocity $V\sb{ns}$ in a dimensional argument to write $\sqrt {\C L}\sim V\sb{ns}/\kappa $ or:
\begin{subequations}\label{gamma}\begin{equation}\label{gammaA}
\sqrt{{\cal L}}= \frac{\Gamma}{ \kappa}\,  V\sb{ns}\,,  \qquad \gamma\equiv \frac{\Gamma}{ \kappa}\ .
\end{equation}
Here $\Gamma$ s a dimensionless parameter which in general is temperature dependent.
Na\"ively one expects that $\Gamma$ is of the order of unity.
Numerical and experimental studies (see e.g. our Tab.~\ref{t--3})
give $\Gamma\sim 0.1$.

It is customary to use in relation~\eqref{gammaA} a dimensional parameter  $\gamma(T)$ instead of $\Gamma$. The parameter  $\gamma$ is the subject of intensive experimental, numerical and theoretical studies and will be discussed  in details in Sec.~\ref{sss:gamma}.

In practice most experimental and numerical data of the time averaged steady
state value of $ {\cal L}$ are
approximated by a slightly different form of this equation\cite {Tough82}:
\begin{equation}\label{gam}
 {\cal L} =\gamma^2  \big(V_{\rm ns}-v_0\big)^2\,, \quad
\end{equation}\end{subequations}
which includes an additional fitting parameter, the intercept velocity
$v_0$.

 \subsubsection{\label{sss:RD}Reconnection dynamics and parameter $c_{\rm r}$}
The reconnections between vortex lines lead to the development of a
steady state vortex tangle. The statistics of the reconnections is
therefore important for characterizing the tangle. In a
periodic box only two kinds of reconnection are possible: one vortex
loop splits into two smaller loops, or two loops merge into one larger
loop. The ratio of the number of reconnection
of two types (in a unit volume), $N_1/N_2$ is shown in Fig.~\ref{f6}.

The second  important characteristic of the vortex dynamics is the total reconnection rate $dN\sb r/dt$ ($N\sb r\equiv N_1+N_2$) in a unit volume.  In the steady state the relation between mean reconnection rate $\langle dN\sb r\rangle /dt$ and $\C L$ can be found by a simple dimensional argument:  $[\langle dN\sb r\rangle /dt]=$cm$^{-3}$s$^{-1}$ may be uniquely expressed via $[\kappa]=$cm$^2/$s and $[\C L]=$cm$^{-2}$:
\cite{TsubotaNemir00,BarenghiSam04,Nem2006,Poole2003} as
\begin{equation}\label{eq:dNr}
\frac{dN_r}{d t}= c\sb r \kappa {\cal L}^{5/2}\ .
\end{equation}
Here $c\sb r$ is a temperature dependent dimensionless coefficient. One sees in Fig.~\ref{f:7} that the relation~\eqref{eq:dNr} is perfectly obeyed  in our simulations but the numerical values  of $c\sb r$, given  in Tab.~\ref{t--2}, crucially depend on the reconnection criteria.  The reasons and consequences of this fact for the final steady state tangle are discussed  in Sec.~\ref{ss:RR}.

 \subsubsection{\label{sss:AVT}Anisotropy of the vortex tangle and the indices $I_{||}, I_{\bot}, I_{\ell},I_{\ell\bot} $}
The presence of the counterflow velocity creates a preferred
direction and the vortex tangle is anisotropic.  To measure the degree of
anisotropy of the tangle Schwarz \cite{Schwarz88}  introduced the
anisotropy indices:
\begin{subequations}\label{aniz}
\begin{eqnarray} \label{Ipar}
I_{\|}&=&\frac{1}{ L\sb{tot}}\int\limits_{\C C} [1-({\bm s}' \cdot
\hat {\bm r}_{\|})^2]d \xi \, ,\\
 \label{Iper}
I_{\bot}&=&\frac{1}{ L\sb{tot}}\int\limits_{\C C} [1-({\bm s}'
\cdot \hat {\bm r}_{\bot})^2]d \xi \, ,\\ \label{Iell}
I_{\ell}&=&\frac{\ell }{ L\sb{tot}}  \int\limits_{\C C} \B {\hat r}_\|\cdot
 ({\bm s}' \times {\bm s}'') d \xi \, ,
\end{eqnarray}
where $\hat {\bm r}_{\|}$ and $\hat{\bm  r}_{\bot}$ are unit vectors in the direction parallel and perpendicular to ${\bm V}_{\rm ns}$, respectively. In the steady state these indices averaged over time obey relation $ I_{\|}/2+ I_{\bot}=1 $. The index $I_\ell$ measures the mean local velocity  (in unites $\kappa/\ell$)  in the direction of the countflow. In the isotropic case $I_{\|}= I_{\bot}=2/3,I_{\ell}=0$.

To test the isotropy of the velocity in the direction perpendicular of the counterflow, we also measure
\begin{eqnarray} \label{Ilp}
I_{\ell\bot} &=&\frac{\ell }{ L\sb{tot}}   \int\limits_{\C C} \B {\hat r}_\perp\cdot
 ({\bm s}' \times {\bm s}'') d \xi \, ,
\end{eqnarray}\end{subequations}
which is expected to vanish if the velocity is isotropic in the plane perpendicular to the counterflow velocity, even if $I_{\ell}$ is not small. Our results for the dimensionless anisotropy indices are given in Tab.~\ref{t:4} and discussed in Sec.~\ref{sss:aniz}.

 \subsubsection{\label{sss:Curv}Mean,  RMS curvatures $\overline S$, $\widetilde S$ and parameters $c_1$, $c_2$}

Other important global properties of the vortex tangle are  the
mean and RMS curvatures $\overline S$ and   $\widetilde S$,  which may be expressed as an integral over
the whole vortex configuration $\C C$, occupying a volume  $\cal V$:
\begin{subequations}\label{curv}\begin{eqnarray}\label{curvA}
\overline S &\equiv& \langle |s''|\rangle=\frac{1}{  L\sb{tot}} \int\limits_{\C C}| s''| d \xi \, ,\\
\widetilde S^2&\equiv& \langle |s''|^2\rangle= \frac{1}{  L\sb{tot}} \int\limits_{\C C}| s''|^2 d \xi \ .
 \end{eqnarray}
These objects are expected to scale with the mean density as~\cite{Schwarz88}:
\begin{eqnarray}\label{sL}
\overline S=  c_1  \sqrt{{\cal L}}\, ,\qquad
\widetilde S = c_2  \,\sqrt{{\cal L}}\,,
\end{eqnarray}
 where $c_1$ and $c_2$ are dimensionless constants of the order of unity (see below Tab.~\ref{t:4}).
\end{subequations}

 Similarly we can find the mean and RMS curvature $\overline {s'' _j}$ and  $\widetilde {s'' _j}$ of a particular vortex loop ${\C C_j}$ of length $l_j  \equiv   \int\limits_{\C C_j} \!\!d \xi $:
 \begin{subequations}\label{jcurv}\begin{eqnarray}\label{jcurvA}
\overline {s'' _j} &\equiv& \langle |s''|\rangle_j=\frac{1}{l_j} \int\limits_{\C C_j}| s''| d \xi \, ,\\ \label{jcurvB}
\widetilde {s'' _j}^2&\equiv& \langle |s''|^2\rangle_j= \frac{1}{l_j} \int\limits_{\C C_j}| s''|^2 d \xi \ .
\end{eqnarray}\end{subequations}
 The global (over the entire tangle) PDF of  $|s''|$ and the PDFs of the vortex-loop length, $l_j$, the mean-loop curvature, $\overline{s''_j}$ and the correlations between   $l_j$ and $\widetilde{s''_j}$ are presented and discussed in Sec.~\ref{ss:loc}.

\subsubsection{\label{sss:VVT}Drift velocity of the vortex tangle $V\sb{\rm vt}$  and parameter $C\sb{\rm vt}$ }
The drift velocity of the vortex tangle with respect to the superfluid rest frame is
\begin{subequations}\label{drift}
\begin{equation}\label{DV}
\B V\sb{vt}=\frac1{L\sb{tot}}\int\limits _{\C C}\frac{d \B s(\xi)}{dt} d\xi-\B V\sb s\,,
\end{equation}
where  the  velocity of the vortex line point $d \B s(\xi)/dt$ is given below by Eq.\,\eqref{eq:s_Vel}.
It is natural to expect that  $\B V\sb{vt}$ is proportional to the counterflow velocity $\B V\sb{ns}$ and to introduce a dimensionless parameter $C\sb{vt}$ as their ratio:
\begin{equation}\label{Kvt}
\B V\sb{vt}= C\sb {vt} \B V\sb{ns}\ .
\end{equation}\end{subequations}
The values of $C\sb{vt}$ are  discussed in Sec.~\ref{sss:drift}.

 \subsubsection{\label{sss:FF}Friction force density  and the Gorter-Mellink constant}
 In  discussions of the mechanical balance in superfluid turbulence an important role is  played by the mutual force density exerted by the normal fluid on the superfluid. It may be found from Eq.\,\eqref{eq:s_Vel}  (the term proportional to  $\alpha'$ vanishes by symmetry)\,\cite{Schwarz88}
 \begin{subequations}\label{Fns}\begin{equation}\label{FF}
 \B F\sb{ns}=  \rho\sb s\kappa\, \alpha J \,, \
 J\equiv -\frac{1}{\C V}\int\limits_{\C C}
 \B s'\times [ \B s' \times (\B V\sb{ns}-\B V\sb{si})] d \xi\ .
 \end{equation}
 The integral $J$ [with dimensions $[J]=1/$(s\,cm)] may be uniquely expressed via $\kappa$ and $V\sb{ns}$ as $V\sb{ns}^3/\kappa^2$. This leads to the dimensional estimate for $F\sb{ns}$
 \begin{equation}\label{FnsB}
 F\sb{ns}=\frac{\alpha \rho\sb s}\kappa (C_f V\sb{ns})^3\,,
 \end{equation}
 with a dimensionless temperature dependent constant $C_f $.
This agrees with the Gorter-Mellink~\cite{GM41} result that reads $F\sb{ns}\propto V\sb{ns}^3$:
\begin{equation}\label{GM}
F_{\rm ns}=A\Sb {GM} \rho_s \rho_n V_{\rm ns}^3\, .
\end{equation}
Comparing equations \eqref{FnsB} and \eqref{GM}  one finds the relationship between  $C_f$ and the dimensional Gorter-Mellink constant $A\Sb {GM}$:
\begin{equation}\label{GMc}
A\Sb {GM}=C_f^3 \,\frac{ \widetilde \alpha}{\kappa \rho}\,, \quad  \widetilde \alpha \equiv \alpha  \rho /\rho  \sb n\ .
\end{equation}
\end{subequations}
As is known, the density $\rho$ of $^4$He   varies only weakly with the temperature in the relevant temperature range, while $\alpha$ varies rapidly.  It increases 6 times as $T$ grows from 1.3 to 1.9\,K, see Table~\ref{t:1}. On the other hand, the temperature dependence of the parameter $\widetilde\alpha$, that actually governs the temperature dependence of $A\Sb{GM}$, is much weaker than $\alpha$. Further discussion of the friction force density is given below in Sec.~\ref{sss:fric}.

 \subsubsection{\label{sss:Auto}Autocorrelation  of the  vortex orientations}
To test the relative polarization of the vortex lines we measure an orientation correlation function
\begin{equation}\label{Kr}
 K({\bm r}_1- {\bm r}_2)=\Big \langle {\bm s}'( r_1)\cdot {\bm s}'( r_2) \Big\rangle_{\C C}\,,
\end{equation}
where $ \B r_1$ and $ \B r_2$ are the Cartesian coordinates of the two line
points and we average over all pairs of the line points in the
tangle. $K({\bm r}_1- {\bm r}_2)$ measures the average angle between
line segments as a function of the distance between them. Averaged
over all distances it quantifies the polarization of the tangle $\overline K$.

\section{\label{s:NM}Vortex Filament  Method }

 The vortex filament method and the reconnection criteria were
presented in details, e.g. in
Refs.~\cite{Schwarz85,Schwarz88,deWaeleAarts94,Aarts94,AartsThesis,AdachiTsubota10,Samuels92,KondaurovaAndrNemir10,Baggaley2012}. Nevertheless,
to keep the paper self-contained,  and  to introduce notations and
definitions, we review these criteria with the focus on the underlying  physical processes. The basic equations are presented in Sec.~\ref{ss:WE} and the
criteria of vortex reconnection are discussed in Sec.~\ref{ss:CVR}. A short description of the
implementation details is given in Sec.~\ref{ss:ID}.

\subsection{\label{ss:WE}Basic Equations and their implementation}

\subsubsection{\label{sss:BSE} Equations of motion of the vortex line}
When no external forces act on the vortex core the vortex
line moves with the velocity ${\bm V}_{\rm si}({\bm
s})$ defined by the entire vortex tangle according to the Biot-Savart
equation:
\begin{subequations}
\begin{eqnarray}\label{BSE}
 {\bm V}_{\rm si}({\bm s}) =  \frac{\kappa}{4\pi}\int_{ \C C}
 \frac{({\bm s_1}-{\bm s})\times d s_1}{|{\bm s_1}-{\bm s}|^3}\ .
\end{eqnarray}
Here the vortex line is presented in a parametric form $s(\xi,t)$, where $\xi$
 is an arclength, $t$ is the time and the integral is taken over the entire vortex tangle configuration.

In addition to the self-induced velocity of the superfluid component,
 we have to account for the interaction with the normal component via
 mutual friction, characterized by two dimensionless temperature
 dependent parameters $\alpha$ and
 $\alpha'$~\cite{Schwarz85,Schwarz88}:
\begin{eqnarray}\label{eq:s_Vel}
\frac{d{\bm s}}{dt} &=&{\bm V}_{\rm s}+{\bm V}_{\rm si}+\alpha  {\bm s}'\times ({\bm V}_{\rm ns}-{\bm V}_{\rm si})\\ \nonumber
&-&\alpha' {\bm s}' \times\Big[{\bm s}' \times({\bm V}_{\rm ns}-{\bm V}_{\rm si})\Big] +{\bm v}_{\rm bc}\ .
\end{eqnarray}\end{subequations}
Here ${\bm V}_{\rm s}$ is the macroscopic super-fluid velocity,   and   the counterflow velocity
 ${\bm V}_{\rm ns}={\bm
V}_{\rm n}-{\bm V}_{\rm s}$ is the relative velocity of the superfluid
component. In the reference frame co-moving
with the superfluid component, $V_{\rm s}=0$ and the relative velocity
equals to the velocity of normal fluid ${\bm V}_{\rm ns}={\bm V}_{\rm
n}$  in this reference frame.  In our simulations ${\bm V}_{\rm ns}$ is oriented towards the
positive $z$-direction. The prime in $\B s'$ denotes derivative with
respect to the instantaneous arc-length $\xi$, e.g.  $\B s'=d\B s/d\xi$.
The mean velocities obey a mass conservation law $\rho_s
{\bm V}_s+\rho_n{\bm V}_n=0$, where $\rho_n$ and $\rho_s$ are the
densities of normal and superfluid components respectively. The density
$\rho=\rho_n+\rho_s$ refers to the density of $^4$He.  The
 term ${\bm v_{\rm bc}}$ describes the influence of the boundary
 conditions.  For the periodic boundary conditions used in this
 work, the line points leaving the box from one side were
 algorithmically brought back to the computational volume by
 appropriately shifting their coordinates without changing their velocity
 $\dot {\B s} (z)=\dot {\B s}(z+H)$,\   $\dot {\B s}(y)=\dot {\B s}(y+H)$,\   $\dot {\B s}(x)=\dot {\B s}(x+H)$,  where $H$ is the  size of the computational domain.

\begin{figure*}
\begin{tabular}{|c|c|}
  \hline
$  A$ \includegraphics[width=8 cm]{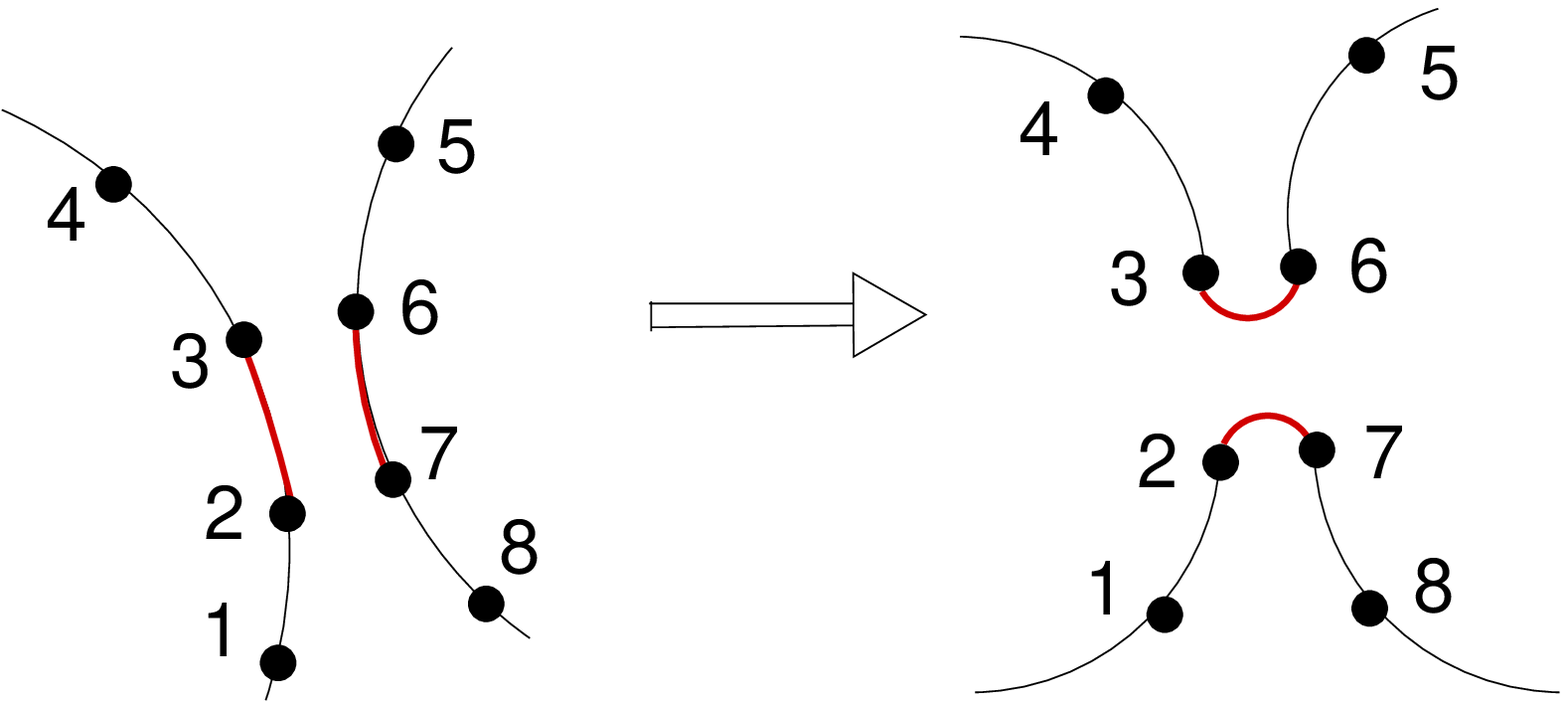} ~~~   &
B~~~\includegraphics[width=8 cm]{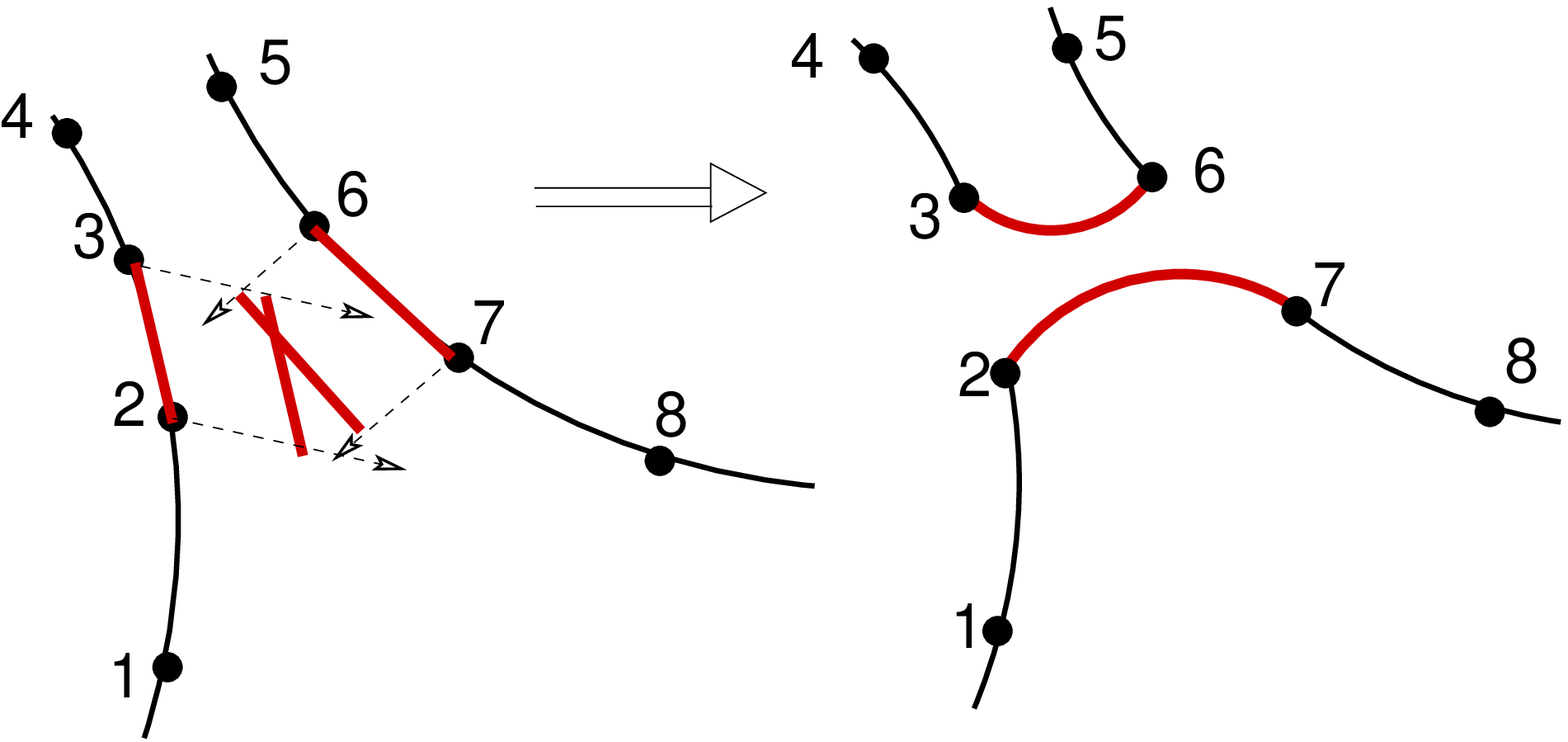}~ \\
  \hline
\end{tabular}

\caption{\label{f:Grec} Topology change for the geometric (GC) and
geometric-energetic reconnection criteria (GEC) (Panel $ A$)
and ``dynamical reconnection criterion (DC) (Panel B).  The GC
requires $l_{2,7}<\Delta \xi$  regardless the value of other
distances, while GEC requires  $l_{2,7}<\Delta \xi$ and, in
addition, $l_{2,3}+l_{6,7}>l_{2,7}+l_{3,6}$.  The DC requires that the
segments (2,3) and (6,7), moving with constant speed, will cross in
space during next time step.  }
\end{figure*}


 \subsubsection{\label{sss:LIA} Local Induction Approximation }
 Eqs~\eqref{BSE} implies that the vortex line is infinitely thin.
 Attempting to calculate the velocity at a particular point $\B s$ on
the vortex line one finds that the integral logarithmically diverges
as $\B s_1\to \B s$. To resolve this diffculty one has either to
cut the integral at $|\B s_1- \B s|=a_0$ or to  account for the
particular form of the vortex core structure. Physically it means that
$\B V\sb{si}(\B s)$ in the integral~\eqref{BSE} is dominated by the
local contributions from the vortex line for which $  a_0 \le |\B s_1-
\B s|\le c R$ The upper limit of integration is about the mean curvature
of the tangle $R$ determined up to a
dimensionless constant $c$ of the order of unity. Neglecting nonlocal
contribution one arrives to the Local Induction Approximation (LIA)\cite{HamaLIA63,Schwarz82}:
\begin{subequations}\label{eq:LIA}
 \begin{eqnarray}\label{LIAeq}
 \B V\sb{si}\Sp{LIA}&=& \beta  \B s^\prime \times \B s^{\prime\prime}\,,
 \quad \beta \equiv \kappa \widetilde \Lambda\,,\quad  \widetilde {\Lambda} \equiv \frac{\Lambda}{4\pi}\,, \\
\label{betaLIA}
  \Lambda&=& \ln
 \Big [\frac{c R}{a_0} \Big ] \approx
  \ln\Big [\frac{\ell }{a_0} \Big ]\ .
 \end{eqnarray}
\end{subequations}
The value of the ratio of mean local to mean nonlocal contributions to the velocity
is about $\Lambda$.  Besides the traditional  parameter $\Lambda$ we introduce also a frequently used combination  $\widetilde {\Lambda}$. The values of $\widetilde {\Lambda}$ found numerically are very close to unity, see Tab.~\ref{t:1}.

Notice that Eq.\,\eqref{LIAeq} is integrable, having an infinite
number of integrals of motion, including the total line
length. Therefore numerical simulations with the full BSE~\eqref{BSE} are
not a question of accounting for a small (about 10\%) nonlocal
contributions to the line velocity, but are required by necessity to
account for the violation of infinitely many conservation laws.

Nevertheless one can exploit the fact that the local contribution
\eqref{LIAeq} to the vortex velocity does dominate
the non-local one and to use the simple local relation
\eqref{LIAeq} in analytical studies of the vortex tangle characteristics,
for example, in the way developed by Schwarz \cite{Schwarz88}.
He established a  set of bridge
relations between different mean characteristics of the
vortex tangle. In Secs. \ref{sss:phen-an}, \ref{sss:drift} and \ref{sss:fric} we demonstrate
that these relations are well obeyed by
the mean vortex characteristics  found directly from numerical simulations in the framework of the VFM with full Biot-Savart equations.

\subsubsection{\label{sss:FullBSE} Implementation of the full Biot Savart velocity }

To implement  the Biot-Savart equations in the vortex filament methods we discretize the parametric curve by
 a large and variable number of points $s_i, i=1 \dots N$ at initial space
 resolution $\Delta \xi$, see Fig \ref{f:1}.
Then the velocity of the point $\B s$ is given by Eqs. \eqref{BSE}
 and desingularized according to Schwarz \cite{Schwarz85}:
\begin{eqnarray}\label{eq:BSE}
 {\bm V}_{\rm si}({\bm s})&=&\beta\Sb{VFM}~ {\bm s}' \times {\bm s}''+ \frac{\kappa}{4\pi}\int_{ \C C}
 \frac{({\bm s_1}-{\bm s})\times d s_1}{|{\bm s_1}-{\bm s}|^3}\,, \\\nonumber
\beta\Sb{VFM}&=&  \frac{\kappa}{4\pi}\ln\Big[\frac{2\sqrt{l_+ l_-}}{e^{1/4}a_0}\Big]\ .
\end{eqnarray}

The integral accounts for the influence of the whole vortex
configuration $\cal C$, excluding the segments adjacent to ${\bm
s}$. Here ${\bm s_1}$ is a the point on the filament.  The contribution of
the line elements adjacent to ${\bm s}$ is accounted for by the local
term $\beta\Sb{VFM}~ {\bm s}' \times {\bm s}'' $. Here $l_{\pm }$ are the
lengths of two line elements connected to ${\bm s}$, $e=2.71...$ is
the base of natural logarithm and $e^{1/4}$ corresponds to the
arbitrary chosen Rankine model of the vortex core
\cite{RayfiledReif64}.

 The distances between adjacent line points change during
 evolution. The space resolution affects the accuracy of the
 derivatives $\B s'$ and $\B s''$\cite{AartsThesis}. To keep $l_\pm$
 of the same order of magnitude we remove a line point whenever two
 points come closer than $\Delta \xi_{\rm min}$ and add a point by a
 circular interpolation \cite{Schwarz88} if the distance between two
 adjacent points become larger than $ \Delta \xi_{\rm max}$. Here $\xi_{\rm min}$ and $\xi_{\rm max}$ are the chosen smallest and largest interpoint distances.


\subsection{\label{ss:CVR} Criteria of vortex reconnection}
In vortex filament methods the reconnections are introduced algorithmically.  When
some criterion is satisfied the vortex line topology is changed
as shown in Figs.~\ref{f:Grec}.  These criteria are based on numerous
studies of the vortex reconnections in the framework of the Biot-Savart and
the Gross-Pitaevskii equations and on the resulting physical intuition.

\subsubsection{\label{sss:RC} Schwartz's geometrical criterion  in LIA}
 Historically the first criterion was suggested by
 Schwarz~\cite{Schwarz85,Schwarz88} in the context of the local induction approximation. He noticed
 that when two vortices approach each other closer than 2 $ R/\Lambda$ (the distance at which the self-induced velocity, given by Eq.~(\ref{eq:LIA}), is of the order of the non-local
 contribution) the vortex-vortex interaction dominates the local contribution, which in the framework of Eqs. \eqref{BSE} leads to a local instability. During this
 process the velocity field of each vortex deforms the other in such a way that
 the vortices are moved toward each other and finally reconnect.
Clearly, all this dynamics cannot be captured by the local induction approximation, which completely
 ignores the intervortex interactions. Thus Schwarz suggested a criterion that can be referred to as a ``geometric criterion"
 for the local induction approximation, or LIA-GC: the vortices are reconnected when they approach each other closer than the minimal distance  $\delta\Sb{LIA-GC}$
\begin{equation}\label{G-LIA}
\delta< \delta\Sb{LIA-GC} \equiv 2 R  /\Lambda\,,
 \end{equation}
 i.e. the distance at which the nonlocal interactions exceed the local interactions.

\subsubsection{\label{sss:G} Other geometric criteria for full Biot-Savart equations}

In the framework of Biot-Savart equations the LIA-GC criterion leads to many spurious reconnections.
 On the other hand, conceptually these equations provide an adequate description of the vortex
 dynamics in the reconnection processes up to the stage when
 $\delta\sim a_0$.  Therefore the vortex filament method with the full Biot-Savart equations describes the vortex line
 motion for distances limited by its resolution $\delta> \Delta \xi$.

 During the last decade several reconnection criteria were
proposed, in which the closeness of the reconnecting points were
related to the space resolution with or without additional physical
requirement. Similar to \cite{Baggaley2012} we consider here two such criteria.

 A natural extension of  LIA-GC~\eqref{G-LIA} was suggested in Refs.~\cite{TsubotaNemir00,AdachiTsubota10}:
\begin{equation}\label{G-BSE}
\delta< \delta\Sb{BSE-GC}\simeq \Delta \xi\,,
 \end{equation}
 By analogy with LIA-GC criterion~\eqref{G-LIA} we  call this rule
``BSE-geometrical criterion" (BSE-GC), see Fig.\,\ref{f:Grec}$\,\C A$.

 Unfortunately the simple BSE-GC~\eqref{G-BSE} ignores energy
dissipation during reconnection events, e.g. due to phonon emission.
Since the vortex length approximates the kinetic energy of the tangle,
it cannot increase during reconnections. A more restrictive criterion was suggested in
Ref.~\cite{Samuels92}, requiring a total vortex line length reduction in
addition to the  geometrical proximity Eq.\,\eqref{G-BSE}. We will refer
to this criterion as to BSE ``geometric-energetic" criterion
(BSE-GEC). In the present work we deal only
with full Biot-Savart simulations and therefore we skip hereafter the notation``BSE-" from the
reconnection names and abbreviate them shortly as GC and GEC (or G-criterion and GE-criterion).

\subsubsection{\label{sss:D} Dynamical  criterion}
The authors of Refs.~\cite{KondaurovaNemir05,KondaurovaAndrNemir08,KondaurovaAndrNemir10}
 approached the problem of reconnection criterion completely differently, by considering the dynamics of vortex line points.
 Their approach is equally applicable to the local induction approximation as well as the Biot-Davart dynamics. Under the assumption that both ends of a
line segment are moving at the same velocity during a time step, the
reconnection is carried out if the reconnecting line segments cross in space
during the next time step. We will refer to this criterion as the "dynamical" criterion
(DC or D-criterion). Note that unlike   GC  and GEC, the  DC
involves reconnecting segments and not points. The assumption of the
same velocity of the two ends of a segment implies sufficiently
high space resolution (small values of $\Delta \xi$) -see Fig.\,\ref{f:Grec}\,B.

To find whether the line segments will meet during the next time step, the  set of equations
\begin{eqnarray}\label{K}
&&{\bm s}_i + V({\bm s}_i)\tau+ ({\bm s}_{i+1}-{\bm s}_i)\theta=\\  \nonumber
&& {\bm s}_j + V({\bm s}_j)\tau+({\bm s}_{j+1}-{\bm s}_j)\phi
\end{eqnarray}
is solved for $ 0\leq \theta\leq 1;\quad 0\leq \phi\leq 1;\quad 0\leq \tau
\leq\Delta t$. If such a solution is found, the segments will collide.
Here  ${\bm s}_i=(x_i,y_i,z_i)$, ${\bm s}_{i+1}=(x_{i+1},y_{i+1},z_{i+1})$ and ${\bm s}_j=(x_j,y_j,z_j)$, ${\bm s}_{j+1}=(x_{j+1},y_{j+1},z_{j+1})$ (in Cartesian coordinates) denote the first and the second reconnecting
pairs of points and $\Delta t$ is the time step. The velocities $V({\bm s}_i)$ and $V({\bm
s}_j)$ remain the velocities of the line points ${\bm s}_i$ and
${\bm s}_j$. Alternatively, the velocities of the midpoint of the
segments ($i, i+1$) and ($j, j+1$) may be used.

 \begin{table}
\begin{tabular}{| c||c|c|c|}
\hline
 $T$, K& 1.3 & 1.6  & 1.9  \\
 \hline\hline
$\alpha$ & 0.036  & 0.098 & 0.210  \\
\hline
  $\alpha'$ & 0.014 & 0.016 & 0.0009 \\
 \hline
   $\rho\sb n/\rho$&0.045& 0.162 & 0.420\\ \hline\hline
 $\alpha\rho/\rho\sb n$&0.8& 0.6 & 0.5\\ \hline
  $\widetilde \Lambda\equiv \Lambda/(4\pi)$ &1.05 & 1.03& 1.02 \\
\hline
\end{tabular}\caption{\label{t:1}Friction parameters $\alpha$ and $\alpha'$ used in simulations,  relative density of the normal component \cite{DonnelyBarenghi98}, combination $\alpha\rho/\rho\sb n$ [which  weakly depends temperature and is responsible for the mutual friction density in Eq.~\eqref{GMc}] and LIA parameter $\widetilde \Lambda$ calculated for $c=1$.}
 \end{table}

\begin{figure*} [t]
\begin{tabular}{ c c c }
 $  A$: $T=1.6\,$K,  $V\sb{ns}=1\,$cm/s  & $  B$: $T=1.3\,$K, different $V\sb{ns}$& $ C$: $V_{\rm ns}=0.5\,$cm/s, different $T$ \\
 \includegraphics[width= 5.8 cm]{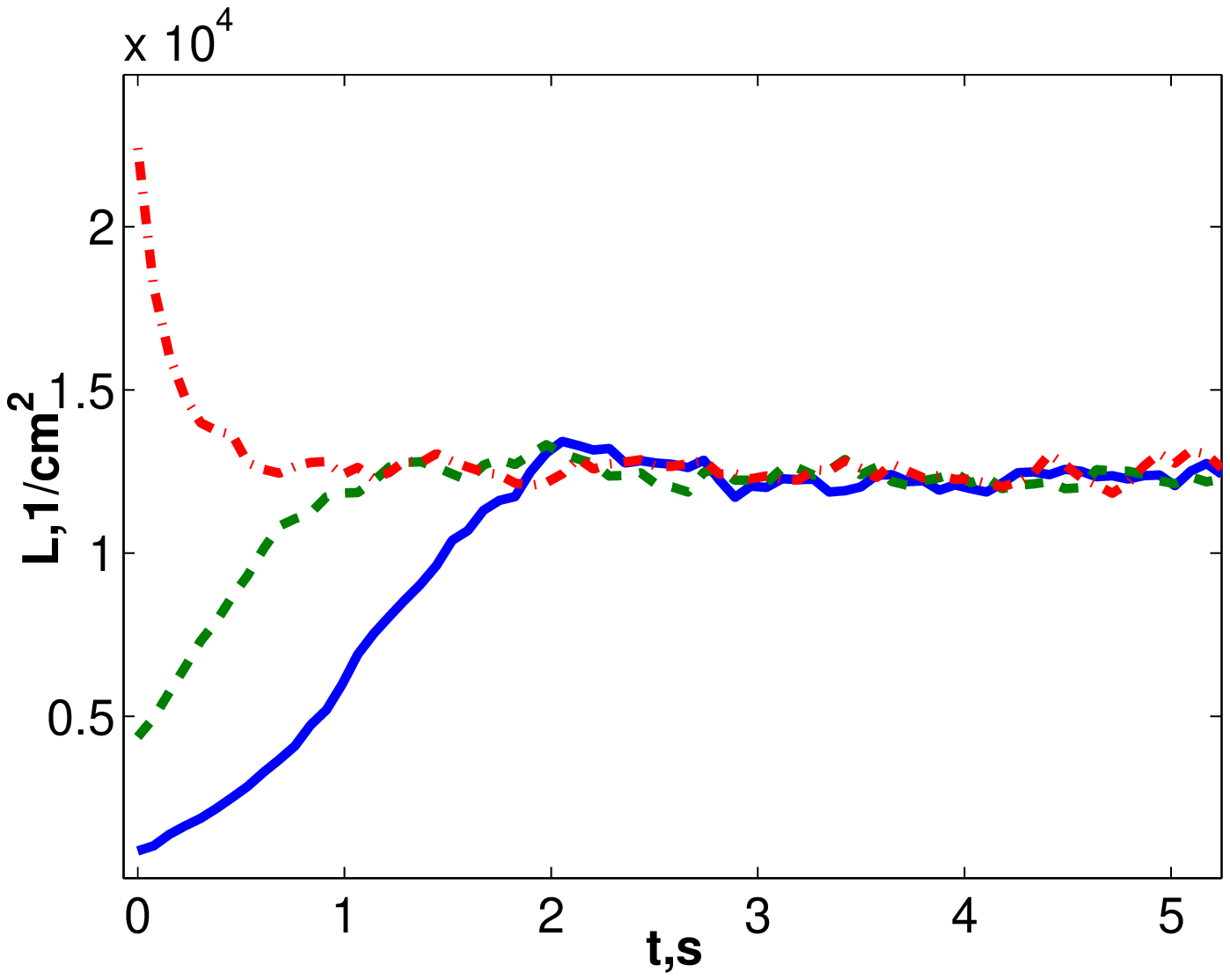} &
   \includegraphics[width=5.6  cm]{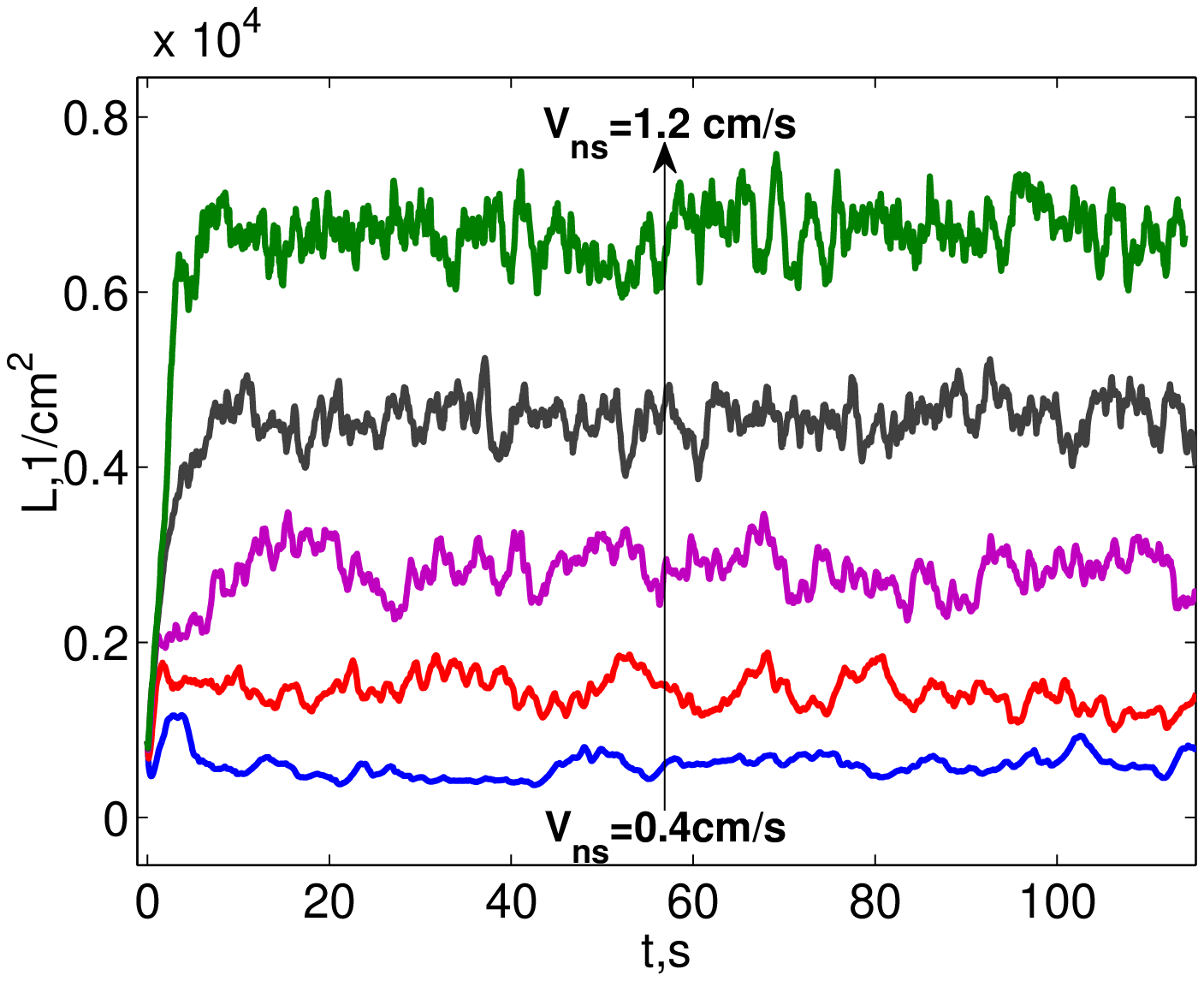}~~~~  &
  ~\includegraphics[width=5.8 cm]{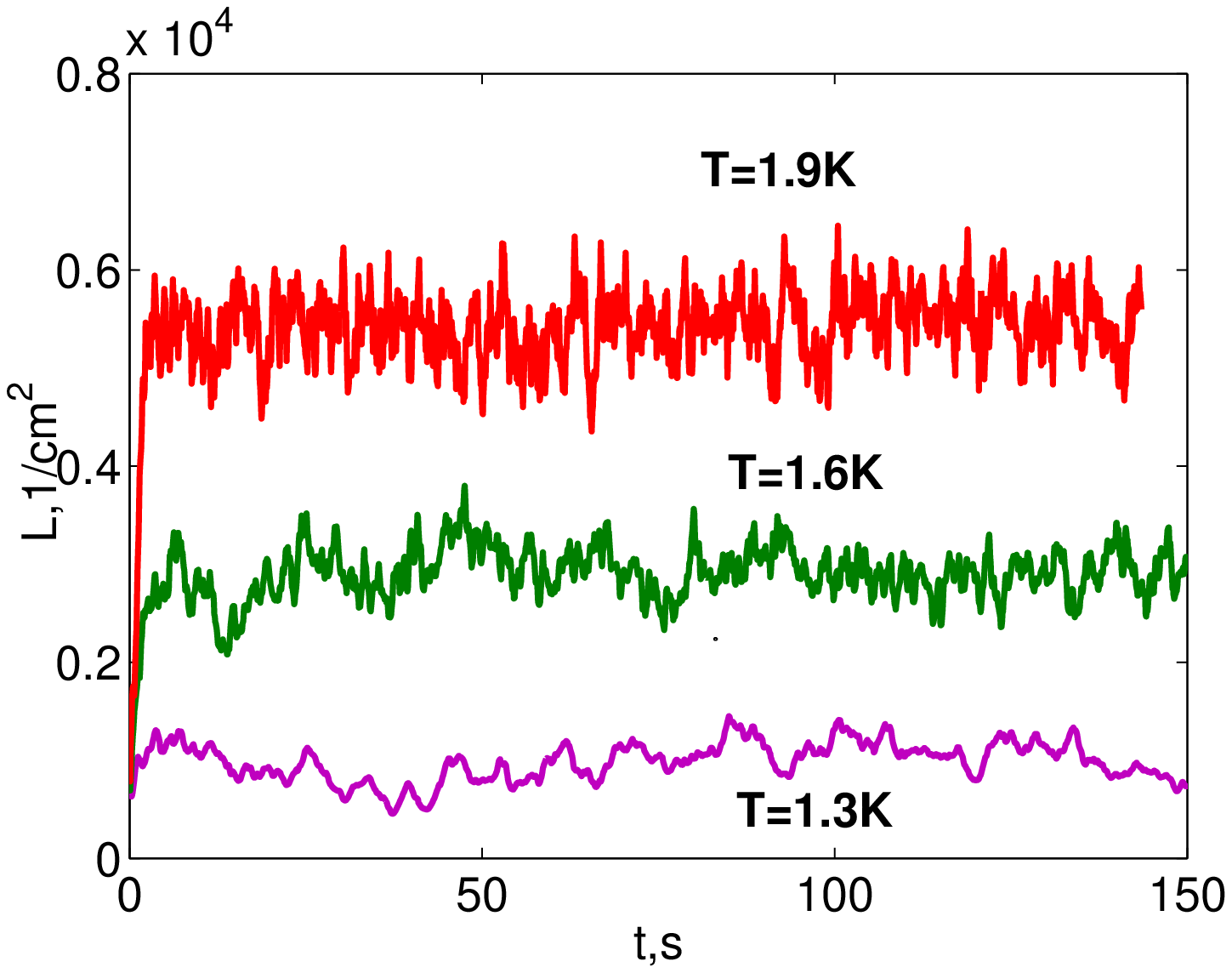}\\

\end{tabular}
\caption{\label{f:3} Color online.
Panel $ A$: Transient regime (with GEC)
  started from 20-ring configuration (blue solid
line), from the steady state configuration obtained at $T=1.3$K (green
dashed line) and from the steady state configuration obtained at
$T=1.9$K (red dot-dashed line).  Panel $ B$:
VLD evolution $T=1.3$ K for
counterflow velocities $0.4, 0.6,0.8,1.0$ and $1.2$ cm/s (from bottom
to top). Panel $ C$: VLD evolution for  different temperatures and $V_{\rm ns}=0.5$ cm/s. }
\end{figure*}

\subsection{\label{ss:ID}Implementation Details}
 The simulations were carried out in the cubic box
 $H=0.1$ cm for temperatures $T=1.3$ K, $1.6K$ and
 $1.9$K and counterflow velocities $V_{\rm ns}$ from $0.3$ cm/s to $1.2$
 cm/s.  The parameters $\alpha$ and $\alpha'$ are given in Table \ref{t:1}. The initial condition consisted of 20 circular rings of radius
 $R_0=9 \times 10^{-3}$ cm oriented such that the total momentum of the
 system vanished. The radius of the rings was chosen to exceed the critical radius of the surviving
 loop\cite{Schwarz85,deWaeleAarts94} $R_{\rm cr}\approx 3 \times 10^{-3}$ for the
 weakest thermal flow ($T=1.3$K, $V_{\rm ns}=0.3$cm/s).

 The initial space resolution $\Delta
 \xi=8 \times 10^{-4}$cm for  D-criterion and $1.6 \times
 10^{-3}$cm for GC and GEC was used. At these values the
 results were insensitive to the resolution as was verified by
 simulations with larger and smaller values of $\Delta \xi$.
As it was  mentioned above, the line points were removed or added during
 evolution to keep $\Delta \xi/1.8\le l_{\pm}\le 1.8\Delta \xi $.

 We use the 4-th order Runge Kutta method for the time marching
 with the time step related by the stability condition to the line
 resolution. For simulations with  GC and GEC
 $\Delta t=3.8 \times 10^{-4}$s, while for DC $\Delta t=9.5 \times 10^{-5}$s was used.
The time evolution was followed for 150
 seconds for GC and GEC and for 75 seconds for DC.

The directionality of the vortex lines is conserved during the
 reconnection procedure. The candidate points for reconnections are
 sought within $1.1 \Delta \xi $ distance for  GC and GEC and within the distance defined by a maximum velocity in the
 tangle at the reconnection time $2\Delta t V_{\rm max}$ for DC. Note that in \cite{Baggaley2012} the candidate pairs for
  similar criterion were sought within distance $\Delta \xi$.

Similar to \cite{AdachiTsubota10}, we remove the small loops and loop
fragments with three or less line segments that are expected to disappear
due to the mutual friction. The maximum length of the removed loops is $8.6\times
10^{-3}$~cm for DC and $1.7 \times 10^{-2}$~cm for GC and GEC, which is smaller than the length of the loop of
the critical size $1.88 \times 10^{-2}$~cm. This procedure was applied in all simulations.

An additional requirement that the angle between reconnecting segments
is at least 10 degrees ($\cos({\bm s}_i,  {\bm s}_j) < 0.9848$), applied
to  GEC, was introduced similar to
\cite{Baggaley2012}. We performed simulations without this additional
requirement as well and did not find any difference in the results.

In BSE simulations, the main computational domain is
surrounded by 26 replicas that take care of the boundary conditions.
 We have verified that the influence of the
replica domains touching the cube edges and corners is negligible. The
influence of the replica domains bordering the faces of the main
domain was studied and discussed below. All results below are
calculated using only the main computation domain.

At each time step we propagate the line points, adjust the space
resolution, perform the reconnections, remove small loops and then
adjust the resolution again. Unlike \cite{Baggaley2012} we reconnect
all pairs of points and segments that satisfy the reconnection
criterion, and not just the closest ones. This may lead to slightly
larger number of reconnection than in  \cite{Baggaley2012}.

\begin{figure*}
\includegraphics[width=5.5 cm]{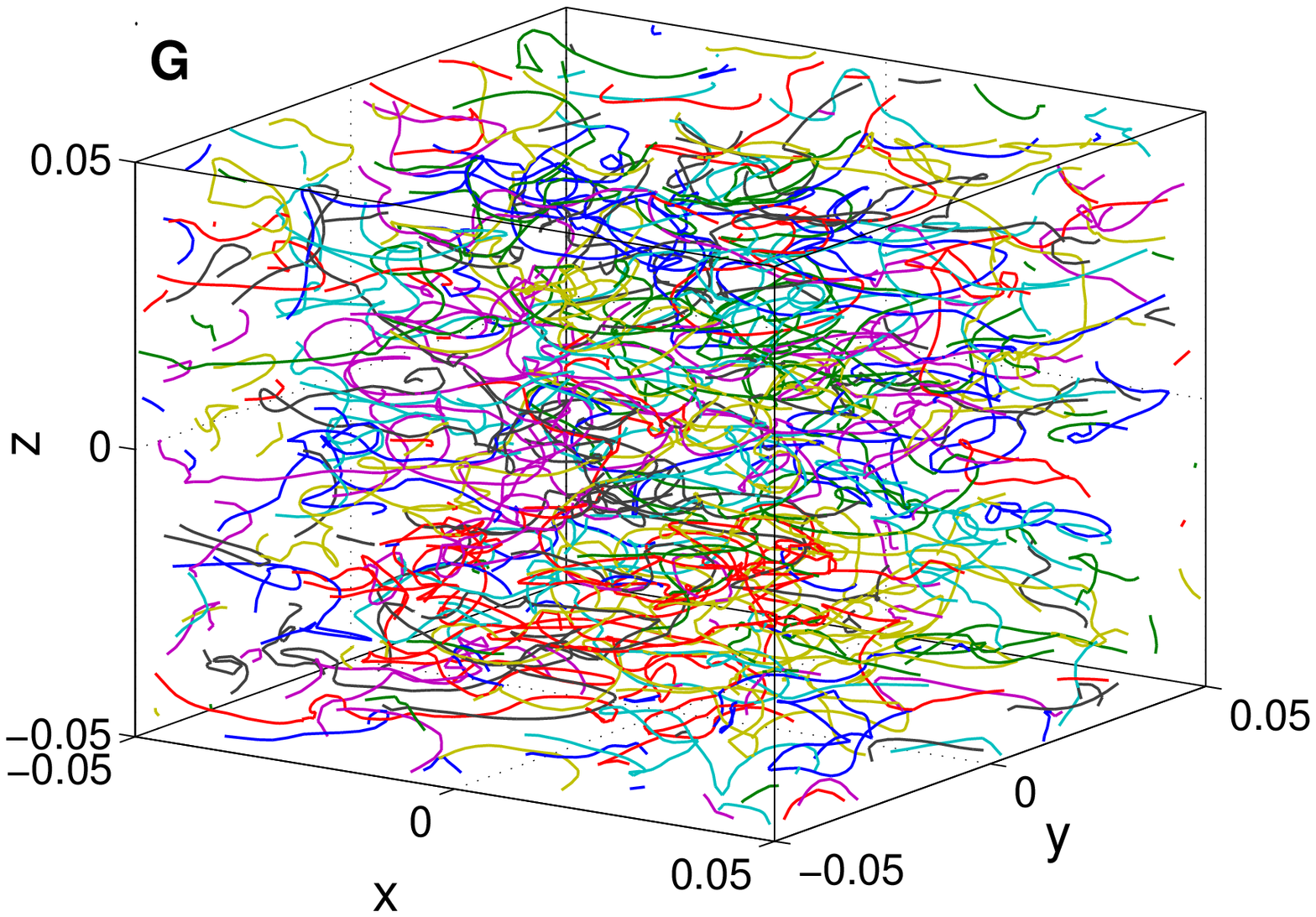}
\includegraphics[width=5.5 cm]{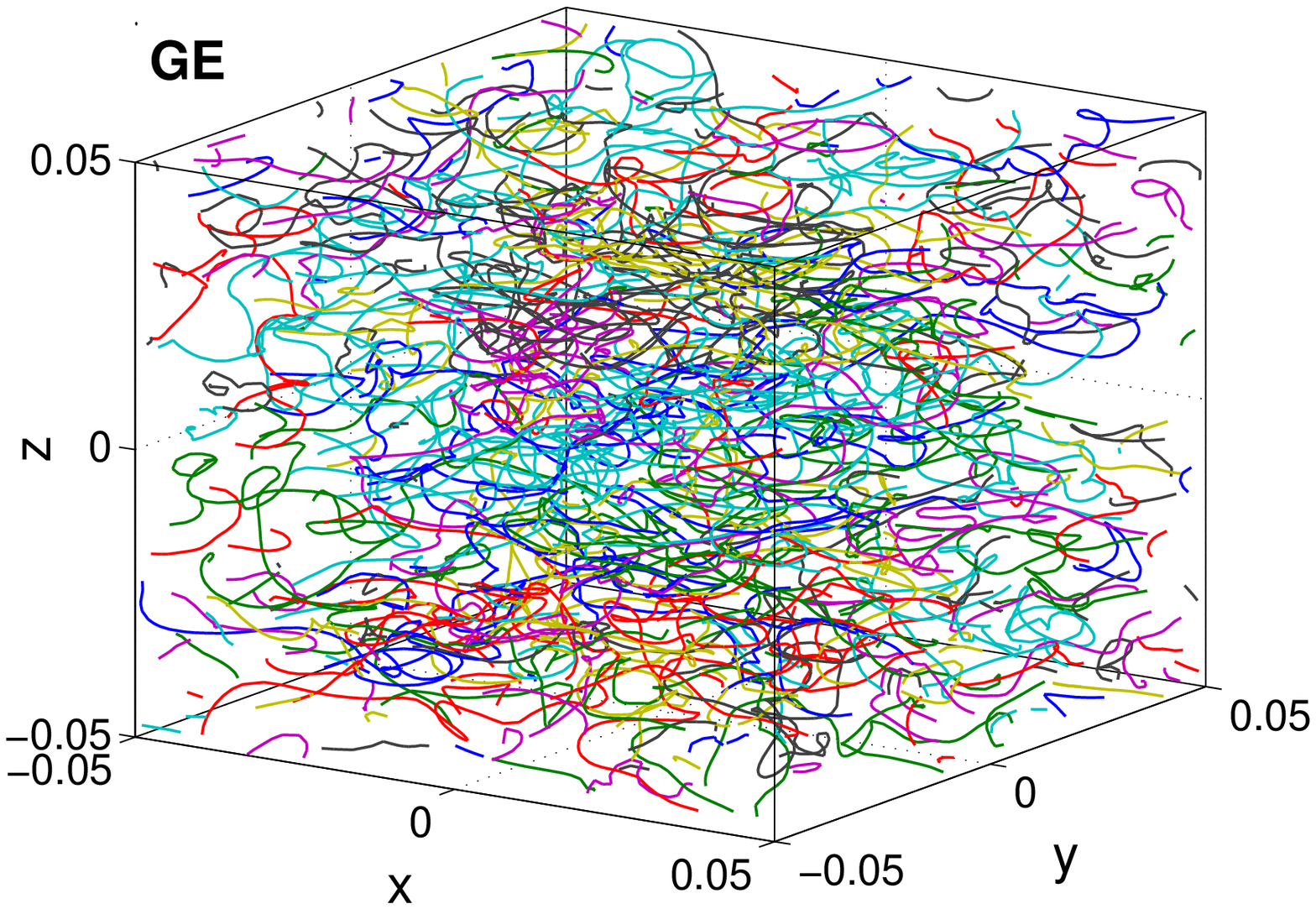}
\includegraphics[width=5.5 cm]{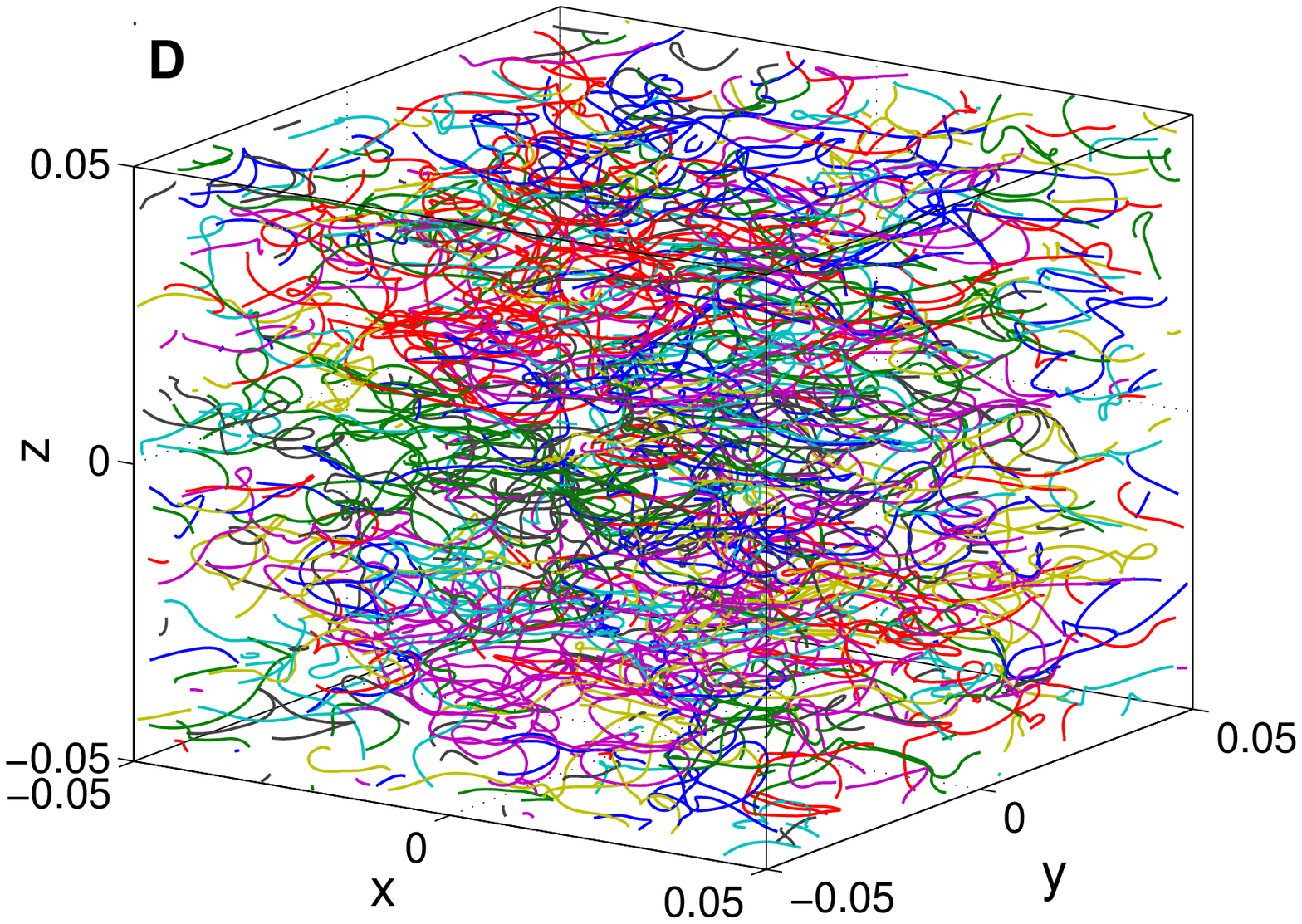}
\caption{\label{t-4}Color online. Typical tangle configurations for different reconnection criteria at $T=1.9$~K and $V_{\rm ns}=1$~cm/s.  }
\end{figure*}

\begin{figure}[t]
\includegraphics[width=8 cm]{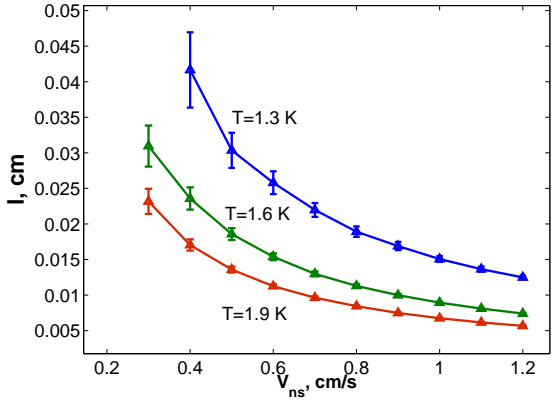}
\caption{\label{f:5}Color online. The  intervortex distance  $\ell={\cal L}^{-1/2}$ as a function of $V_{\rm ns}$ for different temperatures. (with GEC).}
\end{figure}


\section{\label{s:RD1}Dynamics of the vortex tangle}
\subsection{\label{sss:evol}Evolution of  the tangle toward steady state}
A typical time evolution of the vortex tangle is shown in Fig.~\ref{f:3}.
Panel $A$ illustrates that the steady state is
independent of the initial conditions: the evolution at $T=1.6$~K and
$V\sb {ns}=1$~cm/s, started from the 20-ring configuration (blue solid line)
as well as from the steady state configurations for $T=1.3$K and
$T=1.9$K, (green dashed and red dot-dashed lines, respectively) all
give the same steady state vortex line density. This (expected) result allows us
to preform all the simulations starting from the same simple
20-ring configuration.

As one sees in Fig.\,\ref{f:3}A,  the transient time
$\tau\sb{tr}$ it took for the initial configuration to reach the steady state is the shortest for the most dense initial configuration
(steady state at $T=1.9$K -- red dashed line) and the longest for the most sparse one (20-rings -- blue solid line). This can
be easily rationalized by a dimensional reasoning according to
which
\begin{equation}\label{tau-tr}
\tau\sb{tr}\sim 1/ (\kappa \, \C L)= \ell^2/\kappa\ .
 \end{equation}
 This dependence also agrees with our observations that $\tau\sb{tr}$
 is longer for low temperatures (for which the resulting $\C L$ is
 smaller), and shorter for large $T$, at which the  tangle is more dense. For
 moderate values of $\C L\simeq 3\cdot 10^3\,$ cm$^{-2}$ the
 estimate~\eqref{tau-tr} gives $\tau\sb{tr}\sim 0.3\,$s. This is
 slightly shorter than the values observed numerically.

 The values of $\tau\sb{tr}$, deduced from Fig.\,\ref{f:3}A, agree surprisingly well with the experimental results of Vinen~\cite{Vinen58}(Fig.2, panel d) for $T=1.6$~K.  At this temperature the transient time descreases continuously with the inscreasing amount of initially present turbulence. When helium was not exited initially, the time to reach the steady state was about 1.9~sec. It decreased to about 1 sec for moderately exited and to less than 0.5 sec for strongly excited heluim, similar to our results.

Another important characteristic of the vortex dynamics,
clearly seen in Figs.\,\ref{f:3}B and \ref{f:3}C, is the
large amplitude of fluctuations in density in the steady state, which reach up
to $12\%$ of the steady state vortex line density for weak counterflow velocities. One
sees also that the mean line density increases both with $V_{\rm ns}$
and temperature, such that the same line density may be obtained at
lower temperatures and stronger counterflow velocity or at higher $T$
and smaller $V_{\rm ns}$.

\begin{figure}
\includegraphics[width=9 cm]{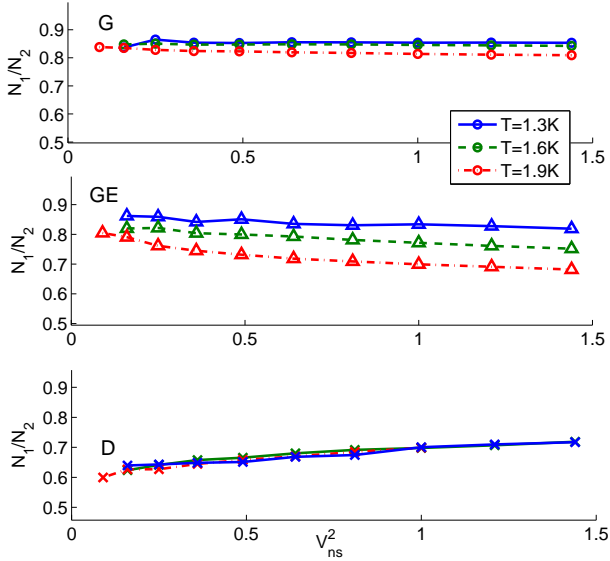}
\caption{\label{f6} Color online.  Ratio of  reconnection rates of two types  for different conditions. Circles denote DC, up-triangles- GEC, crosses- DC. Solid lines correspond to $T=1.3$K, dashed lines- to $T=1.6$K and dot-dashed lines- to $T=1.9$K. Lines serve to guide the eye only.}
\end{figure}

\begin{figure}
\includegraphics[width=9 cm]{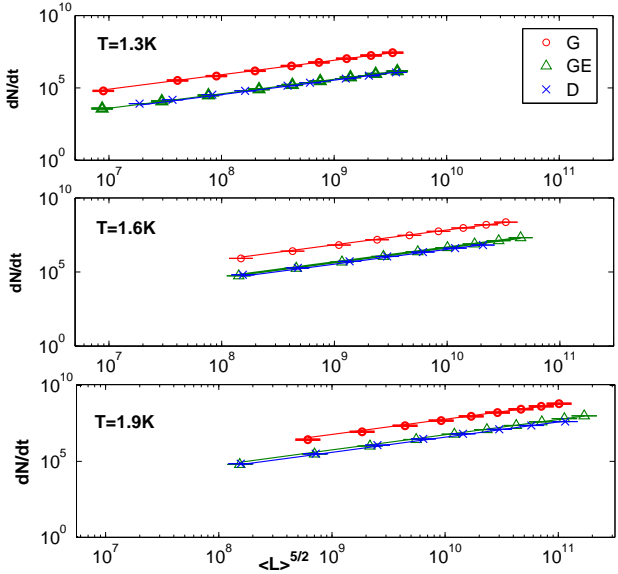}
\caption{\label{f:7} Color online.  The mean reconnection rate $dN_r/dt$ as a function of ${\cal L}^{5/2}$. The symbols with errorbars are data, the lines are linear fit according to Eq.\,\eqref{eq:dNr} (the fit passes through ${\cal L}^{5/2}=0$). }
\end{figure}

\subsection{Tangle vizualization and intervortex distance}
The typical dense steady state tangles obtained with different reconnection criteria are shown in Fig.\,\ref{t-4} for $T=1.9\, $K and $V_{\rm ns}=1$~cm/s. At these parameters the difference in density is visible to the naked eye: the most dense tangle is obtained with DC and the most sparse with GC.
An important characteristic of the developed tangle  is the intervortex distance $\ell$ that quantifies the typical distance between the vortex lines. As seen from Fig.\,\ref{f:5}   the vortex lines come closer with increasing both counterflow velocity and temperature and  $\ell$ becomes comparable with the space resolution $\Delta \xi$ at $T=1.9$ K and the largest $V_{\rm ns}$  used in  our simulations.


\subsection {\label{ss:RR} Reconnection dynamics}
As said above, periodic boundary conditions allow only two types of reconnections:
the merging of two loops into one ($2\to1$) and one loop splitting into two ($
  1\to2$). We denote below the reconnection rate per unit volume of the first type as
  $N_1$ and of the second type as $N_2$ and plot the ratio $N_1/N_2$
  as a function of $V_{\rm ns}^2$ in Figs.~\ref{f6} for
  different conditions. The reconnections leading to splitting one
  loop into two are more frequent in all cases. For GC
  the ratio is almost independent of $V_{\rm ns}$ and the merging of loops is
  even less frequent at $T=1.9$K. For DC  the ratio is
  temperature independent, but the first type of reconnection becomes more
  frequent with increasing $V_{\rm ns}$. For GEC  the loops
  merging becomes less frequent with increasing both the temperature and
  the counterflow velocity. On the average only about $35-45\%$ of
  reconnections lead to loops merging.

In Figs.~\ref{f:7} we show the mean reconnection rate $dN_r/dt$ as
a function of ${\cal L}^{5/2}$. One sees that the linear relation~\eqref{eq:dNr}
is well obeyed throughout the parameter range and for all three
criteria of reconnections. The values of the coefficient $c\sb r$ are
given in Table \ref{t--2}. Note that for all temperatures the
reconnection rate for GC  is several times higher than
for GEC and DC, which are close to each other and
their scaling coefficients fall within the range $0.1<c\sb r<0.5$, as
predicted  by Nemirovskii~\cite{Nem2006}. The much larger number of reconnections
for GC  is in agreement with the results of Baggaley
\cite{Baggaley2012} who found that the time between reconnections for
this criterion was much shorter than for GEC. We conclude that geometric-energetic and dynamic criteria give reliable values of the reconnection rates, while pure geometric criterion overestimates it by an order of magnitude.

Previously the scaling coefficient $c\sb r$ was calculated in
\cite{KondaurovaAndrNemir08,KondaurovaAndrNemir10} using DC within the local induction approximation. They found $c\sb r=2.47$ for $T=1.6$~K and $V_{\rm ns}=6, 8$ and $12$~cm/s. This is much larger
than our current result. The counterflow velocity in their case was
much stronger than we use and their vortex line density was also much larger.  Thus the
difference may stem from  the usage of the local induction approximation, which is doubtful
for these values of $\C L$.

 \begin{table}
\begin{tabular}{|c|c|c|c|c|} \hline
 &           & \ T=1.3K           & \ T=1.6K       & \ T=1.9K \\ \hline
 &   GC      &  $ 8.0 \pm 1.0$    &$6.80 \pm 0.5$     &  $6.0 \pm 0.3$  \\
$c\sb r$ &GEC    &  $ 0.39\pm 0.05$   &$0.45 \pm 0.03$   &  $0.57 \pm 0.03$  \\
 &  DC        &  $0.34 \pm 0.05$   &$0.34 \pm 0.03$   &  $0.40 \pm 0.13$  \\
 \hline
   \end{tabular}
   \caption{The reconnection rate scaling
  coefficient.  The errorbars  were calculated from standard deviations of $dN_r/dt(t)$ and ${\cal L}$. }\label{t--2}
  \end{table}

\begin{table}[t]

\begin{tabular}{||c|c||c|c|c||}
 \hline\hline  &  & T=1.3K & T=1.6K & T=1.9 \\ \hline
   & GC & $  68.6\pm0.1$ &$105.8\pm0.2$ & $128.6\pm0.7$ \\
 $\gamma,$ & GEC & $ 72.1\pm0.2$ &$115.7\pm0.1$ & $148.0\pm0.2$ \\
 ${\rm s/cm}^2$ & DC & $67.1\pm 0.4$ &$120.2\pm0.7$ & $171.2\pm2.6$ \\
  \hline
      $\Gamma \simeq $ & GEC & $0.07$ &$0.12$ & $0.15 $ \\ \hline
  & GC & $3.1\pm 0.1$ & $-0.8\pm 0.1$ & $-5.4\pm 0.3$ \\  $10^2
  v_0$, & GEC & $6.6\pm 0.3$ & $3.3\pm 0.1$ & $0.2\pm 0.1$ \\   cm/s &
  DC & $1.6\pm 0.4$ & $4.3\pm0.5$ & $4.3\pm 0.4$ \\ \hline\hline
 $\gamma$, Ref.~\cite{AdachiTsubota10}& GC & 53.1 & 109.6 & 140.1 \\
  \hline
  &   GC & & 116.9 & \\   $\gamma$, Ref.~\cite{Baggaley2012} & GE& &
  114.35 & \\   &DC & & 112.3 & \\ \hline\hline
    $\gamma\Sb V$, Eq.~\eqref{gam0} &   & 82 & 151 & 266 \\
  \hline  $\gamma\Sb S$, Eq.~\eqref{gam19}  &\cite{Schwarz88} &80 & 130&  198  \\
  \hline\hline

 \end{tabular}\caption{ \label{t--3}  The values of $\gamma$ (in s/cm$^2$) and $10^2 v_0$ (in cm/s)
obtained from Eq.\,\eqref{gam}  and approximate values of $\Gamma\equiv \gamma \kappa$ for GE-criterion.
The errorbars for $\gamma$ and $v_0$ were calculated from  the standard
deviation of ${\cal L}$ by textbook relations~\cite{NumRec}.
The values of $\gamma$ from numerical simulations of  Refs.~\cite{AdachiTsubota10,Baggaley2012}, from Eq.~\eqref{gam0} and the estmate \eqref{gam19} with $c \Sb L$ from \cite{Schwarz88} are given for comparison.}
 \end{table}

\begin{figure*} [t]
\begin{tabular}{ c c c }
 \hskip -.3cm\includegraphics[width=6.2 cm]{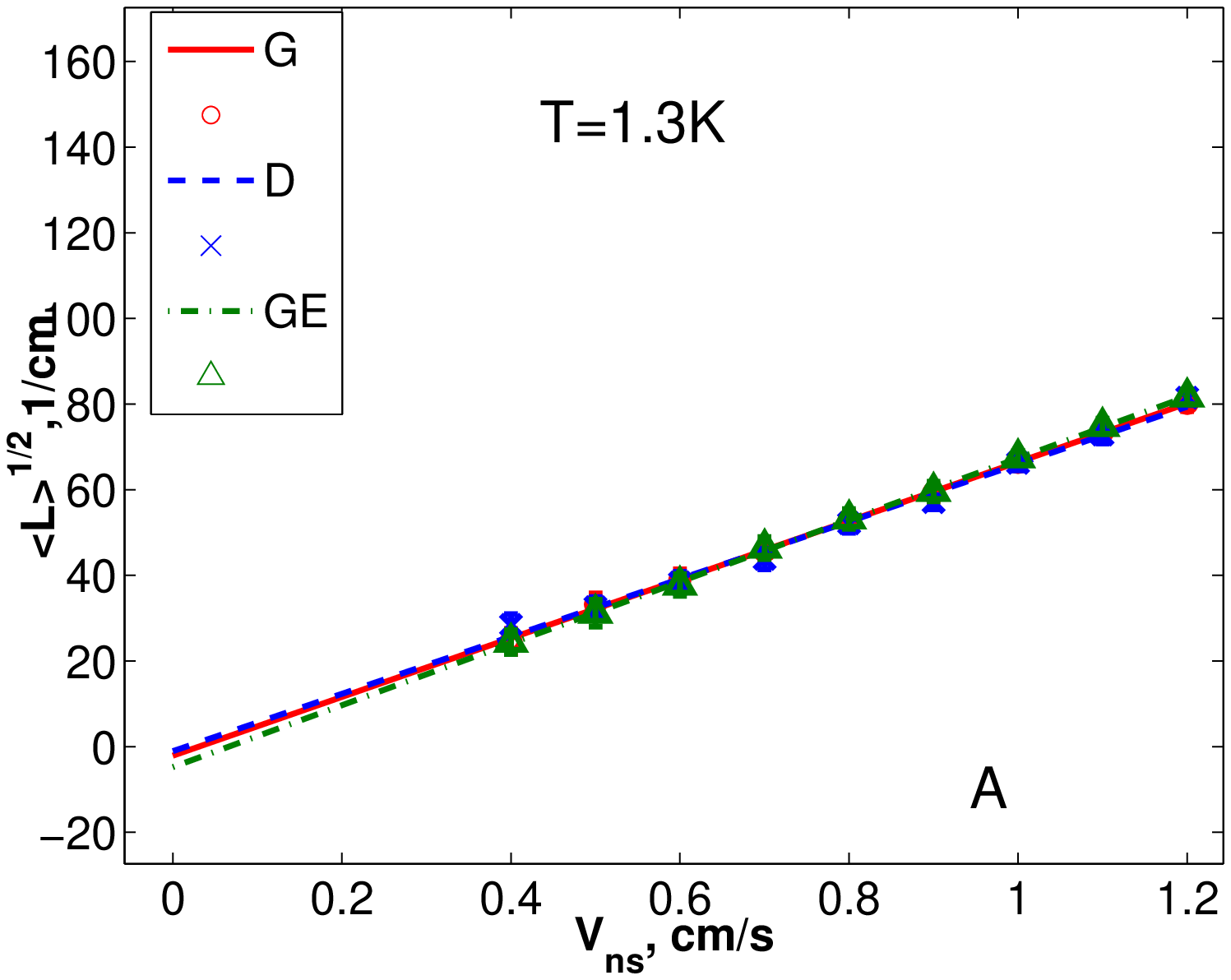}&
 \hskip -0.2cm\includegraphics[width=6.2 cm]{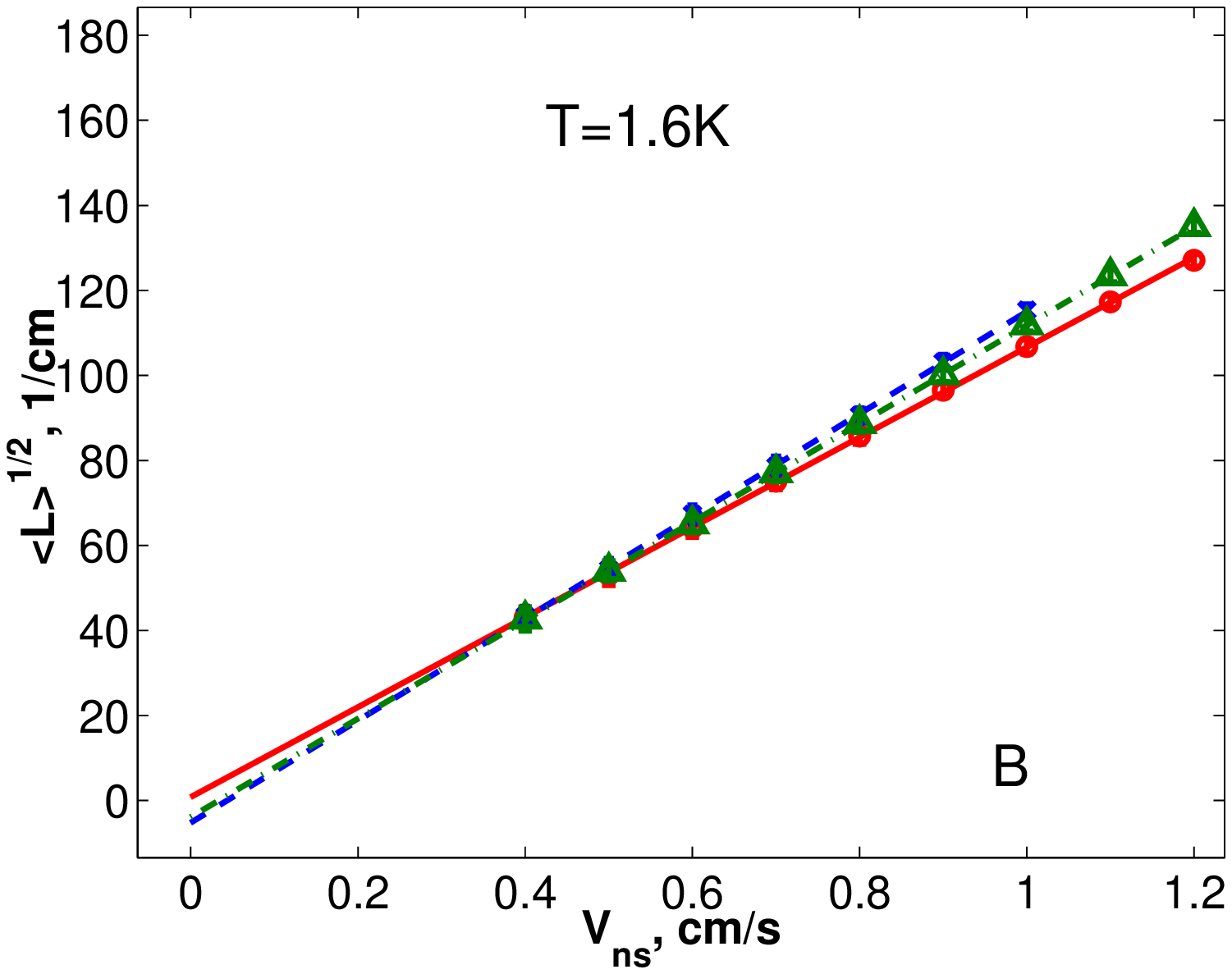}&
 \hskip -0.2cm \includegraphics[width=6.2 cm]{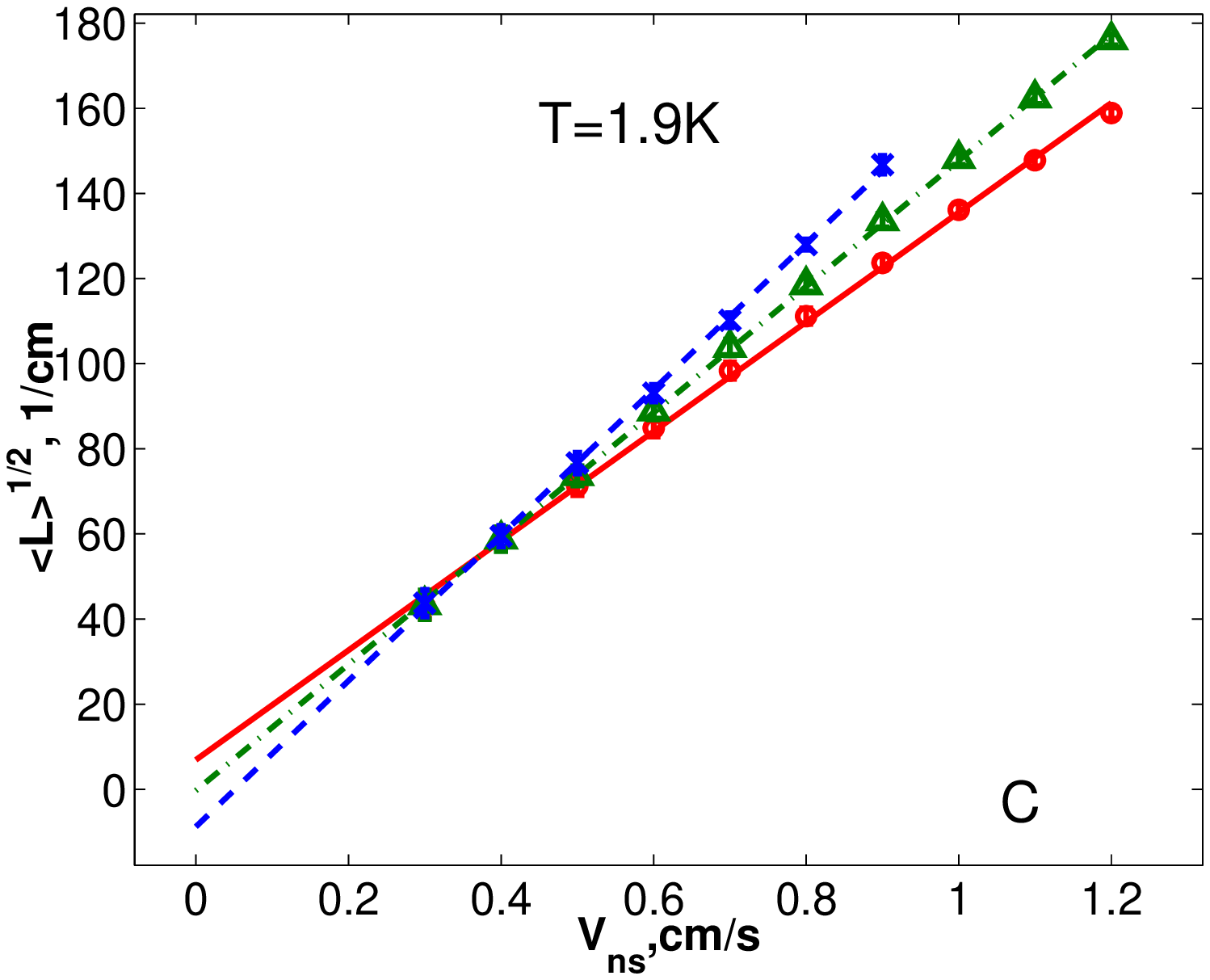}\\

\end{tabular}
  \caption{\label{f:8} Color online. $\sqrt{{\cal L}}$ as a function of $V_{\rm
ns}$ for $T=1.3\,$K (Panel A), $T=1.6\,$K (Panel B) and $T=1.9\,$K (Panel C).
Symbols with errorbars are the numerical results,
lines are fits according to Eq.\,\eqref{gam}. Three sets of symbols and
lines correspond to different reconnection criteria as shown in the
legend in Panel A.}
\end{figure*}

\begin{figure}[t]
\includegraphics[width=9.4 cm]{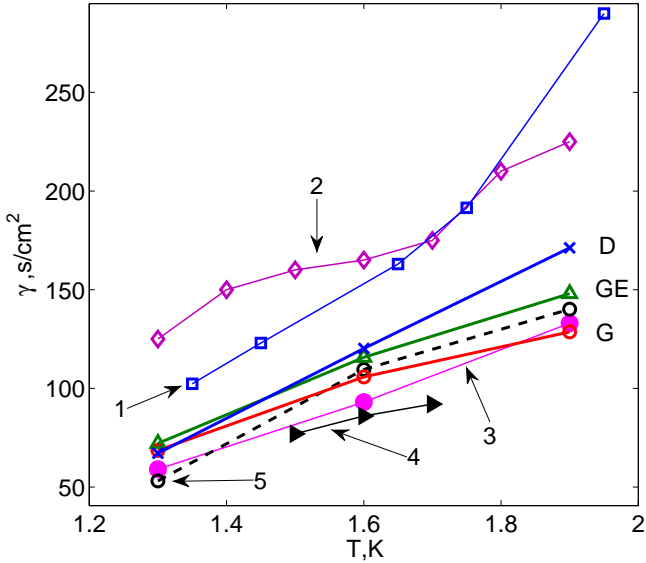}
\caption{\label{f9} (Color online). The numerical and experimental
values of the parameter $\gamma(T)$.  Thick solid lines with open
symbols, marked with letters D(blue line with crosses), GE(green line
with up-triangles) and G (red line with circles) are the results of
our simulations with different reconnection criteria(DC, GEC and GC,
respectively). Line 5, black open circles with dashed line- the
simulations of Adachi el at \cite{AdachiTsubota10}, G-criterion. Thin
lines with open symbols are the results of experiments in pure
superflow: line 1 - Babuin at al \cite{Ladik-2012}(7x7 mm channel);
line 2 -Ashton et al \cite{Ashton81}, 0.13- mm diameter glass
channel.  Thin lines with filled symbols are the results of
experiments in counterflow: line 3- Childers and
Tough\cite{ChildersTough76}, TI state, 0.13-mm diameter glass
channels, line 4- Martin and Tough\cite{MartinTough83} TI state, 1-mm
diameter glass channel.}
\end{figure}

\section{\label{s:stat} Mean  characteristics of the tangle }
\subsection{\label{sss:gamma} Vortex line density $\C L$}

\subsubsection{Numerical results for $\C L$ vs. counterflow velocity }
The steady state value of $\C L$ was obtained by averaging ${\cal
L}(t)$ over the plateau values for  $10-150\,$s for $T=1.3\,$K and for $5-150\,$s
for $T=1.6$ and $1.9\,$K for GC and GEC
and up to $75\,$s for DC. The error bars in the
figures were calculated by the standard deviation over the same time
period. In Fig.~\ref{f:8} we present $\C L$ as a
function of the counterflow velocity and the fit according to
Eq.\,\eqref{gam}. Clearly the data follow this linear relation faithfully and the
corresponding $\gamma$ and $v_0$ are given in Table \ref{t--3}. A
measurable difference between the results with only the main computational
domain and with additional 6 replicas touching its faces was found
only for $T=1.9\,$K and $V_{\rm ns}>0.5$ cm/s, resulting in
$\gamma=(146.2\pm0.2)$ s/cm$^2$  compared to $\gamma=(148.0\pm0.2)$ s/cm$^2$ (for GE-criterion) for the main
domain (about $1\%$ difference). Similar corrections were obtained for the other criteria.  We therefore conclude that
 for the parameter range used in our work  it is
sufficient to calculate the Biot-Savart velocities in the main domain only.

The values of $\gamma$ which was calculated in \cite{Baggaley2012} were obtained for counterflow velocities $0.35< V_{\rm ns}<0.6$ cm/s at $T=1.6$~K, while in \cite{AdachiTsubota10} a similar range of $V_{\rm ns}<0.6$ cm/s was used for  $T=1.9$~K. For these parameters we found that the difference in the computed value of $\C L$ for different reconnection criteria was relatively small. Our simulations with wider range of counterflow velocities demonstrates that the values of $\C L$  for three reconnection criteria are close only for $T=1.3\,$~K,  while for $T=1.6$~K and $1.9$~K they progressively deviate from each other, leading to different values of $\gamma$ -- see Fig.~\ref{f:8} and Tab.~\ref{t--3}. Quantifying the spread of the values as a difference between the largest and the smallest $\gamma$ at each temperature divided by the mean value, we get about $7\%$ for $T=1.3$~K, about $13\%$ for $T=1.6$~K
and about $28\%$ for $T=1.9$~K.
\subsubsection{Comparison of numerical and experimental results}
In Fig.~\ref{f9} we compare the values of $\gamma$ obtained in simulations with the experimental results. This  is an issue that requires careful analysis of particular experimental conditions including the dependence on the channel width, the roughness of the walls,  the finite value of the temperature difference with respect to the mean temperature and problems with temperature stabilization.

Additional uncertanty arises from the fact that the thermal counterflow turbulence
in square channels  of width smaller than 1 mm may exist in two turbulent regimes\cite{Tough82}. The regime TI immediately
follows the laminar state. The regime TII is found  above some critical
line density, usually at higher counterflow velocities. In both regimes  ${\cal L}^{1/2} = \gamma (V\sb{ns}-v_0)$ with $\gamma$ in TII state larger than that in TI.

All these problems lead to a wide spread of experimental values of $\gamma$ -- see lines 1, 2, 3, 4  in Fig.~\ref{f9}.

The values of $\gamma$ for pure superflow (thin lines  with open symbols -- lines 1 and 2) are significantly larger than those for counterflows in TI state (thin lines with filled symbols -- lines 3 and 4).
As was discussed in  Ref.~\cite{Ladik-2012}, the values of $\gamma$ in superflows are close to the results of  counterflow in TII state.

Ignoring these differences in the experimental conditions, we note that i) the spread of numerical results (ours and from Ref.\cite{AdachiTsubota10}) is smaller than that of the experimental data; ii) the numerical results lie within the spread of experimental values of $\gamma$.

More experimental work is needed to better measure the values of $\gamma$ and more numerical simulations are required to account, for example, for the boundary conditions with strong vortex pinning and laminar velocity profile of the normal components, which is not expected to be a constant even for pure superflow.
Nevertheless,  numerical and experimental results demonstrate qualitatively the same kind of behavior that allows us to hope  that the main characteristics of turbulent counterflow are adequately reflected in the numerical simulations.

\subsubsection{Dependence of numerical results for vortex line density on reconnection criteria}
As we  showed above, the  values of $\gamma$ for three different reconnection criteria increasingly differ with
increasing temperature. These differences may be related to  larger values of $\C L$, i.e. to smaller inter-vortex distance $\ell=1/\sqrt{\C L}$. A possible  explanation is that the vortex filament method with any reconnection criterium deteriorates when $\ell$ approaches the inter-point distance $\Delta \xi$, but the degree to which the dynamics of the tangle is affected depends on the reconnection criterion. We performed several control simulations with higher $V_{\rm ns}$ (not shown). Simulations at $T=1.3$~K and $V_{\rm ns}=2$~cm/s resulted in $\C L$ similar to that for $T=1.9$~K and $V_{\rm ns}=0.9$~cm/s and followed the same line as all other results for $T=1.3$~K. On the other hand, at $T=1.9$~K and $V_{\rm ns}=2$~cm/s, $\Delta \xi \approx  \ell $ and the value of $\C L$ was strongly underestimated.   In all our simulations the ratio $\Delta \xi/\ell \le 0.2 $ and we have checked that with twice larger ratio we got practically the same results.

We should stress that the steady state value of $\C L$ is a result of  a delicate balance of all the dynamical processes that are effected by the reconnections, explicitly and implicitly, via details of the resulting tangle characteristics.
 The final steady state value of $\C L$ may be counterintuitive.
For instance, there exists an apparent contradiction: the only
difference between GC and GEC is that in GC the reconnections increasing the length of the vortex line are allowed. Given
a much larger number of reconnections in this case, one expects
that the vortex line density should be larger for GC than for GEC,  while in fact it is smaller. To resolve
this contradiction we analyzed the change in the length of the vortex tangle during
transient evolution, before the steady state tangle was formed.  It
turns out that the reconnection procedure for GC
produce a large number of small loops and loops fragments with a number of
segments not exceeding three. The number of these small loops increases as
the tangle develops.  Since in our procedure such small loops are
removed from the configuration only after all the reconnections were
made, only truly separate loops that did not merge back into larger
loops are removed. The number of such  small loops in GEC
is about 10 times smaller, while with DC they are
hardly created at all.  Removal of these small rings slows down the
growth of the length of the vortex tangle, in particular for GC , and results
in smaller steady state vortex line density in this case. For DC, on
the other hand, no such mechanism exists and the vortex line density grows more before
reaching the steady state value.

In the developed tangle the total length change due to reconnections,
small loops removal and re-meshing is small compared to the total line
length and in a self-consistent manner helps to maintain the density of
the tangle around its steady state value. At this stage the difference
between the reconnection criteria is not significant.

\subsubsection{\label{sss:phen-an}Phenomenological analysis of $\gamma(T)$}
 The na\"ive dimensional estimate $\Gamma\sim 1$ gives $\gamma\sim 1000\,$s/cm$^2$.
 Much better estimates were obtained by using macroscopic properties of the vortex tangle.

 In 1957  Vinen \cite{Vinen57} suggested a phenomenological evolution equation
for the $\C L$:
\begin{equation}\label{eq:Vinen}
 \frac{d {\cal L}(t)}{dt}=\frac{\chi_1 B\rho_n {\cal
 L}^{3/2}}{\rho} \mid V_{\rm ns} \mid-\frac{\chi_2\kappa {\cal
 L}^2}{2\pi}\ .
\end{equation}
It includes the vortex generation and vortex decay terms on its RHS.
Here $B(T)$ is the Hall-Vinen temperature dependent dimensionless
coefficient, describing the interaction between the line and the
normal fluid, while $\chi_1(T)$ and $\chi_2(T)$ are additional
dimensionless phenomenological parameters.
 In the steady state Eq.~\eqref{eq:Vinen} results in the relation~\eqref{gammaA} with
\begin{equation}\label{gam0}
 \gamma\Sb V =\frac{\pi B \rho_n\chi_1}{\kappa\rho\chi_2}\ .
\end{equation}
Estimates of the coefficient  $\gamma\sb v$ with   experimental values for $B$ (for $V\sb {ns}=1$cm/s) and $\chi_1,\chi_2$~\cite{Vinen57,ChildersTough76,DonnelyBarenghi98} are shown in Tab.~\ref{t--3}.  These values are  close to the experimental $\gamma$ measured in superflow\cite{Ladik-2012,Ashton81}.

Another estimate was obtained by Schwarz~\cite{Schwarz88}, who derived the equation of motion for the line density similar to Vinen's equation~\eqref{eq:Vinen} from local induction approximation and  balanced in the  steady state tangle  the mean anisotropy of the self-induced velocity ${\bm s'}\times{\bm s''}$ in the Eq.~\eqref{eq:LIA} against its magnitude:
\begin{equation}\label{gam19}
\gamma\Sb S =c\Sb L/\beta\ ,\quad c\Sb L\equiv I_\ell/c_2^2 \ .
\end{equation}
 Recall that $\beta=\kappa \widetilde \Lambda $ and the values of $\widetilde \Lambda$ are very close to unity, such that $\gamma\Sb S  \approx 10^3 c \Sb L$.

\begin{figure}
\includegraphics[width=9 cm]{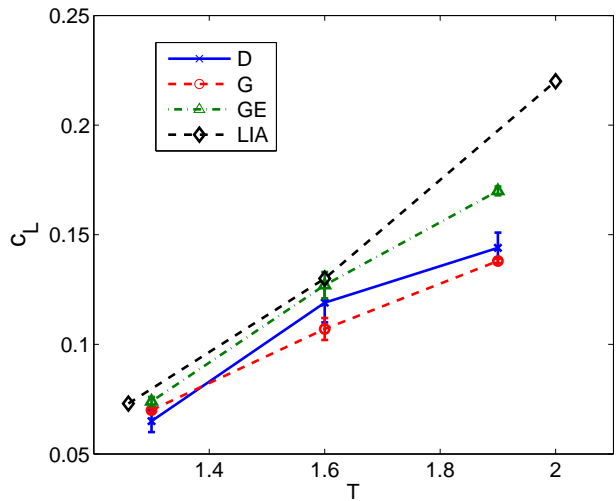}
\caption{\label{f:cl}Color online.The coefficient $c\Sb L$ found in our VFM simulations with full Biot-Savart equations for different reconnection criteria and the LIA-VFM simulations of Schwarz~\cite{Schwarz88}}
\end{figure}

The parameter $c \Sb L$, relating the $V\sb{ns}$-dependence of $\C L$  to the tangle anisotropy and RMS curvature,
 is one of the most important parameters of Schwarz's theory \cite{Schwarz88}. It defines, among other properties, the tangle drift velocity and the mutual friction force, discussed later in Secs.~\ref{sss:drift} and \ref{sss:fric}.  We can calculate $c\Sb L$, using $I_{\ell}$ and $c_2$ given in Table~\ref{t:4}, and see how well the theory based on the local induction approximation works for the vortex tangles, obtained with full Biot-Savart simulations.

As we discuss in Secs. \ref{sss:aniz} and \ref{sss:LC}, the anisotropy index  $I_{\ell}$ is almost independent of both the temperature and the reconnection criterion, while $c_2$ changes  with $T$ and  differ for different reconnection criteria. Therefore, the $T$-dependence of $c\Sb L$ (and consequently, that of $\gamma \Sb S$) is defined by the RMS curvature scaling coefficient $c_2$.

\begin{figure*}
\includegraphics[width=8.9 cm]{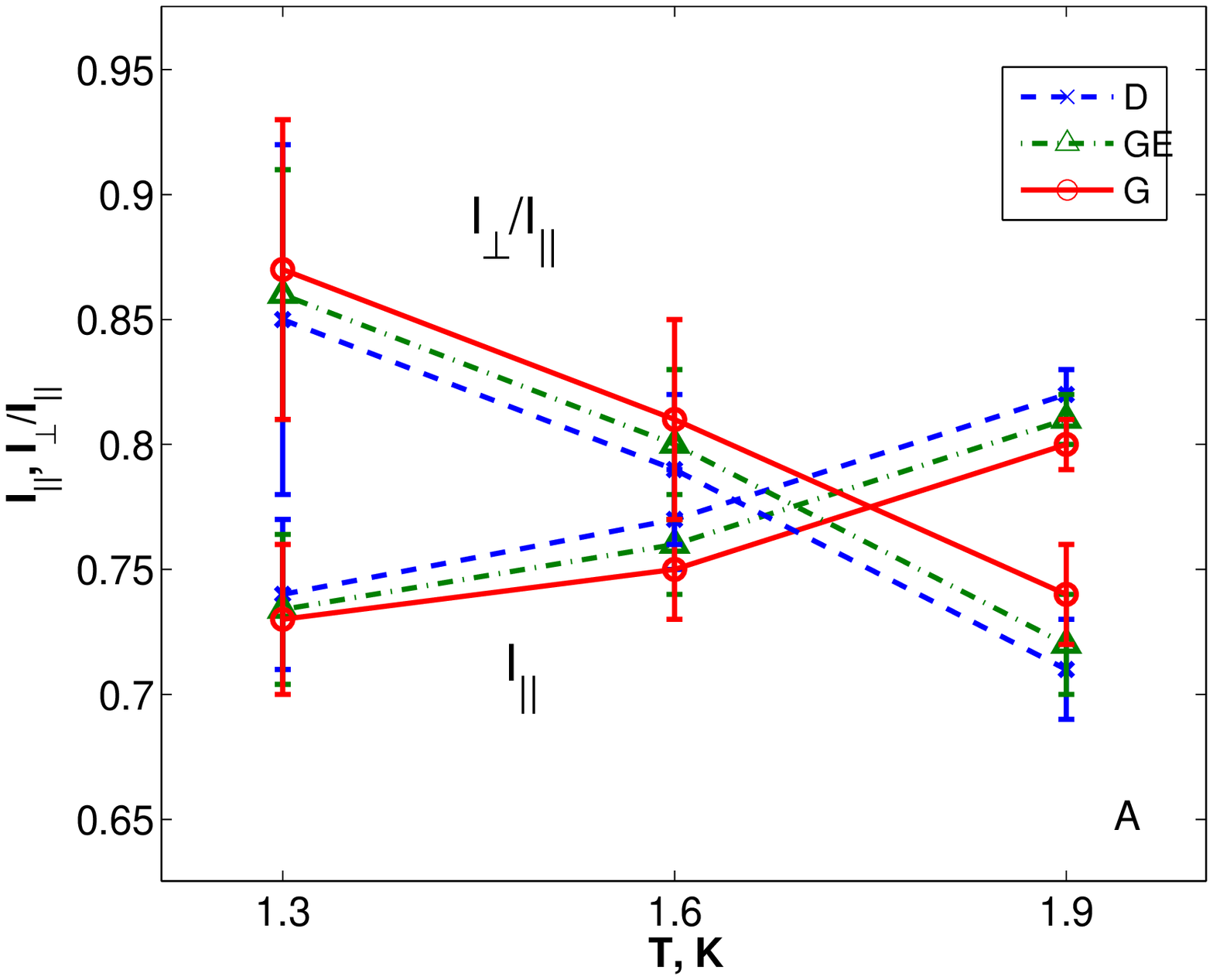}\hfill
\includegraphics[width=8.9 cm]{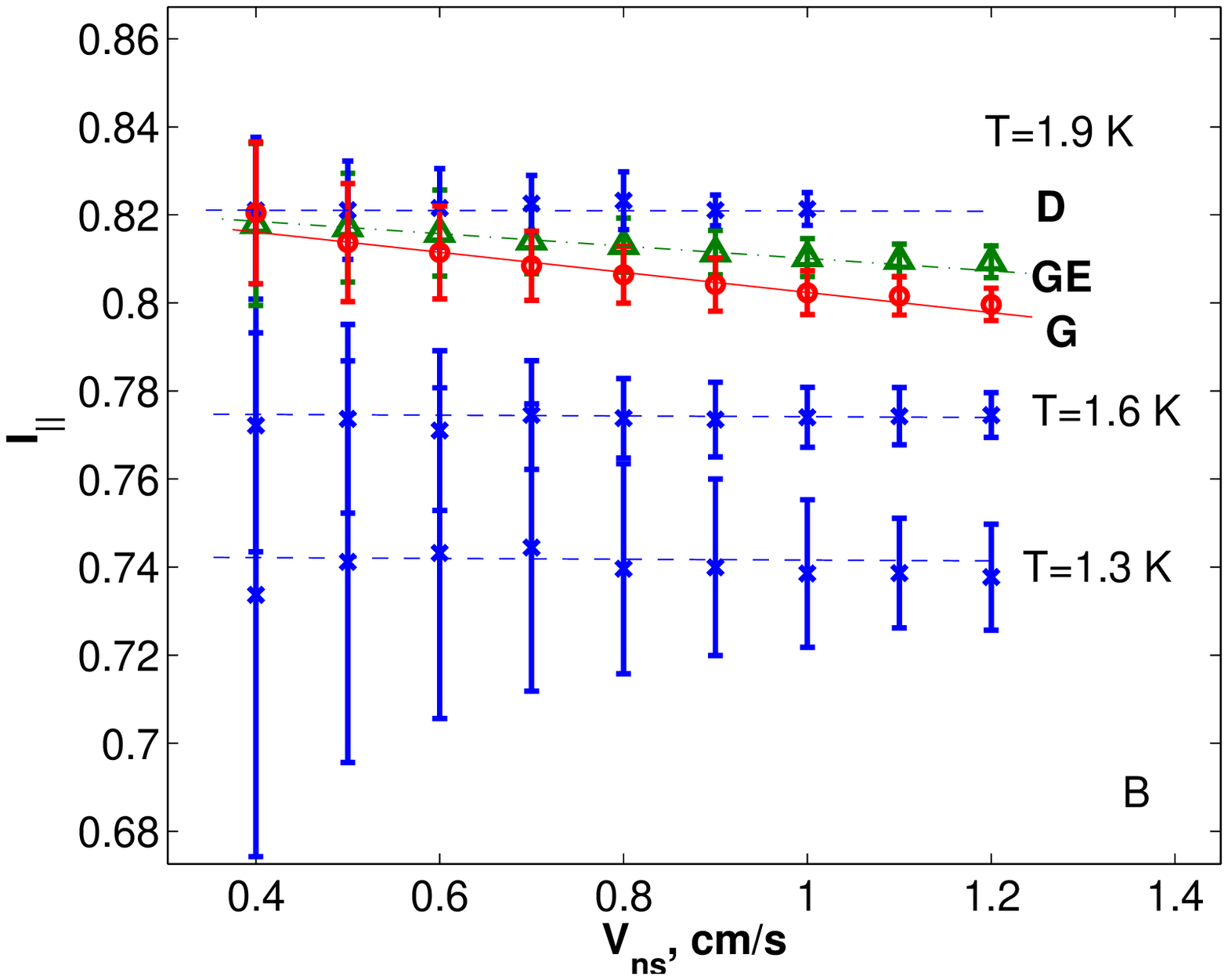}
\caption{\label{f:9}Color online. Panel A: The temperature dependence
of the anisotropy indices $I_{||}$ and the ratio $I_{\bot}/I_{||}$,
Eq. \eqref{Ipar} and \eqref{Iper}. Blue crosses ($\times$) stand for
DC , green triangles ($\Delta$) for GEC
and red circles for GC .  Panel B: The anisotropy index
$I_{||}$ vs. $V_{\rm ns}$.  For $T=1.3$K and $1.6$ K only {\bf D}
results are shown (see text). The lines serve to guide the eye. }
 \end{figure*}

In Fig.~\ref{f:cl} we show the parameter $c\Sb L$ for three reconnection criteria and the results of Schwarz~\cite{Schwarz88}, obtained with local induction approximation simulations.  The overall trend is very similar to that of $\gamma$, shown in Fig.~\ref{f:9}, except that in this case it is the GE-criterion that gives the largest values and not DC, as for $\gamma$. As could be expected, the results for the three criteria differ most at $T=1.9$~K and the estimates for $\gamma \Sb s \approx 10^3 c\Sb L$ give values  slightly larger than $\gamma$, see Table~\ref{t--3}.  The values of $c\Sb L$, obtained by Schwarz from the local induction approximation simulations, are even larger. We therefore conclude that the nonlocal corrections to the line velocity affect the mean tangle properties by decreasing the vortex line density for stronger counterflow velocities.

Schwarz \cite{Schwarz88} also related the phenomenological coefficients $\chi_1$ and $\chi_2$ to the properties of the steady state tangle as $\chi_1=I_{\ell}, \chi_2= 2 \pi \alpha \beta I_{\ell}/\kappa c\Sb L$. Indeed, substituting these definitions to Eq.~\eqref{gam0} and recalling that $\alpha=B \rho_n/2\rho$, we find that within the local induction approximation $\gamma\sb v=\gamma \sb s$.  However, the values of $\gamma\sb s$ obtained with Schwarz's values  of $c\Sb L$ are smaller than  $\gamma\sb v$.

Comparing Eqs.~\eqref{gammaA} and \eqref{gam19} we find that both $\Gamma$ and $c \Sb L /\widetilde \Lambda$ relate the counterflow velocity to the vortex line denisity. Therefore we can expect that the numerical smallness of $\Gamma\sim c\Sb L\sim 0.1$ as well as  their temperature dependence have similar origin. For $c\Sb L=I_{\ell}/c_2^2$  it is   the RMS curvature scaling coefficient $c_2$ that mostly defines the value and $T$-dependence. Some discrepancy in  the behavior of $c\Sb L$ and $\gamma$, calculated from our tangles with different reconnection criteria suggests that there is no one-to-one correspondence between these two parameters. More than one mean property of the tangle is responsible for the value of steady state vortex line density in Biot-Savart simulations. However, for low and moderate temperatures (or low values of $\C L$), the estimates of $\gamma$ via mean tangle properties are quite accurate.

 \subsubsection{Intercept velocity}  The values of $v_0$ found in our simulations are shown in Tab.~\ref{t--3}. The values  are  quite small,  about $(0.03~-~0.06)$~cm/s,  but definitely nonzero within our accuracy of measurement. The values of $v_0$ are very different for different reconnection criteria, including some negative values for GC.  Therefore we tend to consider nonzero values of $v_0$ not as a solid prediction of our simulations, but rather as an artefact stemming from the approximate character of the reconnection criteria. As for the  larger values of $v_0\sp{exp}\sim 0.1\,$ cm/s observed in experiments~\cite{Ladik-2012}  they may be related to the strong pinning of quantized vortices on rough wall surfaces. This effect was not accounted for  in our simulations. Arguments in favor of this statement may by found in Fig.~8 of Ref.~\cite{Ladik-2012} which shows that the experimental values of $v_0$
monotonically decrease for wider and wider channels. Overall, we tend to think that the finiteness of the intercept velocity $v_0$ is a finite size effect.

\subsection{\label{sss:aniz}Mean tangle anisotropy}

The counterflow velocity defines the preferred direction in the
 tangle.  The tangle anisotropy index $I_{ ||}$ and the ratio $
 I_{\bot}/ I_{{\rm ||}}$ are shown in Fig.~\ref{f:9}.
  The temperature dependence ( Fig.~\ref{f:9}A) is consistent with the
 known picture \cite{Schwarz88,AdachiTsubota10} that the tangle become
 more oriented in the direction perpendicular to the counterflow
 velocity with increasing temperature. This may be understood as an
 interplay of two contributions to Eq.\,\eqref{eq:s_Vel}: the term
 proportional to $\alpha$ is oriented in the plane perpendicular to
 $V_{\rm ns}$, while the term proportional to $\alpha'$ is locally
parallel to the counterflow velocity and leads to
 isotropization of the loops orientation. The ratio $\alpha'/\alpha$
 diminishes upon increasing the temperature and so does the relative contribution
 of the this term, and the tangle become more oblate.

\begin{figure}
\includegraphics[width=9 cm]{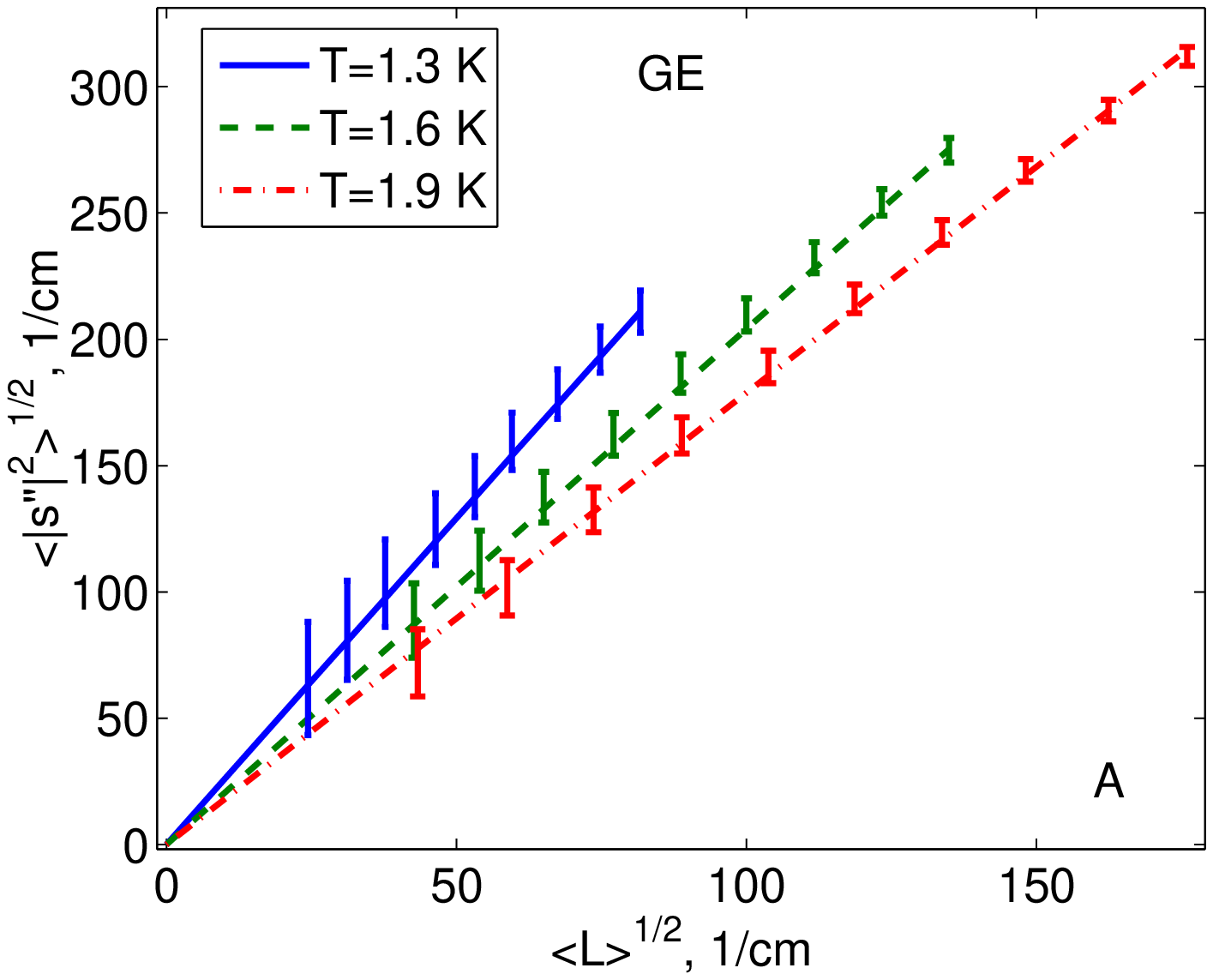}
\includegraphics[width=9 cm]{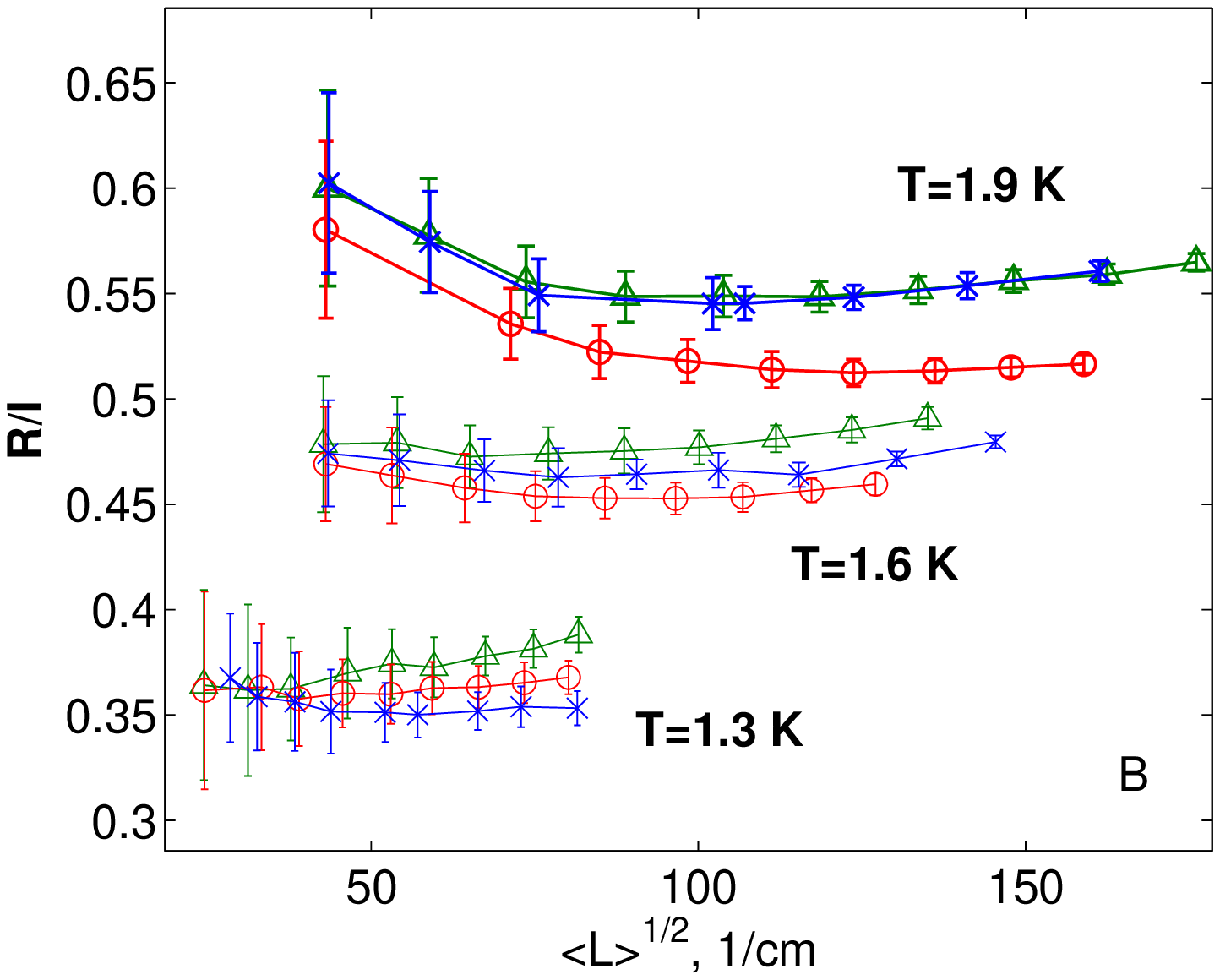}
\caption{\label{f:10}Color online. Panel A: Log-log plot of mean curvature of the tangle versus the tangle density.  GEC. Symbols with errorbars are data, lines are linear fit according to Eq.\,\eqref{sL}. Panel B: Ratio of the mean radius of curvature to the intervortex distance.  Red circles denote data for GC, green triangles -for    GEC and blue crosses - DC. Lines serve to guide the eye only.  }
\end{figure}

\begin{table}
\begin{tabular}{ |c|c||c|c|c| } \hline
       & recon. crit.    & \ T=1.3K          & \ T=1.6K       & \ T=1.9K \\ \hline\hline
 &   GC       &  $ 0.73\pm0.03$   &$0.76\pm0.02$   &  $0.80\pm0.01$  \\
$I_{{||}}$& GEC &  $ 0.73\pm0.03$   &$0.76\pm0.02$   &  $0.81\pm0.01$  \\
 Eq.\eqref{Ipar}&  DC        &  $ 0.74\pm0.03$   &$0.77\pm0.02$   &  $0.82\pm0.02$  \\
 \hline
&  GC                 & $0.86\pm0.06$& $0.81\pm0.04$&$0.74\pm0.02$  \\
$I_\perp /I_{{||}}$ &GEC &$0.86\pm0.05$& $0.80\pm0.03$&$0.72\pm0.02$  \\
 Eq.\eqref{Iper}& DC                  &$0.85\pm0.07$& $0.79\pm0.03$&$0.71\pm0.02$ \\
  \hline
&  GC         & $0.50\pm0.09$& $0.52\pm0.02$   &  $0.53\pm0.03$  \\
$I_\ell$ & GEC  & $0.50\pm0.08$ & $0.53\pm0.03$   &  $0.54\pm0.02$  \\
Eq.\eqref{Iell} & DC         & $0.51\pm0.09$ & $0.53\pm0.03$   &  $0.52\pm0.02$ \\
  \hline
$ I_{{||}}$ &  GC \cite{AdachiTsubota10}  &  $0.738$   & $0.771$   &  $0.820$  \\
  \hline\hline
$I_\perp /I_{{||}} $ &exp.~\cite{Wang87} & $0.85\pm0.05$ & $0.8\pm0.1$ & $0.7\pm0.1$ \\
\hline\hline
& GC & $2.26 \pm 0.01$ & $1.85 \pm 0.003$ & $1.68\pm 0.002$ \\
 $c_1$ &GEC & $2.09 \pm 0.01$ & $1.68 \pm 0.003$ & $1.48 \pm 0.001$ \\
Eq.\eqref{sL}  & DC & $2.28 \pm 0.01$ & $1.64 \pm 0.003$ & $1.48 \pm 0.004$ \\
\hline
 & GC & $2.70 \pm 0.10$ & $2.19 \pm 0.05$ & $1.96 \pm 0.03$ \\
$c_2$ &GEC & $2.60 \pm 0.10$ & $2.04 \pm 0.04$ & $1.78 \pm 0.02$ \\
Eq.\eqref{sL} & DC  & $2.80 \pm 0.20$ & $2.11 \pm 0.05$ & $1.90 \pm 0.07$ \\
\hline
 \end{tabular}
  \caption{\label{t:4} Anisotropy indices and mean curvature scaling
  coefficients. The values and errorbars for anisotropy indices are
  the time averages (same time interval as for $\cal L$) and the
  standard deviation over the same period of time,respectively. The
  errorbars for $c_1$ and $c_2$ were calculated from standard
  deviations of $|s''|$ and $|s''|^2$. }
\end{table}

In our simulations, the DC  systematically gives the
 most anisotropic tangle, while GC  gives the most
 isotropic tangle. For
 all simulations $I_{ ||}$ is almost independent of $V_{\rm ns}$,
 however for both GC and GEC a slight dependence on
 $V_{\rm ns}$ exceeding the error bars was observed (see Fig.\,\ref{f:9}B)
 for $T=1.6$ K ( not shown) and $T=1.9$ K. This is at variance with
 the results of \cite{AdachiTsubota10} where no such dependence was
 observed ( for $V_{\rm ns}\le 0.6$ cm/s), while the values of $I_{\rm ||}$ and of the ratio $ I_{\bot}/
 I_{ ||}$ agree well with their results, as well as with the numerical results of
 Schwarz\cite{Schwarz88} and the experimental data by Wang \emph {et al} \cite{Wang87}.

The values in Fig.\,\ref{f:9}B are the time averaged values and the
 error bars are defined by the standard deviation of $I_{||}$ for
 $I_{||}$ and as a sum of standard deviations of $I_{||}$ and $I_{\bot}$
 for $ I_{\bot}/ I_{ ||}$ over the same time period. The values in
 Fig.\,\ref{f:9}A  are the average values
 for a given temperature and the error bars are the largest error bars for
 $V_{\rm ns} \ge 0.4$ cm/s.

 The anisotropy indices $I_{\ell}$ and $I_{\ell\bot}$ are practically independent of
$V_{\rm ns}$ for all temperatures and all criteria. $I_{\ell\bot}$ is close to
zero in all the simulations indicating that the tangles are isotropic in
the direction perpendicular to the counterflow velocity.
$I_{\ell}\approx 0.5$ and slightly increases with temperature (see Table
\ref{t:4}). No measurable difference for different reconnection
criteria was observed.

\subsection{\label{sss:LC} Mean and RMS vortex-line curvature}
Next important mean characteristic of the tangle is its RMS curvature  $\widetilde S=\sqrt{\< |s''|^2\>}$, plotted in   Fig.\,\ref{f:10}A,  as a function of
the $\C L$  for different temperatures. One sees that the curvature $\widetilde S$ is increasing with tangle density as $\sqrt{\C L}$ according to Eq.\,\eqref{sL} with the numerical prefactor $c_2$ that decreases as temperature grows. In other words, for the same density of the vortex lines the tangle is more curved at lower temperatures. Table~\ref{t:4} shows that the scaling is well obeyed in simulations with all
reconnection criteria and the coefficients $c_1$ and $c_2$ are quite
close. The value of $c_2$, calculated at $T=1.9$~K with GC agrees well with the result of \cite{Mineda2013} ($c_2=1.99\pm 0.38$).

 However some differences in the fine structure may be seen in Fig.\,\ref{f:10}B, showing the
 ratio of the mean radius of curvature $R =1/ \widetilde S $ to the intervortex distance $\ell$. In this way we compensate the $\sqrt{{\cal L}}$ dependence of the curvature and the
 lines are almost flat. This ratio is distinctly
 different for different temperatures - the mean radius of curvature
 is about a third of the intervortex distance at $T=1.3\,$K and it
 grows to more than a half of $\ell$ for $T=1.9\,$K.

The strongest change in the structure is for  DC - it has the smallest $R/\ell$ at $T=1.3$K, while for $T=1.9$K it appears
 smoother and the ratio coincides with that for GEC.
  For moderate and high temperatures the DC local
 structure appears the most sinuous.

The ratio $R/\ell$, shown in Fig.\,\ref{f:10}B,  gives interesting \emph{global} information about the
relation between the RMS curvature and the mean intervortex distance.
However it does not allow to distinguish whether the
small values of $R$ are due to dominant contributions of small loops with
large curvature while the large loops are smooth, or because the large loops are
fractal. To answer this and similar questions we need to have more detailed information on the vortex tangle, not only its mean characteristics.

\subsection{\label{sss:drift} Drift velocity of the vortex tangle $V\sb{vt}$}
\begin{figure*}~~~~~~~~~~~~~~~~~~~~~~~~~~~~~~~~~A \hfill B~~~~~~~~~~~~~~~~~~~~~~~~~~~~~~~~~~~~~~\\
\includegraphics[width=8.9 cm]{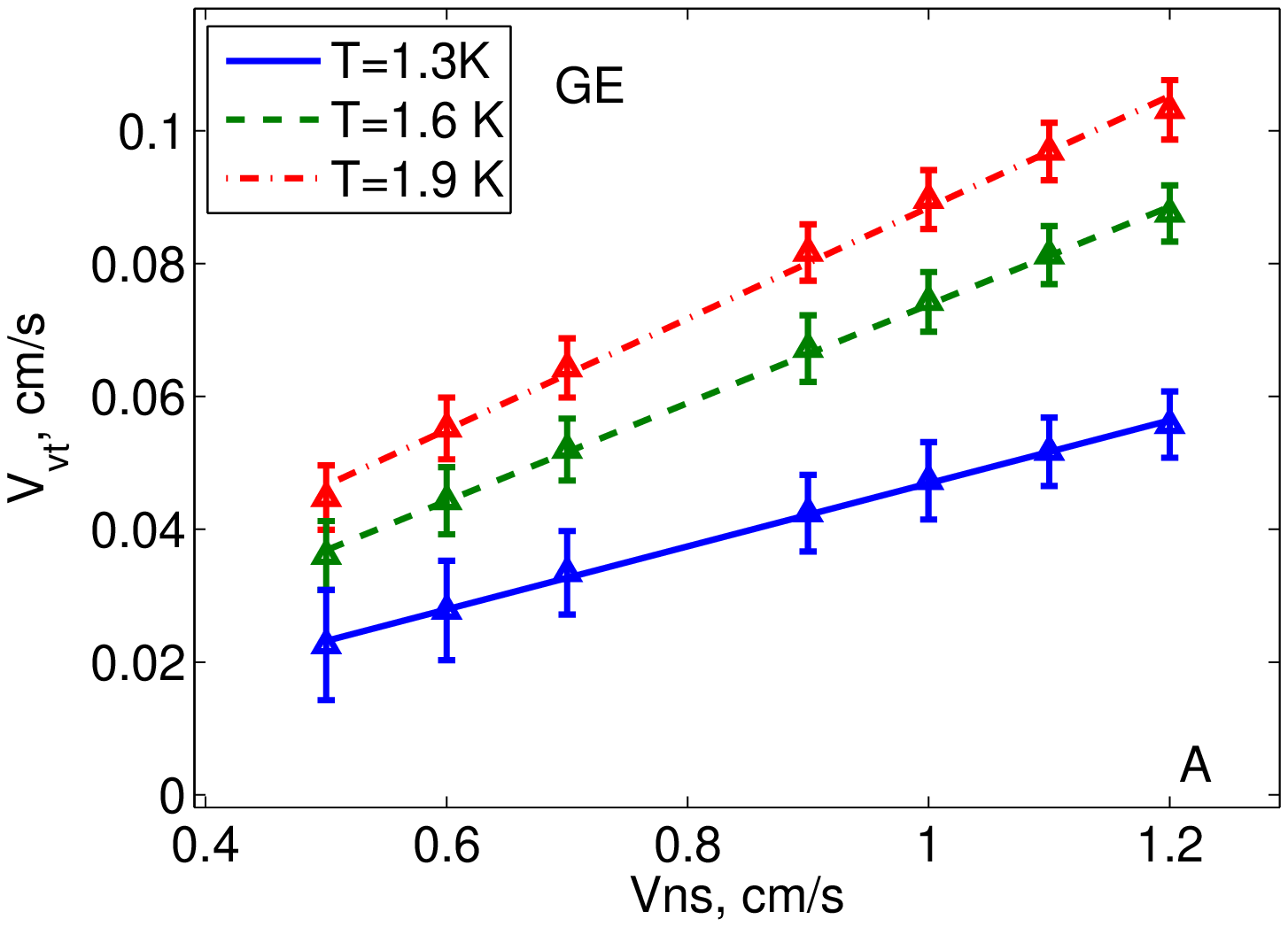}
\includegraphics[width=8.9 cm]{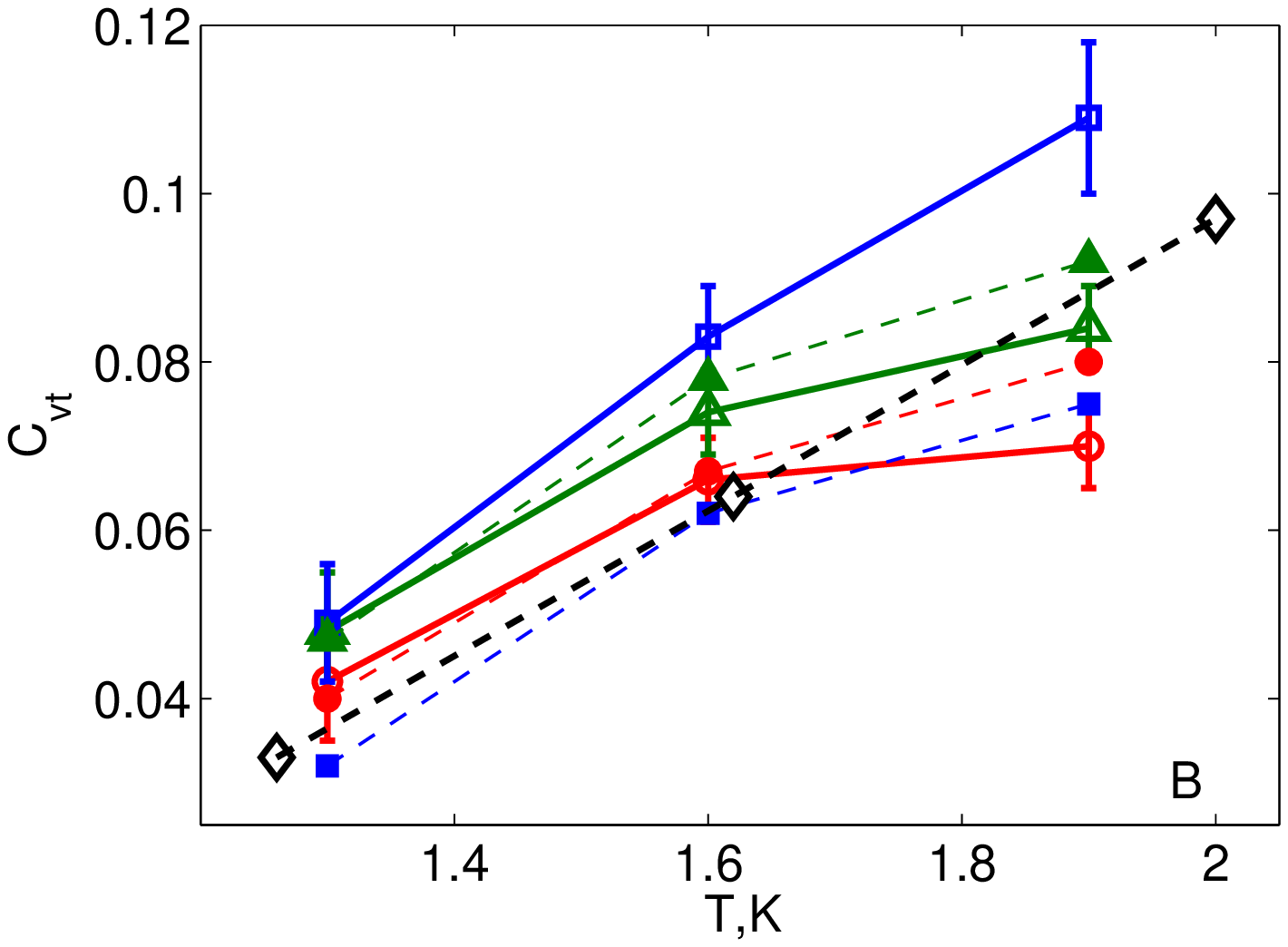}
 \caption{\label{f--drft}Color online. Panel A: Drift velocity vs $V\sb{ns}$ for three temperatures. The symbols with errorbars are data, the lines are the fit according to Eq.~\eqref{Kvt}.  GE-criterion. Panel B: The drift velocity coefficient $C\sb{vt}$ as a functon of temperature for different reconnection criteria and their LIA estimates Eq.~\eqref{CLIA}. Thick solid lines with open symbols -$C\sb{vt}$ obtained by fit Eq.~\eqref{Kvt} (red cicles -- GC, green triangles -- GEC, blue squares -- DC). Thin dashed lines with filled symbols are the LIA estimates $C\sb{vt}\Sp{LIA}$ Eq.~\eqref{CLIA} (the symbols and colors are the same as for  $C\sb{vt}$). The thick dashed line with open diamonds -- the results of Schwarz\cite{Schwarz88}.}
 \end{figure*}

 In some physical problems, like the evolution of neutron-initiated micro-Big Bang in superfluid $^3$He-B~\cite{Bunkov}, an important role is played by the drift velocity $V\sb{vt}$ of the tangle with respect to the superfluid rest frame.
The natural expectation is that $V\sb{vt}$ is proportional to and oriented along $V\sb{ns}$. In Fig.~\ref{f--drft}A we plot $V\sb{vt}$ calculated according to Eq.~\eqref{DV}. We see that the linear relation \eqref{Kvt} is well obeyed. The value of $V\sb{vt}$ is fully defined by its $z$-component, parallel to the direction of counterflow velocity, while two other components are zero within our accuracy of measurement.

As in case of $c\Sb L$, the coefficient~$C\sb{vt}$  may be  analytically  related to the structural parameters of the tangle in the Local-Induction Approximation by
plugging   $d\B s/dt$, Eq.~\eqref{eq:s_Vel}, with $\B V\sb{si}\Sp{LIA}$  into \eqref{DV} and  by considering different contributions to the integral~\cite{Schwarz88}:
\begin{equation}\label{CLIA}
C\sb{vt}\Sp{LIA}\approx [c\Sb L(1-\alpha')I_{\ell}+\alpha' I_{||}] \ .
\end{equation}
The superscript ``~$\Sp {LIA}$ " stresses the fact that this relation is not exact, but obtained in the Local-Induction Approximation.
The terms proportional to $\alpha$ vanished  in this equation  by symmetry.  Note that $\alpha'=O(10^{-2})$ and therefore $C\sb{vt}\Sp{LIA} \approx c\sb L I_{\ell}$, which is the value plotted in Fig. 31 of \cite{Schwarz88}.

As we mentioned, the local contribution~\eqref{LIAeq} provides up to 90\% of the total vortex velocity. Therefore we can expect that Eq.~\eqref{CLIA} will be valid with accuracy about 10\%. To check this we compare  in Fig.~\ref{f--drft}B  the coefficients $C\sb{vt}$, obtained directly by fitting plots $V\sb{vt}$ vs. $V\sb{ns}$, presented in  Fig.~\ref{f--drft}A,  and   coefficients $C\sb{vt}\Sp{LIA}$, given by Eq.~\eqref{CLIA}, in the RHS of which we used mean vortex parameters, found in our full Biot-Savart simulations. First of all, we note that the drift coefficients  for different criteria  are close at $T=1.3$~K but differ significantly at $T=1.9$~K with $C\sb{vt}$  for DC being almost twice larger than that for GC. Also, for DC the coefficients increase almost linearly with temperature, while for both GEC and GC the growth slows down at $T=1.9$~K. One sees that  $C\sb{vt}$ and $C\sb{vt}\Sp{LIA}$ are very close for GC and GEC, except for  $T=1.9$~K, where  $C\sb{vt}$ is   larger. Surprisingly, $C\sb{vt}\Sp{LIA}$ for DC is \emph{smaller} than Biot-Savart results for all temperatures, and in fact smaller than most of the other values. The possible reason is that the drift velocity is very sensitive to the nonlocal effects on the local tangle structure for more dense tangles (DC always gives denser tangles).

The Schwarz's $C\sb{vt}\Sp{LIA}$   found from Eq.~\eqref{CLIA} in the RHS of which the mean vortex parameters are found by simulations in the LIA~\cite{Schwarz88}
is very close to the GC results (both $C\sb{vt}$  and $C\sb{vt}\Sp{LIA}$) for $T=1.3$ and $1.6$~K, but is somewhat larger than $C\sb{vt}\Sp{LIA}$ for  $T=1.9$~K. Comparing with Fig.~\ref{f:cl}, we see that the difference in the tangle structure ($I_{\ell}$ in this case) between Biot-Savart and LIA simulations is important: for $c\Sb L$ the GEC results were closer to Schwarz's values. Therefore the particular closeness of different Schwarz's results to our results with different reconnection criteria is not systematic and should be taken with caution.


 The main and well expected physical message is that $C\sb {vt}$ is  small (below  upper limit of
$V\sb{vt}/V_{\rm ns}=0.2$, suggested in \cite{Awschalom84} and in accord with results of \cite{Wang87}). This means that the tangle velocity is close to the superfluid velocity and its slippage is about 5\% at $T=1.3\,$ K and close to 10\% at $T=1.9\,$K.

\subsection{\label{sss:fric}Mutual friction force $ F\sb{ns}$}

\begin{figure*}~~~~~~~~~~~~~~~~~~~~~~~~~~~~~~~~~A \hfill B~~~~~~~~~~~~~~~~~~~~~~~~~~~~~~~~~~~~~~\\
\includegraphics[width=8.9 cm]{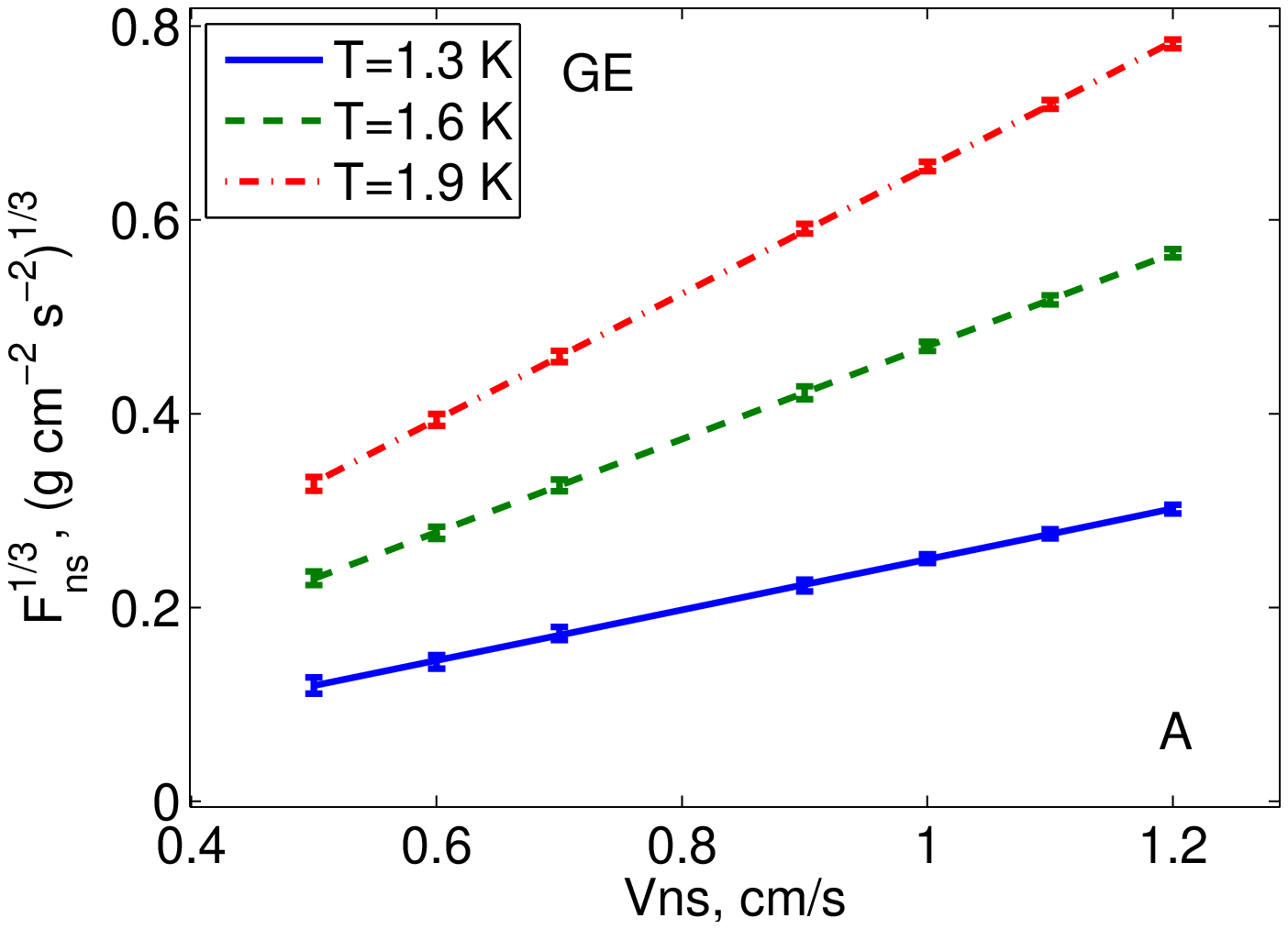}
\includegraphics[width=8.9 cm]{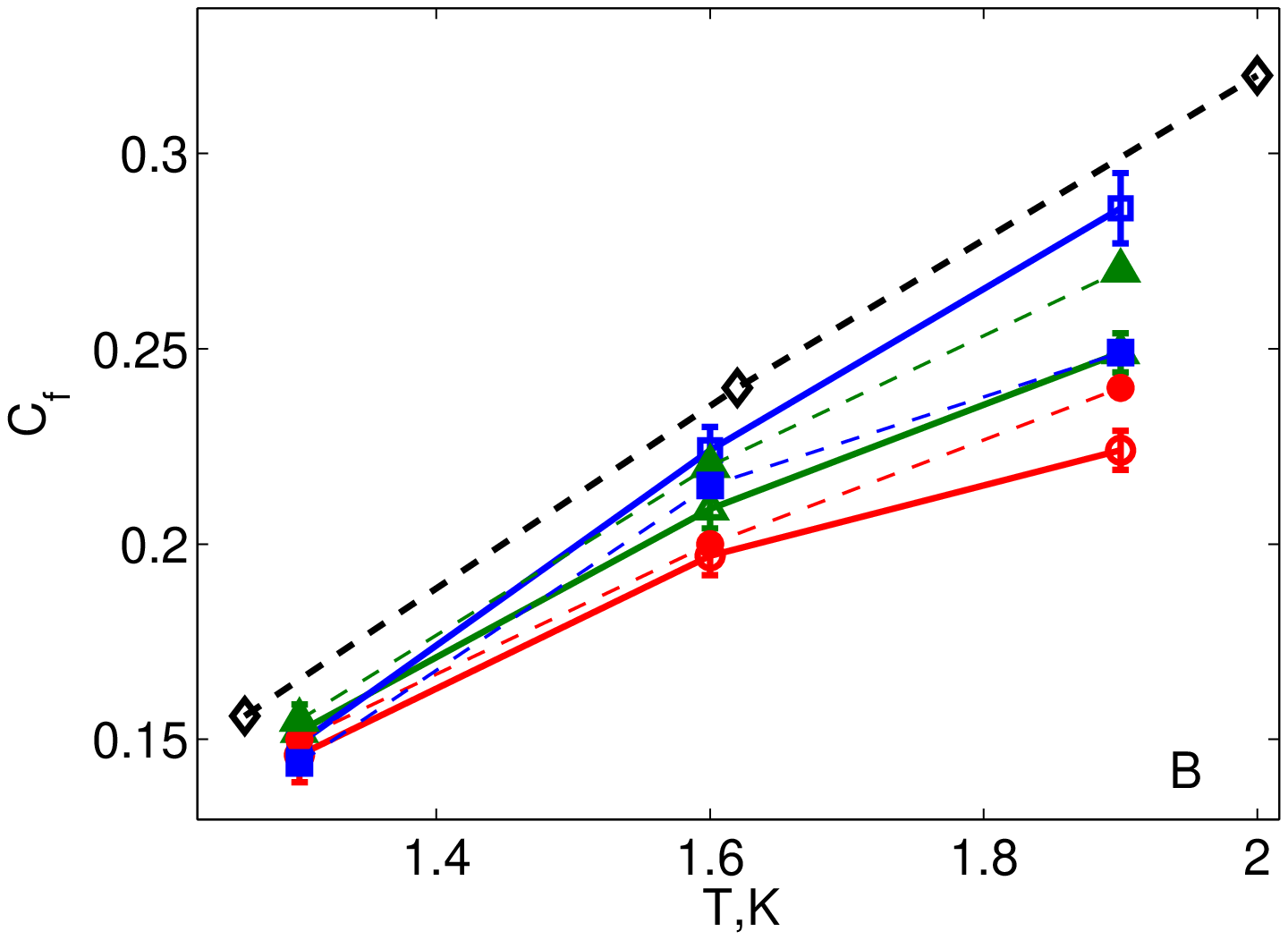}\\

 \caption{\label{f--fric}Color online. Panel A: The friction force density  \eqref{FF} vs counterflow velocity $V\sb{ns}$. The symbols with errorbars are data, the lines are fit according to  Eq.~\eqref{FnsB}. GE-criterion.  Panel B: The friction force coefficients $C\sb{f}$  as a functon of temperature for different reconnection criteria and their LIA estimates. Thick solid lines with open symbols -$C\sb{f}$ obtained by fit according Eq.~\eqref{FnsB} (red cicles -- GC, green triangles -- GEC, blue squares -- DC). Thin dashed lines with filled symbols are the LIA estimates $C\sb{vt}\Sp{LIA}$ Eq.~\eqref{CFL} (the symbols and colors are the same as for  $C\sb{f}$). The thick dashed line with open diamonds -- the results of Schwarz\cite{Schwarz88}. }
 \end{figure*}

The scaling of the mutual friction force  $ F\sb{ns}\propto V\sb{ns}^3$ is well obeyed in all simulations with all three criteria, as we illustrate in Fig.~\ref{f--fric}A for GE-criterion.  Their fit allows to find   coefficients $C\sb f$ plotted in Fig.~\ref{f--fric}B.

 Notice that analytical expression for $C_f$ can be found by    considering different contributions to the integral $J$  in Eq.~\eqref{FF}  with  the only local-induction contribution to the vortex velocity~$V\sb{si}\Sp{LIA}$\cite{Schwarz88}
\begin{equation}\label{CFL}
 C_f\Sp{LIA}\approx\Big (\frac{ c\Sb L}{\widetilde \Lambda}\Big) ^{2/3} (I_\parallel-c\Sb L I_\ell)^{1/3}\ .
\end{equation}
 Like in Eq.~\eqref{CLIA} we have added here superscript `` $\Sp {LIA}$ " to stress approximated character of the relation obtained in the Local-Induction Approximation.

In Fig.~\ref{f--fric}B we compared  the coefficients $C\sb f$  and $ C_f\Sp{LIA}$  for different reconnection criteria. One sees that they  almost coincide for $T=1.3$~K. At $T=1.9$~K our results show significant spread of about 25\% with $C\sb f=0.22$ for GC and $0.29$ for DC. There is again a discrepancy in the behaior of $ C_f\Sp{LIA}$: while for GC and GEC   $ C_f\Sp{LIA} > C\sb f$, especially for $T=1.9$~K, the LIA estimate is smaller than $C\sb f$ for DC at all temperatures.  The results of Schwarz\cite{Schwarz88} are larger than our values of $C\sb f$ and their LIA estimates. Interestingly, here the results for DC are the closest to Schwarz's results, including linear in $T$ behavior, albeit the largest VLD and, therefore, worst conditions for comparison with LIA results. This again confirms that the closeness of LIA and Biot-Savart results should not be taken too seriously.

\begin{figure}
\includegraphics[width=8.9 cm]{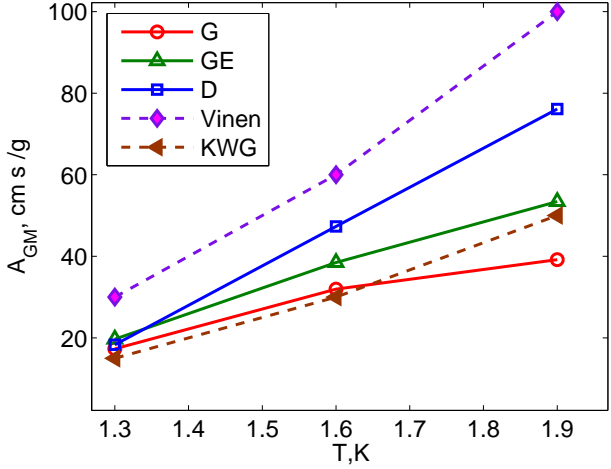}

 \caption{\label{f--gm}Color online. The Gorter-Mellink coefficients $A\sb{GM}$  as a function of temperature for different reconnection criteria. Solid lines with open symbols -$A\sb{GM}$ obtained by fit according to  Eq.~\eqref{GM} . The dashed lines with filled symbols are the exerimental results of Vinen~\cite{Vinen57} and  Kramers, Wiarda, and van Groenou~\cite{KWG61}. }
 \end{figure}

The coefficient $C_f$ is directly related by Eq.~\eqref{GMc} to the more
experimentally used Gorter-Mellink constant $A_{\rm GM}$. We plot in Fig.~\ref{f--gm} the values of $A_{\rm GM}$ obtained as a fit according to Eq.~\eqref{GM} as well as some experimental data.  The experimentally measured values of $A_{\rm GM}$, summarized by Arp~\cite{Arp70}, show significant spread. We only plot the results of Vinen\cite{Vinen57} and Kramers el al. \cite{KWG61}(as cited by Arp~\cite{Arp70}).  As it is clearly seen, all our values (except for GC at $T=1.9$~K) fall between the representative experimental results. This means that i) we get the correct order of magnitude and correct $T$-dependence of the Gorter-Mellink coefficient; ii) the direct comparison with particular experimental results is complicated and subject to the same difficulties as for $\gamma$.


\begin{figure}
\includegraphics[width=8.9 cm]{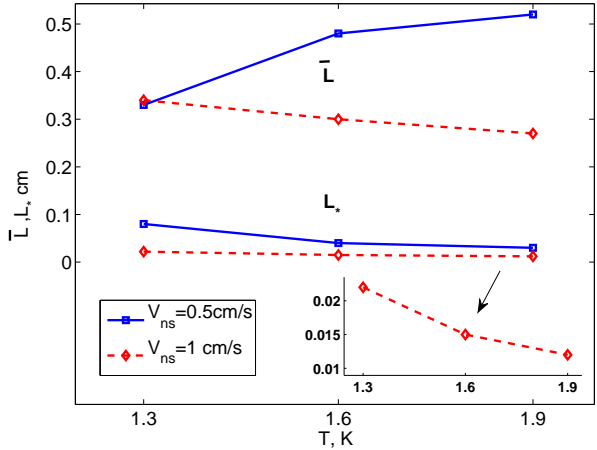}
 \caption{\label{f:11}Color online. Comparison of the mean loop length $\overline L\sim (0.3 \div 0.5)\,$cm  and the most probable loop length $L_*\sim (0.08 \div 0.01)\,$cm  at different temperatures  and values of $V\sb{ns}$. Inset shows details of  $L_*$ for $V\Sb{ns}=1$ cm/s. GEC. }
 \end{figure}

 \begin{figure*}
 \begin{tabular}{c c c}

   A & B & C \\
   \includegraphics[width= 5.9 cm]{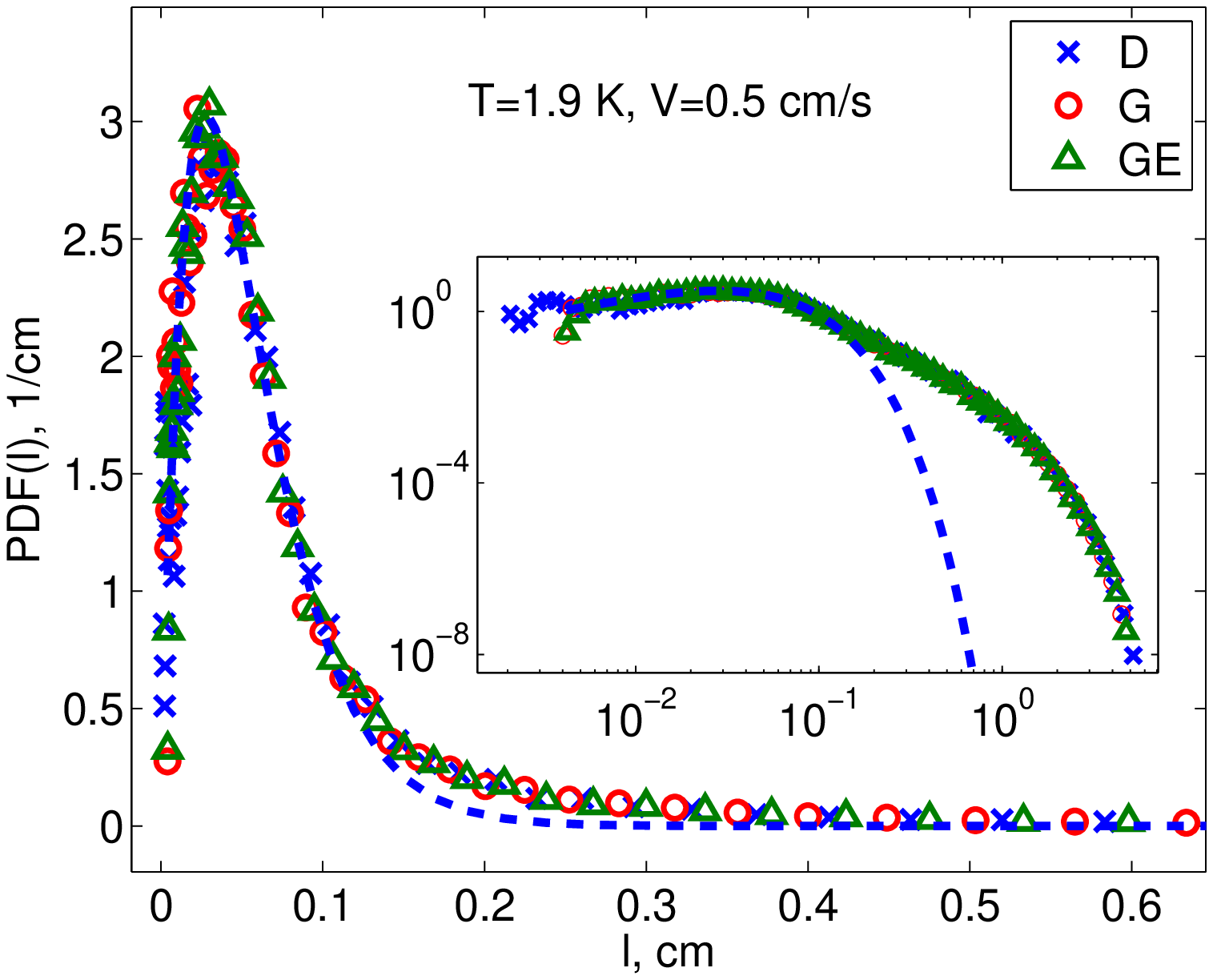} &
   \includegraphics[width= 6.2  cm]{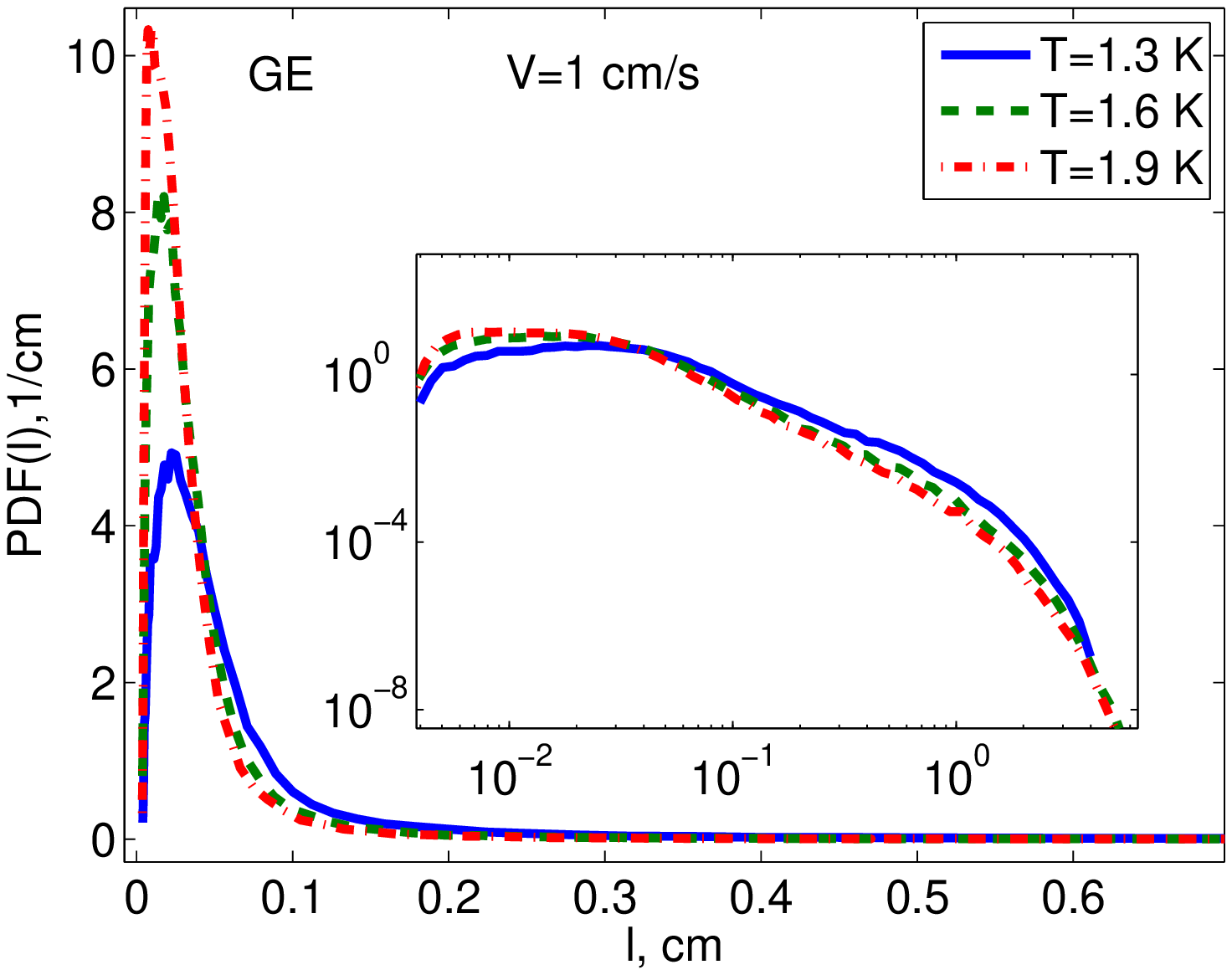} &
    \includegraphics[width= 6.2  cm]{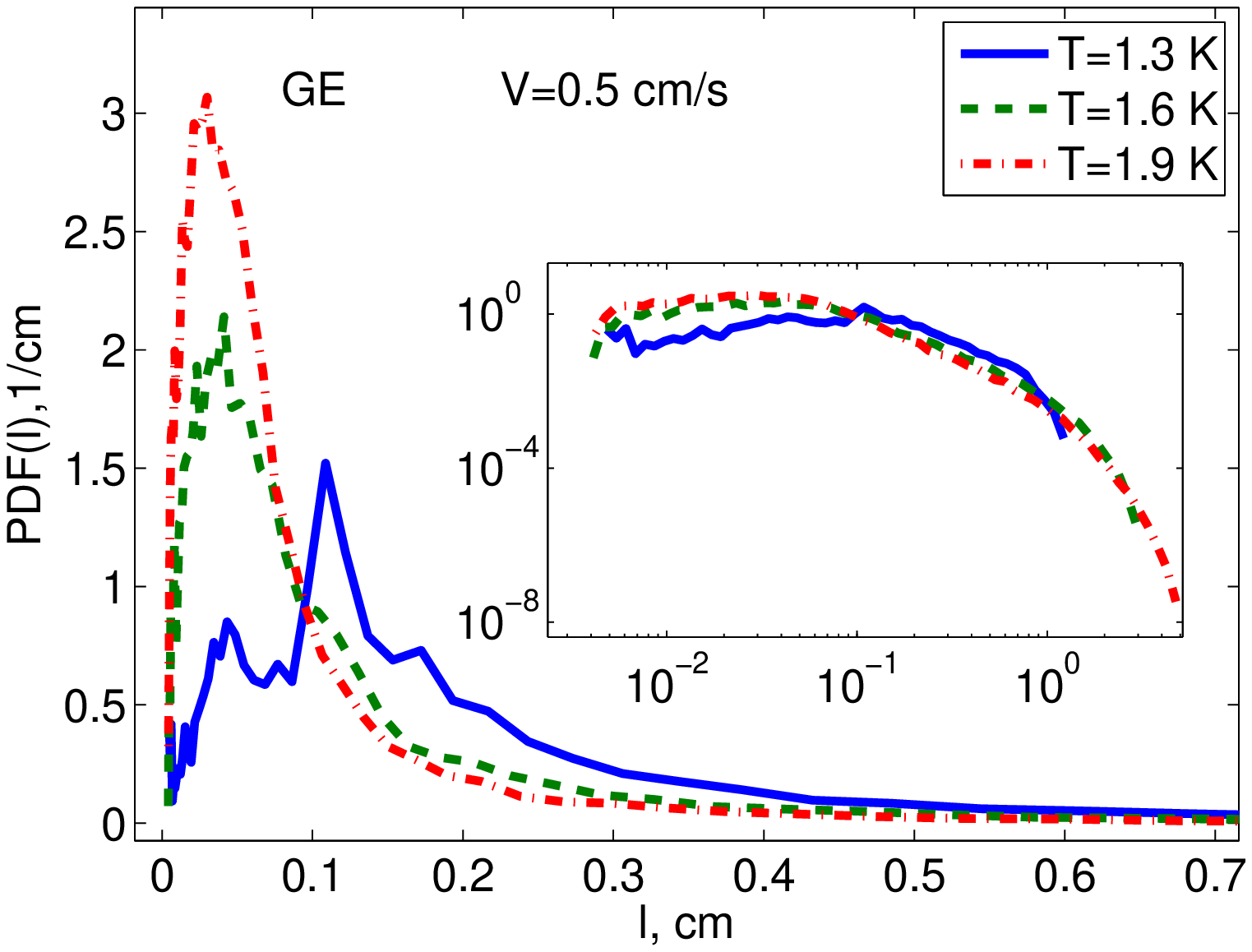} \\

 \end{tabular}
\caption{\label{f:12} Color online.  PDF of the vortex-loop length $l$, $\C P(l)$. Panel A:
 $T=1.9$, $V\sb{ns}=1\,$cm/s, three reconnection criteria. Blue dash lines shows exponential core of the PDF~\eqref{PDFsA}.  Panel B: $V\sb{ns}=1\,$cm/s,
for $T=1.3,\ 1.6,\ 1.9\,$K, GEC.  Panel C:  $V\sb{ns}=0.5\,$cm/s,
for $T=1.3,\ 1.6,\ 1.9\,$K, GEC.  Insets show the same PDFs in log-log scale.}
\end{figure*}

%
\begin{figure}
\includegraphics[width=9.2 cm]{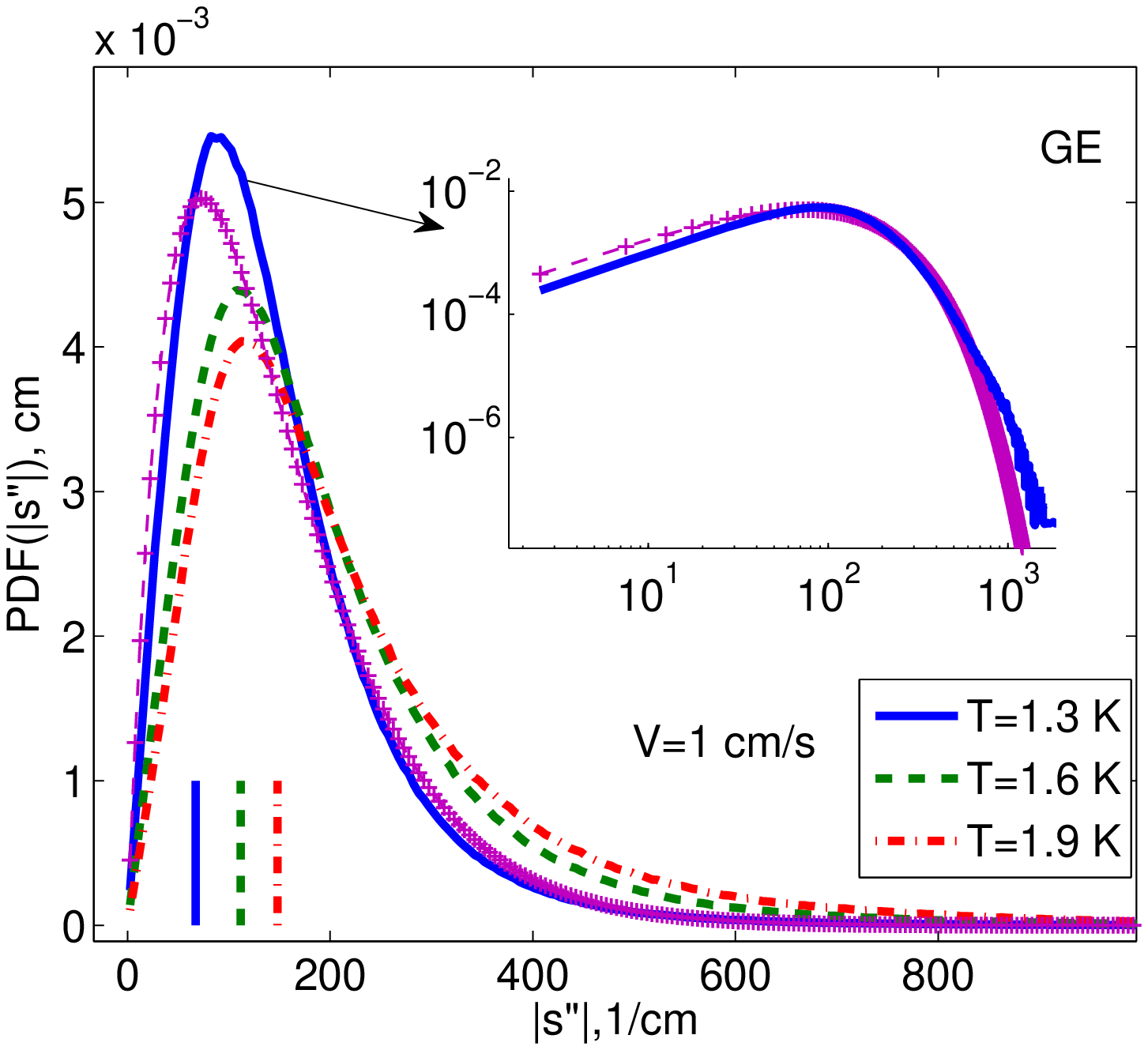}
\caption{\label{f-13}Color online.   PDF of local line curvature  for 3 temperatures and $V_{\rm ns}=1$ cm/s, GEC. Vertical lines near horizontal axis show $1/\ell$ at given conditions.
Inset: the PDF(s) for $T=1.3$ K (blue solid line) and the fit by Eq.\,\eqref{PDFsB}
(purple crosses).  }
\end{figure}

\subsection{\label{sss:loops} Mean and most probable loop lengths}
The temperature and $V\sb{ns}$ dependence of the mean loop length $\overline L$ are shown in Fig.\,\ref{f:11}. One sees that at $T=1.3\,$K $\overline L\approx 0.33\,$cm  \ for both $V\sb{ns}=0.5$ and $V\sb{ns}=1$ cm/s but their $T$ dependence is different: $\overline L$ \emph{increases} with $T$  for $V\sb{ns}=0.5$~cm/s and  \emph{decreases} for $V\sb{ns}=1$~cm/s. Remarkably, the most probable loop length defined in Eq.\,\eqref{PDFsA} and also shown in  Fig.\,\ref{f:11}, is essentially smaller, falling below  $0.015$ cm for the most dense tangle ($T=1.9\,$K and $V\sb{ns}=1\,$cm/s).  We return to this fact below in  Sec.~\ref{sss:PDF-l}.

\section{\label{ss:loc}Detailed  statistics of the vortex tangle}
 As we mentioned in the Introduction, the mean characteristics of the vortex tangle, studied in previous Sec.~\ref{s:stat},   provide important but very limited information on the tangle properties. Much more detailed statistical  information on local tangle properties is  required for better understanding of basic physics of counterflow  turbulence as well as  for further advance in its analytical studies. This information may be obtained  from probability distribution functions of local tangle properties (like line curvature), of global vortex-loop characteristics (e.g. their lengths) and from corresponding (cross)-correlation functions (e.g. of vortex line orientations, of loop length vs. mean curvature).  Bearing in mind that this information will not be avaliable from experiments in foreseeable future, the only way to get it today is from numerical simulations. This is the motivation and the subject of present Section.

\subsection{\label{sss:PDF-l} Probability density function (PDF) of vortex-loop lengths}
Turning to a more detailed description of the tangle structure we plot in Figs.~\ref{f:12} the PDF of the vortex-loop length, $\C P(l)$ for $T=1.3,\ 1.6,\ 1.9\,$K and $V\sb{ns}=0.5, \ 1.0\,$cm/s.   Panel A shows that   $\C P(l)$ is practically independent of the reconnection criterion at least for cases with moderate to  large line density.

The second observation is that the core of the PDF $\C P(l)$ may be approximated by a simple formula

\begin{equation}\label{PDFsA}
\C P(l) \simeq   \psi \C P_0(l)\,, \quad  \C P_0(l)\equiv  \frac{  l} {L_*^2}\exp \Big ( - \frac{  l}{L_*}\Big )\,,
 \end{equation}
 shown in  Fig.\,\ref{f:12}A, left, by blue dashed line.
 The function $\C P_0(l)$ is normalized to unity: $\int_0 ^\infty \C P_0(l)\, dl=1$\,.
 The fitting parameter $L_*$ corresponds to the maximum of the core function~\eqref{PDFsA} and simultaneously  to the maximum of $\C P(l)$. Therefore we called it the most probable length as plotted in Fig.\,\ref{f:11}.
    The second fitting parameter $\psi$ shows the fraction of loops that belong to the core and define $L_*$. The value of $\psi\simeq 0.2 $ for $V\sb{ns}=0.5\,$cm/s and $\psi\simeq 0.3 $ for $V\sb{ns}=1\,$cm/s  is only very weakly dependent  on $T$. We conclude that the majority of loops  belongs to the long tail, which is clearly seen in the insets in  Fig.~\ref{f:12}. For loops lengths slightly exceeding 0.1 cm the PDF tails  exhibit a power-law-like behavior over an interval of lengths about half a decade with a non-universal exponent ranging between -2 and -3 for different $V\sb{ns}$ and temperatures. The mean value of the loop length $\overline L$ is determined by the tails and, as we have shown in Fig.\,\ref{f:11}, is much larger than $L_*$.

Panels B and C of Figs.~\ref{f:12} show how $\C P(l)$ varies with temperature and $V\sb{ns}$. As we know, with increasing  $T$ and $V\sb{ns}$ the VLD increases, the intervortex distance becomes smaller and the reconnection rate increases. All that shifts the PDF  $\C P(l)$ toward shorter loops. For the least dense case ($T=1.3\,$K and $V\sb{ns}=0.5\,$cm/s) the PDF looks very indented, probably  because of the lack of statistics.

\subsection{\label{sss:PLC} PDF  of the line-curvature}

The next object of interest is the PDF of local curvatures $\C P(|s''|)$, shown in Fig.\,\ref{f-13} for $T=1.3,\ 1.6,\ 1.9\,$K and $V\sb{ns}=1\,$cm/s.  These PDFs linearly vanish for $|s''|<\widetilde S$ and exponentially vanish for $|s''|>\widetilde S$. We suggest an interpolation formula between these two asymptotes, which is very similar to Eq.\,\eqref{PDFsA}:
\begin{equation}\label{PDFsB}
\C P (|s''|)\simeq \frac{6\,|s''| }{\widetilde S^2}  \exp \Big ( -\frac{\sqrt 6\,|s''|}{\widetilde S} \Big )\ .
\end{equation}
Notice  that Eq.\,\eqref{PDFsB} has no fitting parameters, it just involves the RMS curvature $\widetilde S$. As one sees in inset in  Fig.\,\ref{f-13}, this equation describes reasonably well the entire form of  $\C P (|s''|)$. Accepting Eq.\,\eqref{PDFsB} we can find the ratio $\overline S/\widetilde S=\sqrt {2/3}$. Correspondingly, the ratio $c_1/c_2$ defined by Eqs.~\eqref{sL} is also $\sqrt {2/3}$.  This prediction agrees well with our numerical results for $c_1$ and $c_2$ given in Tab.~\ref{t:4}. For example, for GEC the ratio  $\sqrt 3 c_1 / \sqrt 2 c_2$ is equal to 0.985, 1.009 and 1.018 (instead of the predicted value of unity) for $T=1.3,\ 1.6$ and  $1.9\,$K, respectively.

This equation also allows us to find the most probable curvature $S_*\simeq \widetilde S/\sqrt 6\simeq \overline S/2$. All three characteristic curvatures are determined by the exponential PDF~\eqref{PDFsB} and therefore they are of the same order of magnitude. This is different from the  characteristic loop lengths, where $L_*$ is determined by the exponential core of the PDF~\eqref{PDFsA}, while $\overline L\gg L_*$ is determined by the long power-law tail of the PDF.

\begin{figure}

\includegraphics[width=9.5 cm]{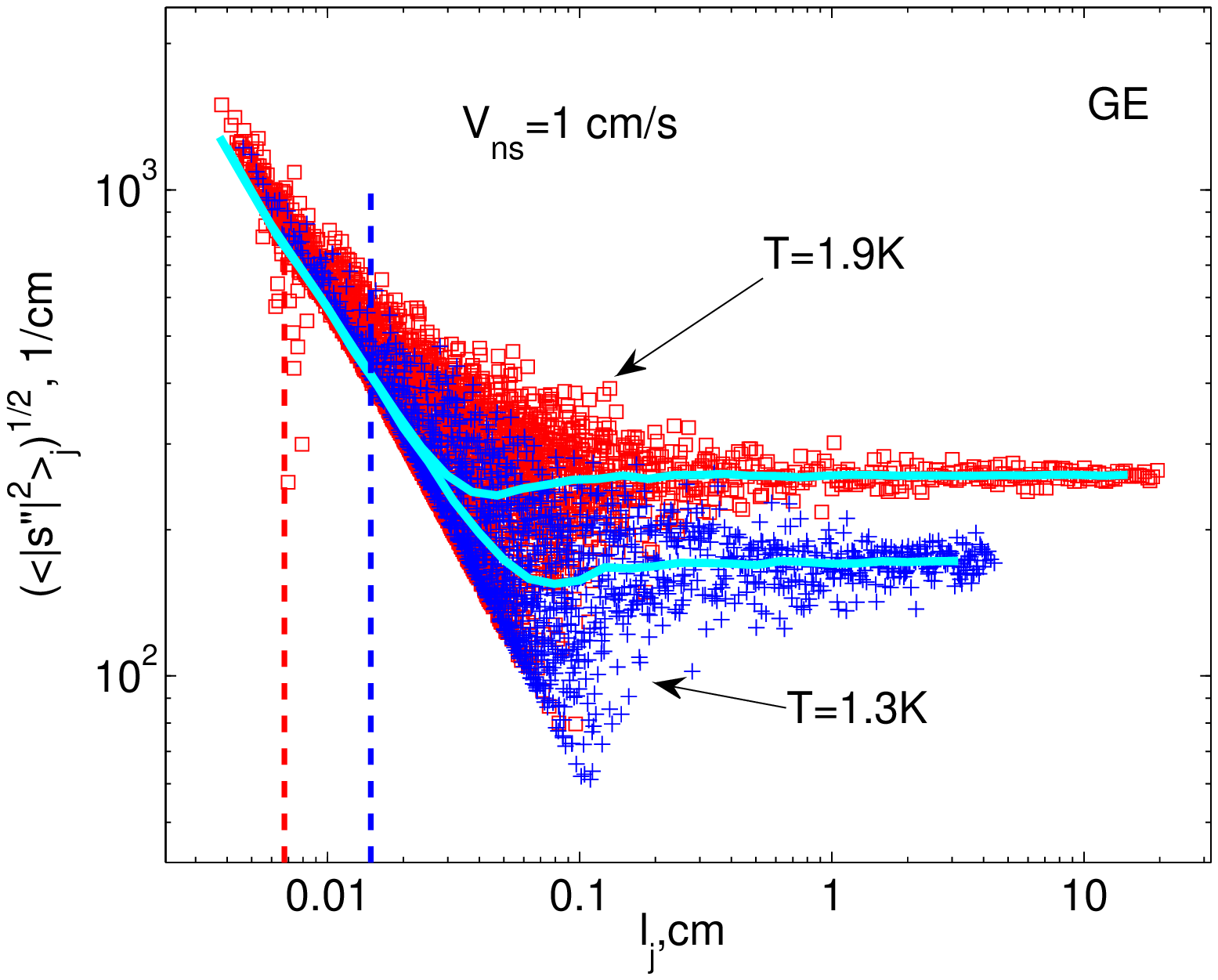}
\caption{\label{f-14}  Diagrams of RMS of loop-curvature $\widetilde {s''_j}$ vs. length-loop $l_j$ for  $V_{\rm ns}=1\,$cm/s with GEC for two temperatures. The plus sign($+$) denote data for $T=1.3$K, the squares -- for $T=1.9$K.   Light blue lines correspond to  the  mean curvature of loops vs their length. The intervortex distances  are  denoted by vertical dashed lines, left -- for  $T=1.9$K, right -- for $T=1.3$K.}
\end{figure}

\begin{figure*}
\begin{tabular}{|c|c||c|}
  \hline
  A & C & E \\
 \includegraphics[width=6.3 cm]{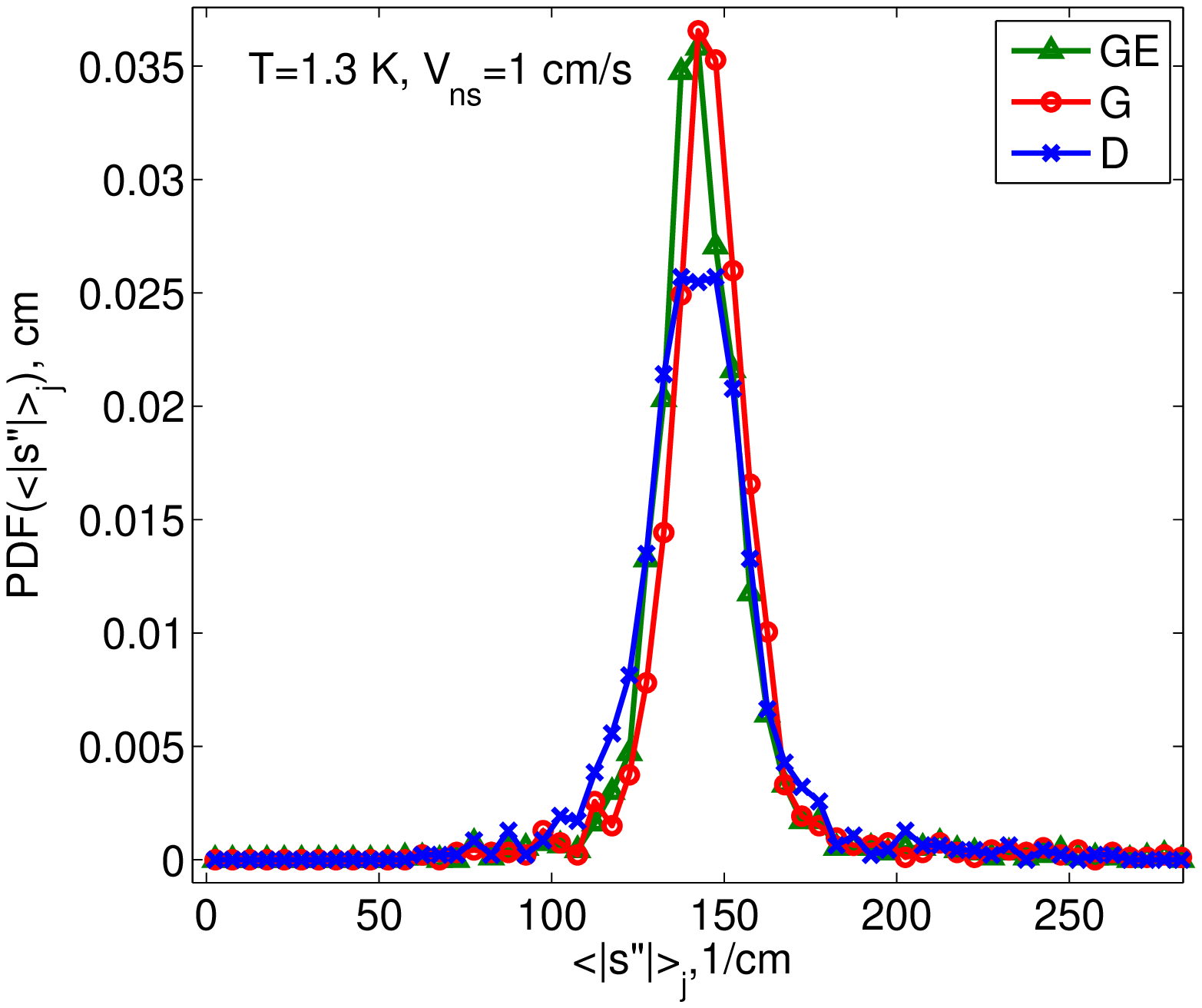}  &
   \includegraphics[width=6.3 cm]{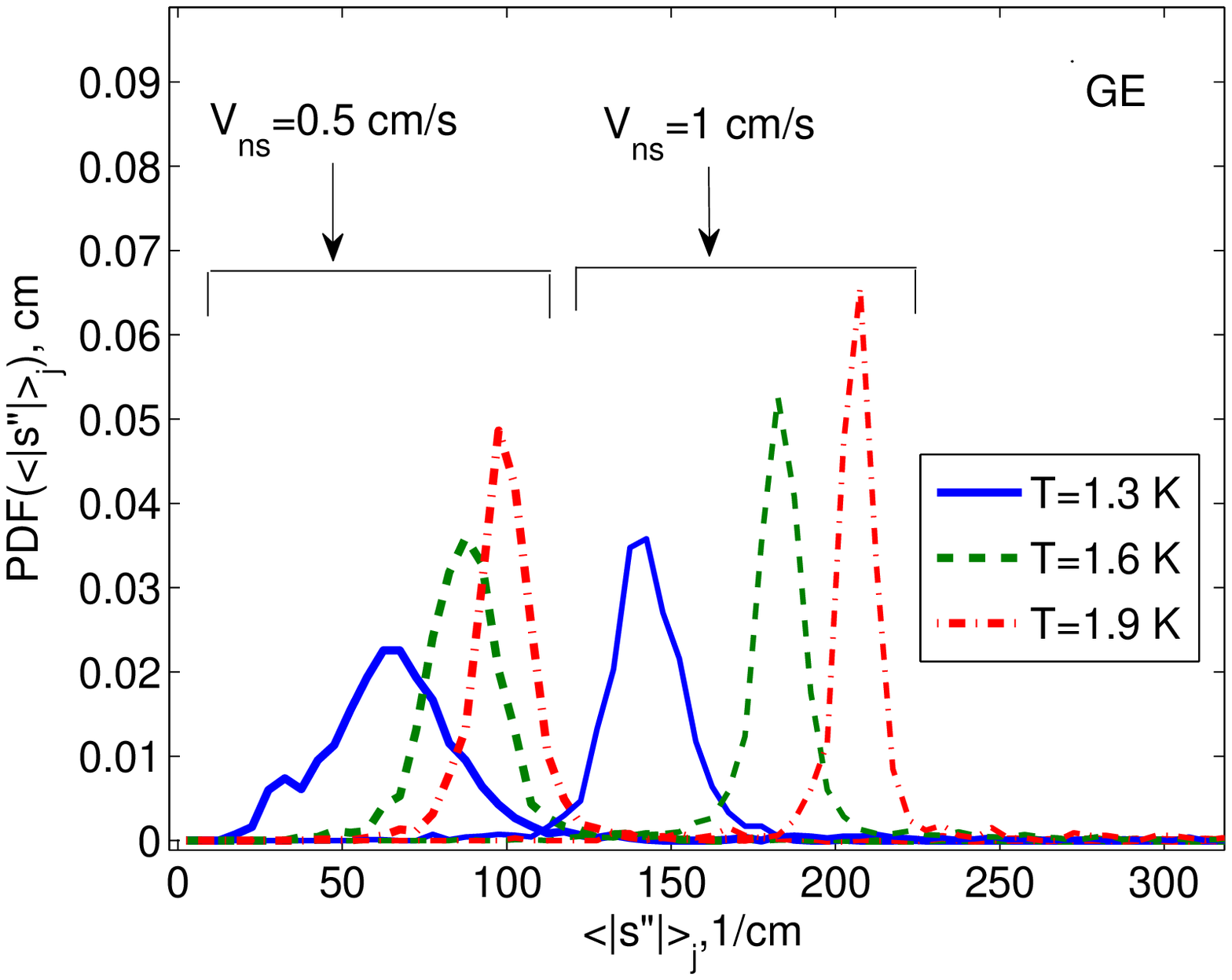} &
  \includegraphics[width=4.7  cm]{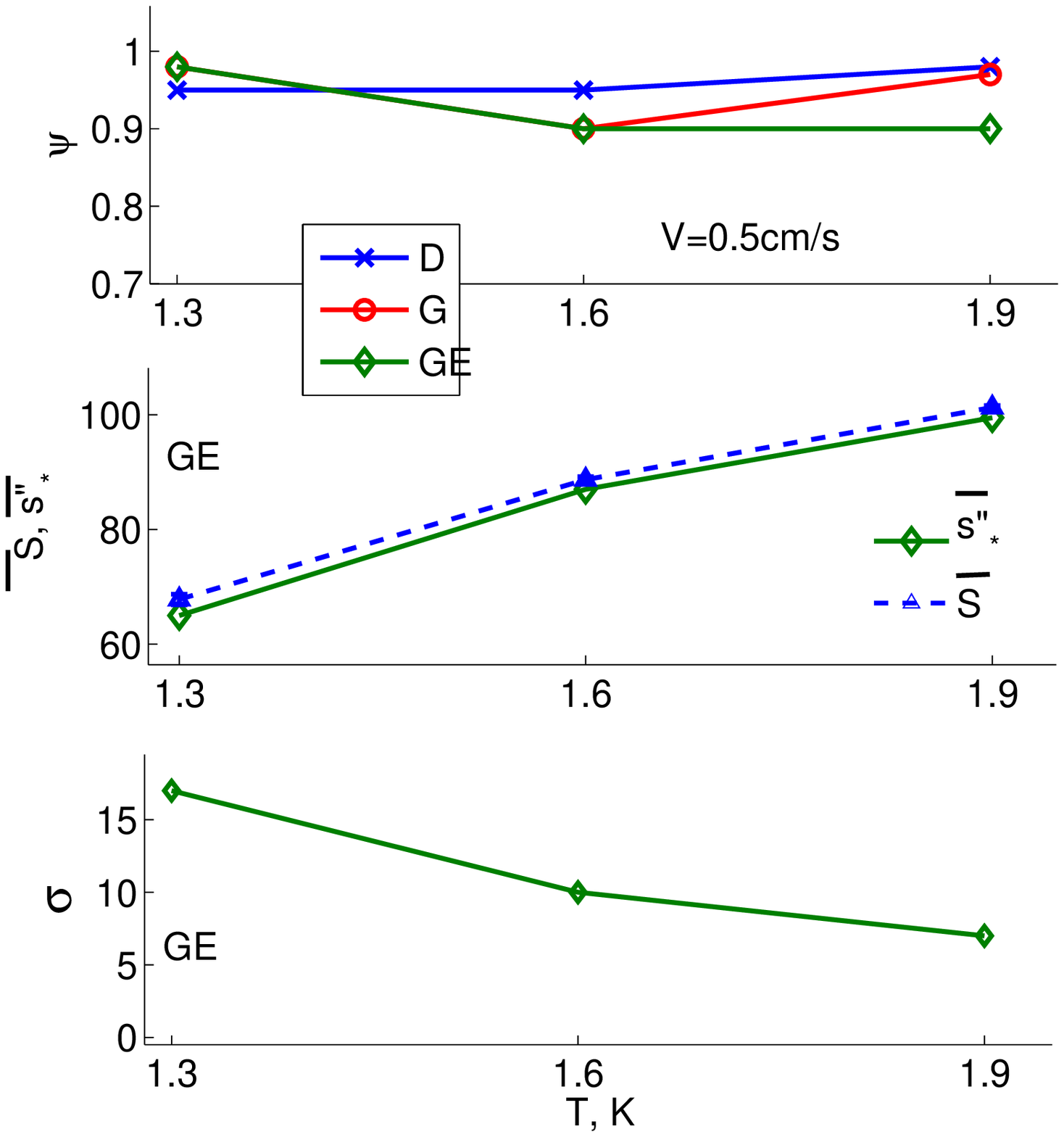}  \\
  \hline
   B & D  & F \\
   \includegraphics[width=6.3 cm]{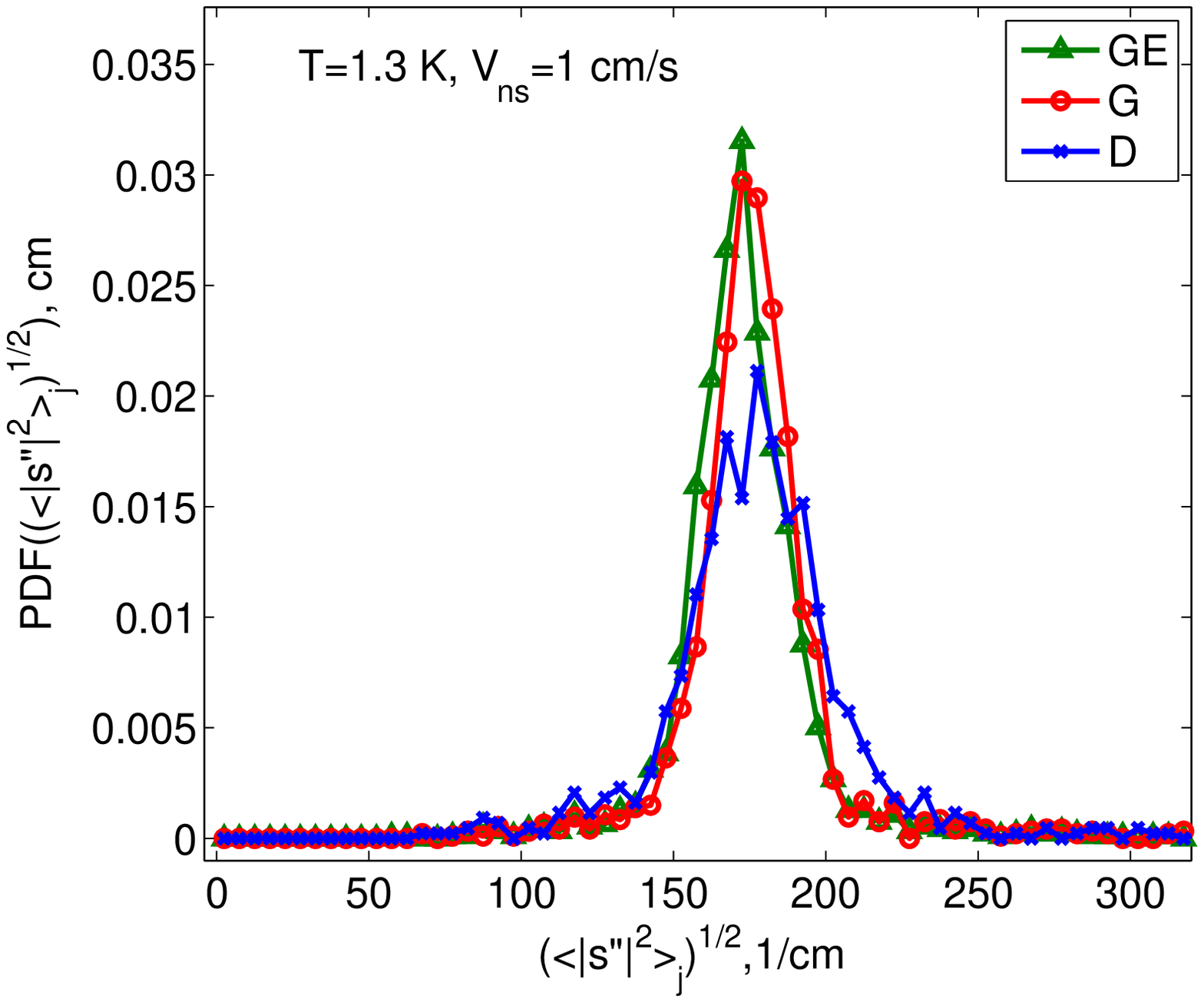} &
  \includegraphics[width=6.3 cm]{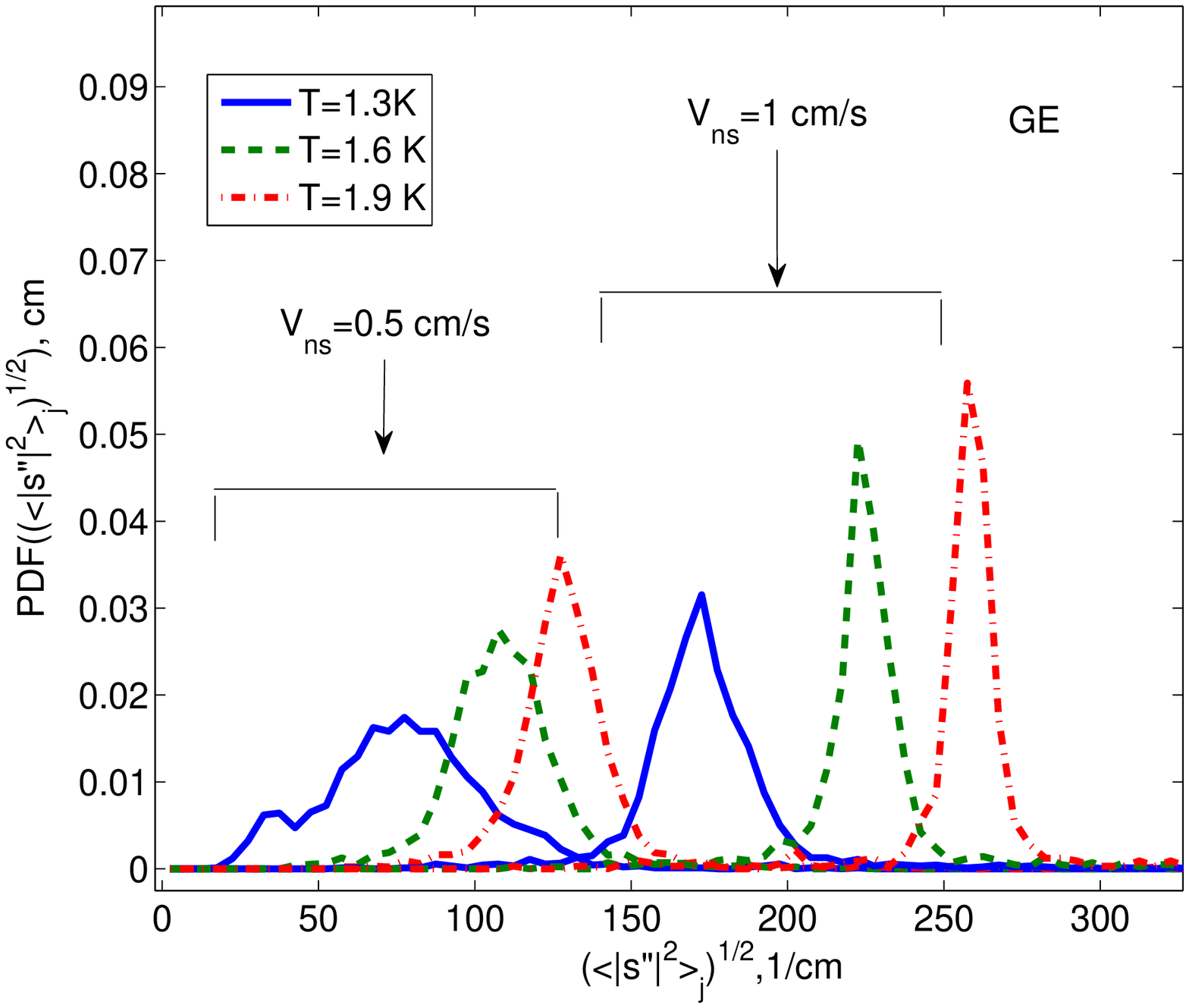} &
  \includegraphics[width=4.7  cm]{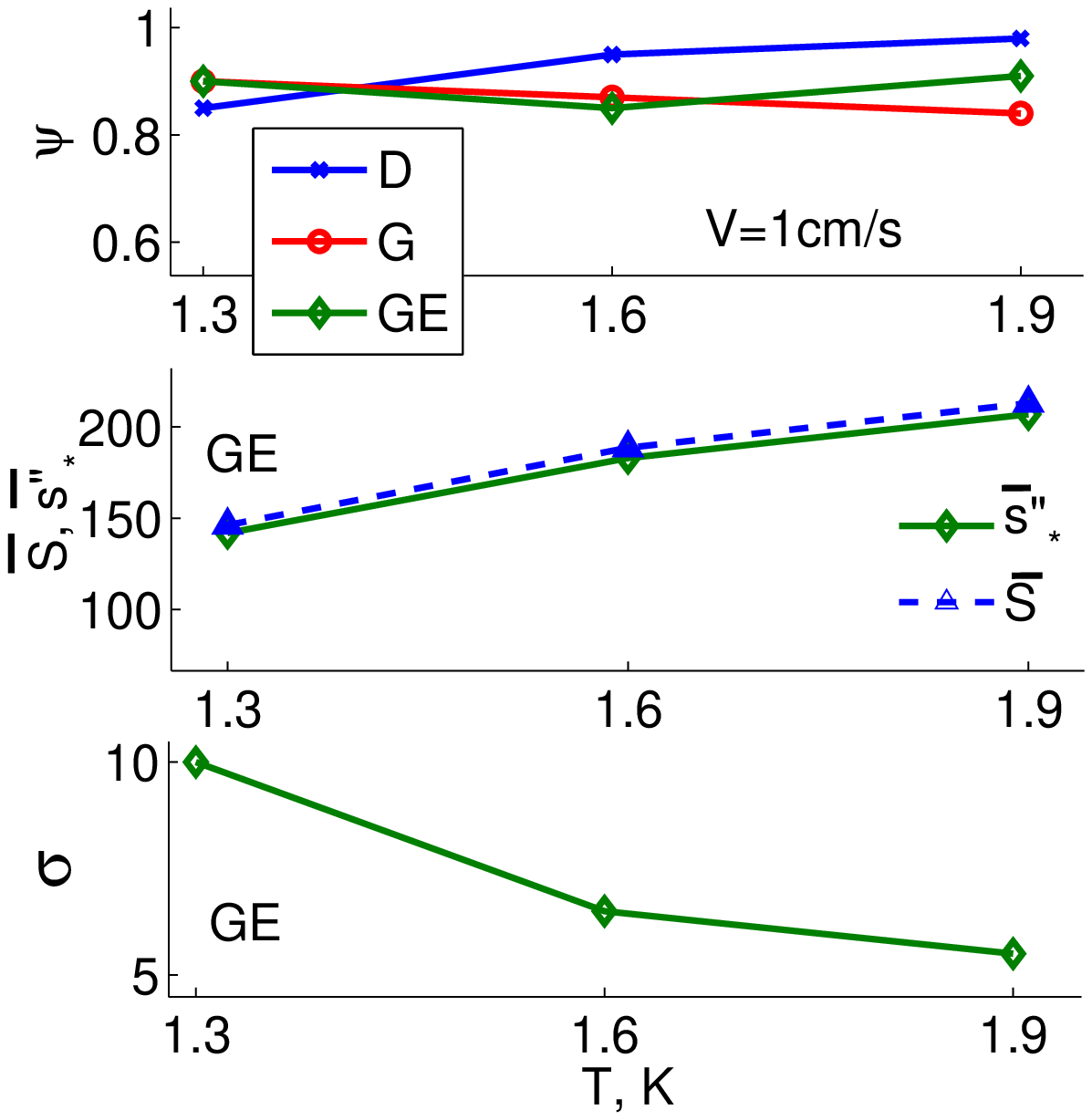}
   \\
  \hline

\end{tabular}
\caption{\label{f-15}Color online.  Comparison of PDFs of the mean (Panel A) and RMS (Panel B)  loop curvature for three reconnection criteria ($T=1.3\,$K and $V\sb{ns}=1\,$cm/s). The lines serve to guide the eye only. Temperature and $V\sb{ns}$ dependence of the PDF of  the mean (Panel C) and RMS (Panel D) loop curvature (for GEC). Parameters of the Gaussian fit Eq.~\eqref{PDF} for the PDF of the mean curvature are shown in Panels E for ($V\sb{ns}=0.5\,$cm/s) and F for ($V\sb{ns}=1\,$cm/s).}
\end{figure*}

\subsection{\label{ss:cor} Correlation between loop length $l_j$ and  RMS of the loop curvature $\widetilde s''_j$}
Knowing the PDFs Eq.~\eqref{PDFsA} and  Eq.~\eqref{PDFsB} of the loop length and line curvature  separately we now come to the next question:  ``How are these objects correlated?" In particular, do all loops (long and short) have more-or-less the same RMS and mean curvatures  $\widetilde {s''_j}$ and $\overline{s''_j}$ [defined by Eqs.~\eqref{jcurv}] or do short loops have larger values of $\widetilde s''_j$?
 To resolve this question we plot
   numerous  $(\widetilde{s''_j}\,, \,l_j)$-points belonging to all loops in the statistical set of the tangle configurations, computed for particular $T$ and $V\sb{ns}$. These points   form a  $(\widetilde{s''_j}-l_j)$-diagram shown in Fig.\,\ref{f-14} for $T=1.3\,$ and $1.9\,$K with $V\sb{ns}=1\,$cm/s.

The majority of points are located to the left of $l_j=0.1\,$cm according to the PDFs $\C P(l_j)$ shown in Figs.~\ref{f:12}. Next,  for small $l_j$ below 0.1\,cm one sees a sharp boundary that restricts  from below available $S_j$ at given $l_j$. This boundary corresponds to the minimal possible  RMS loop curvature $\widetilde {s''_j}=2\pi / l_j$, realized for ideal circle of radius $1/\widetilde {s''_j}$ with  $l_j=2\pi /\widetilde {s''_j}$. Some points below this line for small $l_j$  are the result of the finite space resolution in the continuous vortex-line presentation via a discreet set of points: the smallest loops, displayed in Fig.\,\ref{f-14} are parameterized by only three points. Long loops have curvatures well concentrated around the conditional (with fixed $l$) RMS value
\begin{equation}\label{meanS}
\widetilde S(l)=  \langle \widetilde {s''_j} \rangle   _{l_j=l} \,,
\end{equation}
shown in Fig.\,\ref{f-14} by blue lines (upper line for $T=1.9\,$K and by the lower line for $T=1.3\,$K). One sees that $\widetilde S(l)$ is practically
independent of $l$  for $l$ that exceeds substantially the intervortex distance $\ell$,  denoted by vertical lines. This is an evidence in favor of the natural expectation that local properties of long loops are independent of their total length.

\begin{figure*}
\begin{tabular}{|c|c|c|}
  \hline
  A & B & C \\
  \includegraphics[width=5.75 cm]{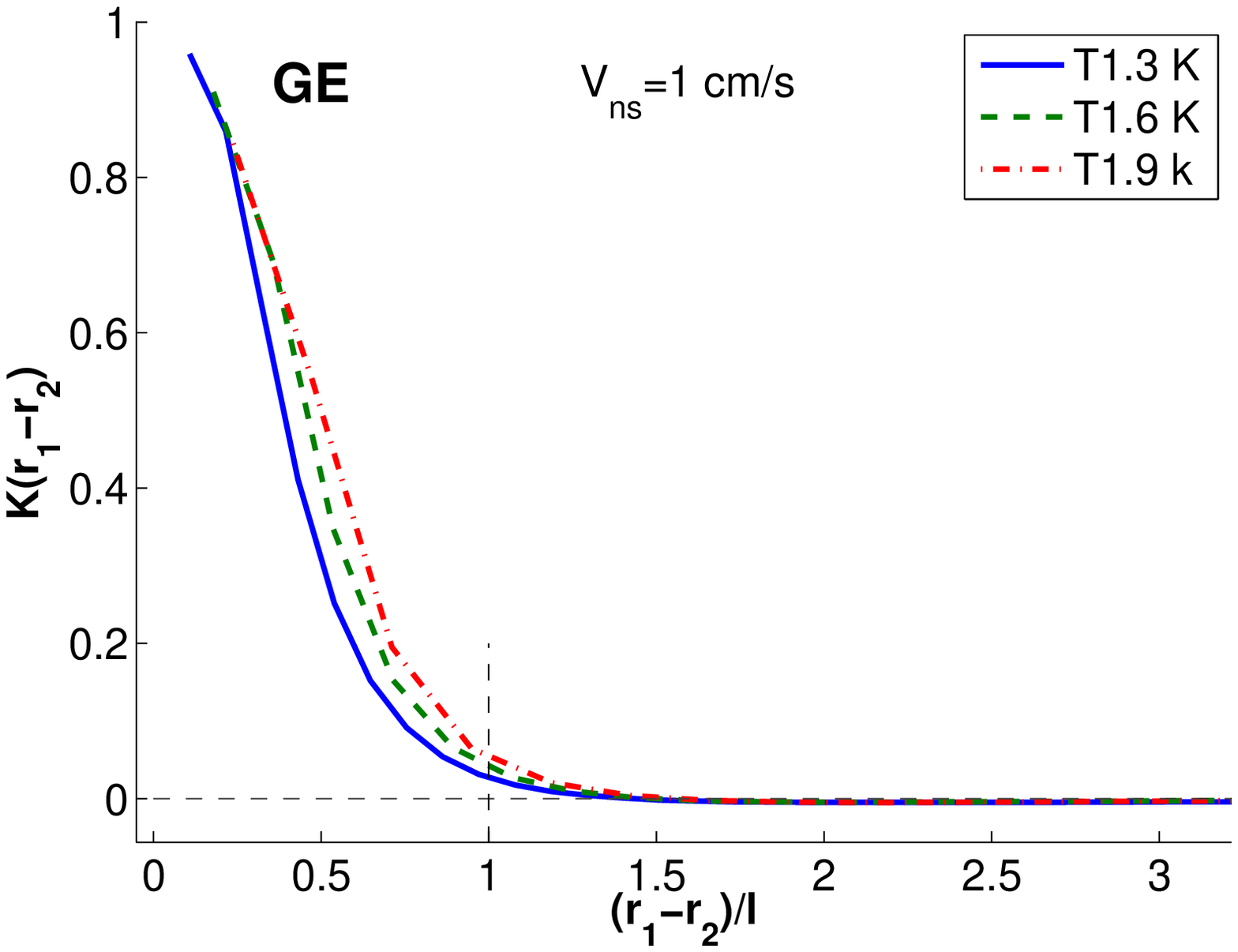}&
\includegraphics[width=5.75 cm]{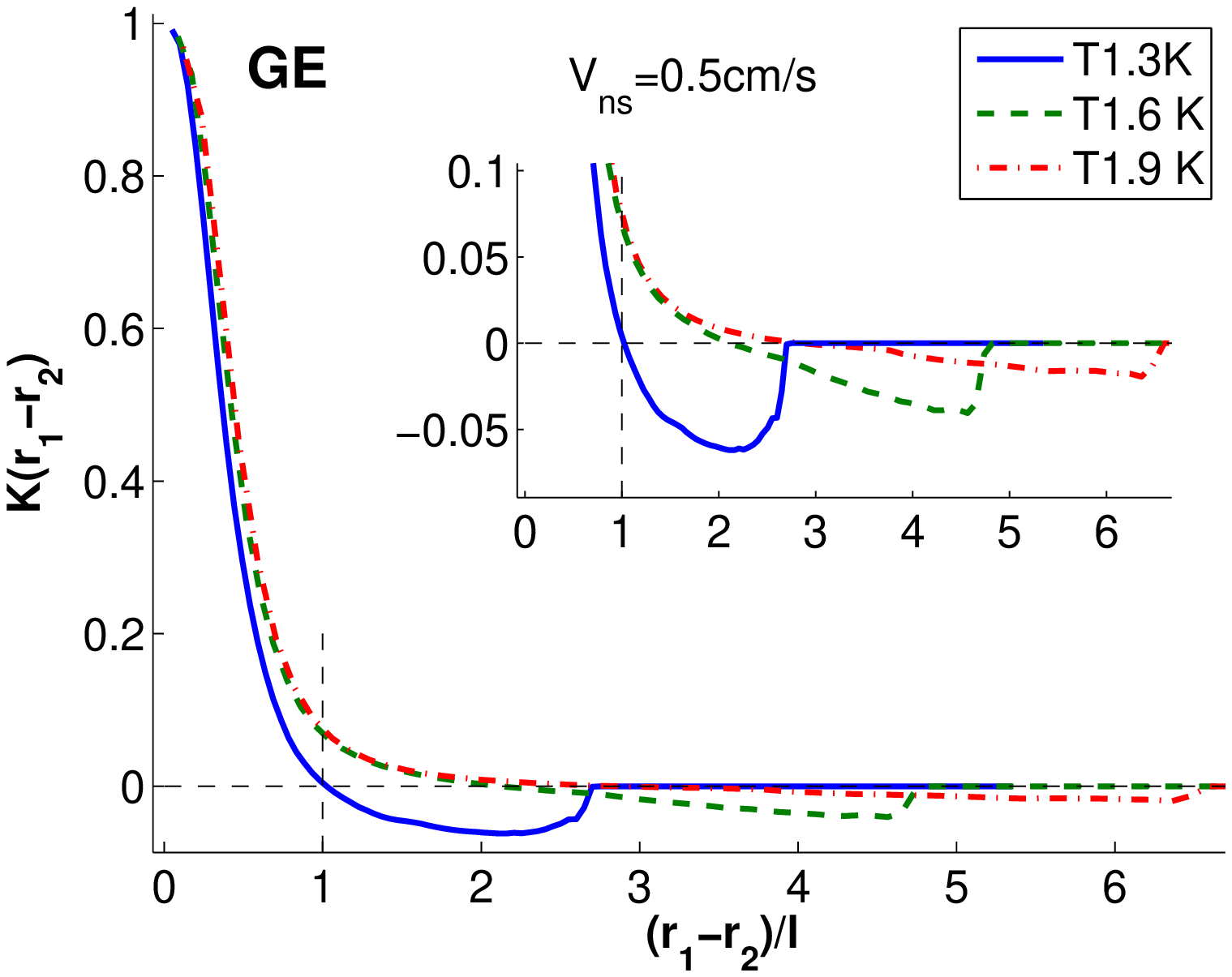}&
\includegraphics[width=5.75 cm]{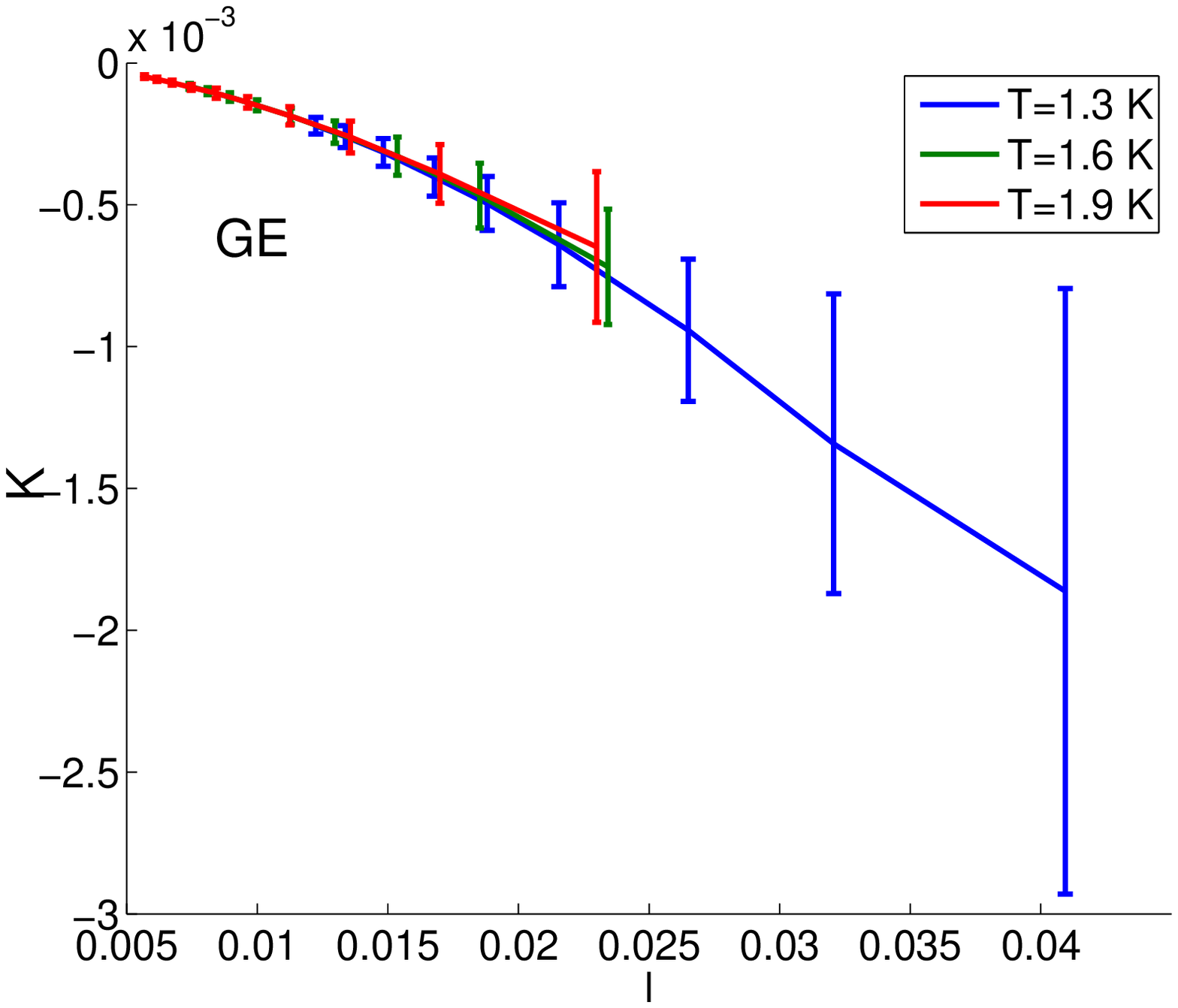} \\
  \hline
\end{tabular}
\caption{\label{f:16}Color online. Panels A and B: The orientation correlation function
for different temperatures and two counterflow velocities. Inset:
details of long distance tail. The distance is measured in
 units of intervortex distance. Panel C: The mean polarization of the tangle
as a function of the intervortex distance. GEC .}
\end{figure*}

\subsection{\label{ss:PDF} PDFs of the mean and RMS loop curvature}
Full information about the statistical distribution of $\widetilde {s''_j}$ for loops with given length $l$ can be found from the conditional PDF, $\C P(\widetilde {s''_j}, l_j=l)$.  In particular, this PDF would describe the difference in properties of short and long loops. Nevertheless, for the beginning  we will restrict ourselves by analysis of less detailed  information:  unconditional PDFs of the mean and the RMS loop curvature, $\C P(\overline{s''})$ and $\C P(\widetilde{s''})$,  shown in Figs.~\ref{f-15}, A-D.  In Panels A and B  we compare these PDFs for the three reconnection criteria with $T=1.3\,$K and $V\sb{ns}=1\,$cm/s. They look similarly, at least on a semi-quantitative level.     Therefore to clarify  how these PDFs vary with $T$ and $V\sb{ns}$ it would be sufficient to analyze the results for GEC only, as shown in Panels C and D. One sees that these  PDFs agree with the fact  that the mean and RMS curvatures of the tangle increase with density (or $V_{\rm ns}$) for a given temperature and  are smaller for  higher temperatures for the same density (Fig.\,\ref{f:10}, top).

Both PDFs, $\C P(\overline{s''})$ and $\C P(\widetilde{s''})$, may be roughly approximated as a narrow peak of some width $\sigma$ which is much smaller than the position of its maximum. PDF of the mean curvature looks more regular. Its core is   approximated well enough by a Gaussian
\begin{equation}\label{PDF}
 \C P(\overline {s''})\approx\frac \psi {\sqrt{2\pi}\sigma}\exp \Big \{ - \frac{(\overline {s''}- \overline {s_*''} )^2}{2\sigma^2}\Big \}\,,
\end{equation}
with three fitting parameters: the position of the maximum $\overline {s_*''}$ (the most probable mean-loop curvature), the width $\sigma$ and the total amount $\psi$ of $\overline {s''}$, described by the core of PDF~\eqref{PDF}: $\int \C P(\overline {s''}) d \overline {s''}=\psi$.
This allows us to quantify the differences
between curvature distributions in a wider range of parameters. The temperature dependences of $\overline {s_*''}$,   $\sigma$ and   $\psi$ are shown in Panels E and F for $V\sb{ns}=0.5\,$cm/s and 1.0 cm/s, respectively. One sees that $\psi$ is quite close to unity: $\psi> 0.9$ at $V\sb{ns}=0.5\,$cm/s and $\psi> 0.8$ at $V\sb{ns}=1\,$cm/s for all three reconnection criteria. This means that Eq.\,\eqref{PDF} describes  reasonably well the entire PDF and not only its core. Therefore at our level of description the contribution of non-exponential tail can be ignored.

 Notice that on a semi-quantitative level  there are no differences in the values and behaviors of  $\overline {s_*''}$ and $\sigma$ for the three reconnection criteria. Therefore in Panels E and F we presented these parameters only for GEC. In addition we show (by blue dashed lines) in the same panels the overall (over the entire tangle) mean value of the curvature $\overline S$ which, by definition, has to coincide with the mean (over different loops) of the mean-loop curvature:  $\overline S=\int \overline {s''} \C P(\overline {s''})d \overline {s''}$.   We see that the mean tangle curvature $\overline S$ practically coincides with  the most probable mean-loop value $\overline {s''_*}$. This means that the role of the PDF tail can be ignored, as we stated above on the basis that $\psi \simeq 1$.

The next observation is that $\overline {s_*''}$ (and $\overline S$) increases with temperature and counterflow velocity, i.e. with the tangle density. This agrees with the well known fact $\overline S\propto \sqrt{\C L}$.  A less expected observation is that $\sigma$
decreases with increasing $T$ and $V\sb{ns}$, i.e. in the denser tangle the mean-loop curvatures are less spread around their mean (or most probable) value.

\subsection{\label{ss:CF} Autocorrelation of the vortex orientation}
It was recognized by Schwarz~\cite{Schwarz82} that the structure of the vortex lines is reminiscent of random walks.
As the vortex segments get further apart, their relative orientation becomes more random. To find out at which distances the
correlation between the segment orientation is lost we plot in
Fig.~\ref{f:16}, panels A and B,  the orientation correlation function $K({\bm r}_1-{\bm
r}_2)$, defined by Eq.\,\eqref{Kr}.

A crucial  observation is that  the correlation falls off very fast being almost zero at the
intervortex distance. This result supports Nemorivskii's Gaussian model of $^4$He-vortex tangle~\cite{Nem1}, in which correlation of the orientations disappears at inter-vortex distance $\ell$ and the mean loop length $\overline L \gg \ell$.

Interestingly, for weak counterflow velocities
$V_{\rm ns}=0.5$~cm/s there is a distinct negative correlation (the
segments are anti-parallel) at distances just beyond $\ell$. This can be related with the tendency of close vortex lines to become antiparallel on the way to reconnection.
For stronger $V_{\rm ns}=1$~cm/s, i.e. in more dense tangles, this tendency is masked by the influence of other  neighboring vortex lines. Therefore  such an anti-parallel orientation is not observed.

Averaging these correlation functions over all distances we find that
on average the tangle is slightly polarized and this polarization $K$
depends on the intervortex distance $\ell$, but not on the temperature: see Fig.\,\ref{f:16}C where we plot $K$ as a
function of $\ell$  for three
temperatures. The value of $\ell$ at given $T$ was varied by the counterflow velocity. Here again there is no noticeable difference in the values and dependencies of $K({\bm r}_1-{\bm
r}_2)$ and $K$ for different reconnection criteria, therefore only GEC case is displayed.

 The most important observation in Fig.\,\ref{f:16}C is that $\overline K\sim 10^{-3}$ i.e. is vanishingly small with respect to unity. This means that there is no coherent contribution (of many vortex lines) to the velocity field at large scales  (much above $\ell$). Therefore the energy spectrum of the turbulent vortex tangle, $E(k)$, has to be determined by contributions of individual vortex lines even for $k\ell \ll 1$, up to the box size.

\section{\label{s:Con}On the physics of $^4$He counterflow turbulence}

In this section we present a summary of the result obtained here and in other studies of counterflow turbulence.

\subsection{\label{ss:idea} Idealizations and relevant parameters}
\subsubsection{Spatial homogeneity}
In analogy to classical hydrodynamic turbulence, the basic models of counterflow turbulence are based on the assumption of  spatial homogeneity of the problem. In laboratory experiments on counterflow in $^4$He this can be realized to some extent  in a wide channel or a pipe of transverse size  $H$ that significantly exceeds the intervortex distance $\ell$. For example,  in the super flow  experiments of Ref.~\cite{Ladik-2012} the largest $H=1\,$cm,  while $\ell$ varies (approximately) from 0.1 to $4\cdot 10^{-4}$cm (for $V\sb{ns}\simeq 20\,$cm/s).

In numerical simulations (like ours) the homogeneity can be simply reached with periodic boundary conditions. Again the size of the box $H$ (cube in our case) should be larger than $\ell$. In our simulations $H=0.1\,$cm, while $\ell$ varies from $0.005\,$cm to $0.04\,$, as seen in Fig.~\ref{f:5}.

One additional  simplifying  assumption made in our study (and many others) is that the flow  of the normal component is laminar. In numerical simulations (including ours) this  simply requires $V\sb n =$const. In experiments this is achieved to some extent in a core of a wide-channel counterflow, when $V\sb{ns}$ is below some critical value $V\sb {cr}$, above which the normal fluid flow is expected to become turbulent. Probably a better realization of the laminarity assumption in laboratory experiments is achieved in the ``pure" super flow, where normal fluid flow is prevented by super leaks, a kind of (e.g. silver) porous medium with sub-micron size pores to prevent a net flow of the viscous normal component through the channel on any experimentally relevant flow time scale, see e.g. Ref.~\cite{Ladik-2012}. Now, if one neglect the $V\sb n$ dependence on the (transverse) distance to the wall, the entire problem can be approximated as spatially homogeneous.

In order to relax the assumption of space homogeneity one has to develop a theory (or a model) of superfluid wall-bounded flow  which will find and account for an actual laminar super- and normal-fluid velocity profiles across a channel. This is still an open problem. Even more sophisticated and challenging open problem is a superfluid wall-bounded turbulence at large   counterflow velocities, when both the normal and the superfluid components are turbulent and their mean-velocity and turbulent-energy profiles have to be found self-consistently, accounting for the mutual friction between the components. Detailed information about the vortex tangle structure, found and analyzed in this paper, is required to successfully approach this problem. This was one of the important motivations for the present study.

\subsubsection{No isotropy, just axial symmetry}
   It is generally accepted that  the  classical hydrodynamic turbulence is almost isotropic at small scales $l \ll H$ due to the isotropization effect that is observed going from the outer scale $H$ toward the small scales $l$. The theory of small scale turbulence then simplifies. In the counterflow case there is no energy cascade and the superfluid counterfow turbulence is inherently  anisotropic due to the built-in direction of the counterflow velocity $\B V\sb{ns}$.   This anisotropy is of principal importance and cannot be ignored at all. Indeed, in the isotropic case there is no friction force between the normal- and superfluid components and the counterflow does not create a vortex tangle.
One can  formally see this from the following argument: consider the parameter $C\sb f$ that quantifies the mutual friction and $\gamma$ that determines the vortex tangle density (Sects.~\ref{sss:phen-an} and \ref{sss:fric}). Both are proportional to  the anisotropy parameter $I_\ell$, which is equal to zero in the isotropic  tangle.

Nevertheless, in a spatially homogeneous case with the  only relevant direction $\B V\sb{ns}$ one expects to see axial symmetry around  $\B V\sb{ns}$. Indeed, in our simulations the coefficient $I_{\ell\perp}$ [defined by Eq.\,\eqref{Ilp}], which is responsible for the axial asymmetry, is close to zero.

\subsubsection{The physical parameters of the problem}
\begin{description}
\item The main parameter in the problem of quantum turbulence is the circulation quantum $\kappa\approx 10^{-3}\,$cm/s$^{-2}$.
\item The second parameter is the vortex core radius $a_0$. In $^4$He $a_0\simeq 10^{-8}\,$cm.  In the theory of counterflow turbulence $a_0$ appears in the combination with the intervortex distance $\ell$ as a dimensionless parameter  $\Lambda\simeq \ln(\ell/a_0)$. More accurate definition of $\Lambda$ is given by Eq.~\eqref{eq:LIA}, where we also introduced $\widetilde \Lambda= \Lambda/(4\pi)$. The parameter $\widetilde \Lambda$ naturally appears in the equations  of motion  for the vortex line in the local induction approximation.  Table~\ref{t:1} shows that in actual experimental situations $\widetilde \Lambda$ is very close to unity.

\item Additional dimensionless parameters are $\alpha$ and $\alpha'$ which determine the mutual friction force [according to Eq.~\eqref{eq:s_Vel}]. Of the two $\alpha$ is more important, being responsible for the dissipative part. As one sees in Tab.~\ref{t:1}, $\alpha$ varies in the relevant temperature range by a favor of 8, being much smaller than unity ($\alpha=0.036$) at $T=1.3$ and approaching unity, when $T$ is close to $T_\lambda$, see Tab.~\ref{t:1}.

\item We have also to mention the ratio of normal and superfluid densities $\rho\sb s/\rho\sb n$. Having in mind that the total $^4$He density  $\rho\equiv \rho\sb s+\rho\sb n$ in the problem at hands can be considered as temperature independent we can use the ratio $\rho\sb n/\rho$ instead of $\rho\sb s/\rho\sb n$ . It varies  about  ten times (from 0.045 to 0.42, see Tab.~\ref{t:1}) in the studied temperature range.

\end{description}

Having so many dimensionless parameters that essentially deviate from unity and vary significantly with temperature, one may think that dimensional reasonings are useless in our problem. However, as we have shown in the paper, they are still useful. Being supplemented with simple physical arguments they give quite reasonable results, for example to determine the $V\sb{ns}$ dependence of the basic tangle characteristics, see below.

\subsection{\label{ss:Vns} $V\sb{ns}$-dependence of the vortex-tangle characteristics}
Using dimensional reasoning with the only parameter $\kappa$ we reproduced a set of relationships that determined the $V\sb{ns}$-dependence of the main characteristics of the vortex tangle. For concreteness, we list them in the order of increasing powers of $V\sb{ns}$ and remind the values of corresponding dimensionless parameters:
\begin{description}

\item The mean and RMS vortex line curvature, $\overline{ S}=c_1/\ell \propto V\sb{ns}^{-1}$, $\widetilde {S}=c_2/\ell \propto V\sb{ns}^{-1}$, see Eqs.~\eqref{curv}, Fig.~\ref{f:10} and Tab.~\ref{t:4}; $c_2 \approx \sqrt{3/2}\, c_1\simeq 2 \div 3$.

\item The anisotropy indices $I_\parallel$, $I_\perp$ $I_\ell$ and   $I_{\ell \perp}$ are practically independent of $V\sb{ns}$, i.e. $\propto V\sb{ns}^0$, see Eqs.~\eqref{aniz}, Fig.~\ref{f:9} and Tab.~\ref{t:4}; $I_\parallel$, $I_\perp\simeq 0.7 \div 0.9$,  $I_\ell\simeq 0.5$,
    $I_{\ell \perp}\approx 0$ (because of the axial symmetry).

\item The drift velocity  $V\sb{vt}= C\sb{vt} V\sb{ns}$,  see Eq.~\eqref{Kvt}, Fig.~\ref{f--drft}; $C\sb{vt}\simeq 0.05\div 0.08$.
\item Vortex line density  $\C L= \kappa^2 \,(\Gamma V\sb{ns})^2$, see Eq.~\eqref{gammaA},  Fig.~\ref{f:8} and Tab.~\ref{t--3};  $\Gamma\simeq 0.07  \div 0.16$.

\item The mutual friction force density, $ F\sb{ns} = \alpha\, \rho\sb s\,\kappa^{-1} (C_f V\sb{ns})^3$, see Eq.~\eqref{FnsB}, Fig.~\ref{f--fric}; The derivation of Eq.~\eqref{FnsB} was based not only on dimensions of $F\sb{ns}$ but also on its explicit expression~\eqref{FF} via configuration of the vortex tangle; $C_f\simeq 0.15 \div 0.25$.
\item The reconnection rate $   dN_r/dt= c_r \kappa \C L^{5/2}\propto V\sb {ns}^5$, see Eqs.~\eqref{eq:dNr} and \eqref{gammaA}, Fig.~\ref{f:7} and Tab.~\ref{t--2}; $c_r\simeq 0.4\div 0.6$.
\end{description}
As one sees in the figures mentioned above, the results of our numerical simulations agree well with all these expected $V\sb{ns}$ dependencies. Notice that the numerical values of the corresponding dimensionless parameters (presented above) are not always of the order of unity but often smaller by an order of magnitude. Some understanding of the reason for that can be obtained with the explicit form of the bridge relations of Schwartz's obtained using the local interaction approximation
as discussed next.

\subsection{\label{ss:Br} Schwartz's bridge relations}
Using the equation of motion~\eqref{eq:LIA} in the local interaction approximation Schwartz~\cite{Schwarz88} analytically  derived   the bridge relations
that connect some of the  parameters mentioned above. In our dimensionless notations these are:
\begin{description}
\item  $\displaystyle \Gamma \Sb  S  \equiv \kappa \gamma\Sb  S   \approx \frac{I_\ell}{\widetilde \Lambda c_2^2}$ [by comparing Eqs.~\eqref{gammaA} and~\eqref{gam19}].
\item Equation~\eqref{CLIA} for  $C\sb{vt}\Sp {LIA}$ that quantifies the tangle drift velocity $V\sb{vt}$ [see Eq.~\eqref{Kvt}].

\item Equation~\eqref{CFL} for $C_f\Sp {LIA}$ that quantifies the mutual friction force density $F\sb{ns}$ [see   Eq.~\eqref{FnsB}].

\end{description}
These equations bridge $\Gamma\Sb  S $, $C\sb{vt}\Sp {LIA}$ and $C\sb f\Sp {LIA}$ with the tangle anisotropy parameters $I_\parallel$, $I_\ell$, defined by Eqs.~\eqref{aniz} and tangle RMS curvature parameter $c_2$, defined by Eqs.~\eqref{curv}.

These bridge relations  are fulfilled with reasonable accuracy in our simulations (with the full Biot-Savart equation), especially for low temperatures.  For higher temperatures the discrepancy increases,   but overall
the order of magnitude of these coefficients is close to those calculated
by Schwarz (with $I_\parallel$, $I_\ell$ and $c_2$ obtained by LIA simulations) and agree with available experimental data.  This allows us to believe that the Local-Induction Approximation may be successfully used in analytical studies of counterflow turbulence in spite of the fact that it fails in numerical simulations.

\subsection{\label{ss:PDFs} Probability distribution and correlation functions in the vortex tangle}
Many of the mean parameters discussed above can be measured experimentally, at least in principle.   However detailed statistical information of the random vortex tangle statistics is  hardly expected in foreseeable experiments. Because of their importance for better understanding the basic physics of counterflow turbulence we put some efforts to clarify it numerically. In particular, we studied:
\begin{description}
\item The PDF of the vortex-loop length $\C P(l)$ and showed that its core (that contains about $20\div 30\%$ of the total loops) can be described by a simple exponential form~\eqref{PDFsA} as seen in Fig.~\ref{f:12}A. It has a peak at some $L_*\sim 0.01\div 0.02\,$cm, which is much smaller than the mean loop length $\overline L\sim 0.3\div 0.5\,$cm, defined by the tail of the PDF,  see   Fig.~\ref{f:11}.

\item The correlation between the length $l_j$ and the RMS curvature $\widetilde {s''}_j$ of loops with a given $l_j$ is demonstrated in Fig.~\ref{f-14}. We show that for long loops $\widetilde {s''}_j$ is practically independent of their length and close to the overall RMS curvature $\widetilde S$, while for short loops its bounded from below by (and concentrated close to) the curvature of a circle with a given length.

\item The PDF of the line curvature $\C P(|s''|)$. We show that $\C P(|s''|)$ may be well described by an exponential form~\eqref{PDFsB} without fitting parameters, just involving the RMS curvature $\widetilde S$. This allowed us to find the ratio between the structural parameters $c_1/c_2\approx \sqrt{2/3}$.

\item The PDFs of the mean and the RMS loop curvature. We show that more that 90\% of the PDF  are close to a Gaussian form~\eqref{PDF} and studied in Sec.~\ref{ss:PDF} the temperature dependence of their maxima and widths.

\item Last but not least: the characteristics of the vortex tangle in the form of the autocorrelation function of the vortex orientation $K(\B r)$, defined by Eq.~\eqref{Kr}. Figures~\ref{f:16} shows that $K(\B r)$ practically vanishes at distances about  $\ell$.  This means that vortex lines is reminiscent of a random walk with a correlation length of the order of the intervortex distance. This fact has many important consequences, e.g. for the energy spectra in counterflow turbulence.
\end{description}

\subsection{ \label{ss:Res} Dynamical and statistical characteristics vs. reconnection criteria}
We carried out full Biot-Savart simulations of the evolution of the
vortex tangle within the vortex filament method in a wide range of parameters. We
compared the statistical and geometrical properties of the
dense tangles  using three different reconnection criteria
(geometrical G, geometrical-energetic GE and dynamical D) and identified
which properties are robust and which are sensitive to the choice of the criterion.
We found the following:
\begin{description}
\item The reconnection rate is a property directly related to the choice of
the criterion. We concluded  that the reconnection rate is similar for
{\bf GE} and {\bf D} criteria albeit their different physical
interpretations. On the other hand, in simulations with GC  the
reconnection rate is significantly higher. The detailed analysis
shows that most of the reconnections according to GC
lead to an increase of the total length and to the creation of a very large
number of small loops and loop fragments. The small loops removal
procedure is therefore an essential part of the algorithm for this criterion.

\item One of the main parameters, $\C L$, depends on the choice of the
reconnection criterion for high temperature and strong counterflow
velocities, when the tangle become dense. GC  lead to
sparser tangle, while DC  gives the most dense tangle for the
same $T$ and $V_{\rm ns}$. As a consequence, the coefficient $\gamma$
differs beyond measurement errors. Our results for $\gamma$ agree
well with available data. Thus the sensitivity of $\C L$ to the choice of
the reconnection criterion may explain the spread of the results for
$\gamma$ as found in literature.

\item In agreement with previous studies we found that the vortex tangle is oblate and
isotropic in the direction perpendicular to the
counterflow.  We observed a slight $V_{\rm ns}$ dependence of $I_{||}$
for $T=1.6$ and $1.9$~K for {\bf G} and {\bf GE} criteria. For all
temperatures the tangle was most oblate for GC  and
least oblate for DC.

 \item
 The tangle drift velocity and the mutual friction force density depend on the choice of the reconnection criterion at moderate and high temperatures, with the corresponding coefficients being largest for DC and smallest for GC. This is similar to the behavior of the vortex line density.

\item
The local tangle structure -- the mean and RMS curvature of the tangle as well as PDFs of the loops length and curvatures -- are only slightly dependent on the reconnection criterion.
\item
The autocorrelation of the vortex orientation is  practically independent of the choice of the reconnection criterion.
\end{description}
Despite some clear differences in some results obtained with different
reconnection criteria, vortex filament methods may be considered robust and well suited
for the description of the steady state vortex tangle in the counterflow
provided the results are interpreted having in mind the found values of the
spread due to particular details of implementation.\\~

\begin{center}
  ***
\end{center}

We   believe  that the numerical results obtained in this paper and their  analysis  will help in further studies of counterflow turbulence.

\acknowledgements
This paper had been supported in part by the Minerva Foundation, Munich, Germany and  by the grant 13-08-00673 from RFBR (Russian Foundation of Fundamental Research). LK  acknowledges the kind hospitality at the Weizmann Institute of Science during the main part of the project. We  are grateful to W.F. Vinen and L. Skrbek for their important comments, criticism  and suggestions. We also thank  S.~K.~Nemirovskii, N.~J.~Zabusky and P.~Mishra for their comments on the manuscript.

\end{document}